\documentclass[a4paper, 11pt]{article}

\usepackage{natbib}

\usepackage{graphicx}
\usepackage{amsfonts}
\usepackage{latexsym}
\usepackage{amsmath, amsthm, amssymb}
\usepackage{setspace}
\usepackage{url}
\usepackage{multirow}
\usepackage{color}
\usepackage[margin=1in]{geometry}
\usepackage{epsfig}
\usepackage{psfrag}

\newtheorem{thm}{Theorem}
\newtheorem{cond}{Condition}

\newtheorem{lem}{Lemma}

\newcommand{\cS}{\mathcal{S}}
\newcommand{\cA}{\mathcal{A}}
\newcommand{\cN}{\mathcal{N}}
\newcommand{\cC}{\mathcal{C}}
\newcommand{\cJ}{\mathcal{J}}
\newcommand{\cK}{\mathcal{K}}
\newcommand{\cE}{\mathcal{E}}

\newcommand{\cD}{\mathcal{D}}

\newcommand{\ep}{\epsilon}
\newcommand{\lam}{\lambda}
\newcommand{\bep}{\boldsymbol{\ep}}

\newcommand{\bA}{\mathbf{A}}

\newcommand{\bX}{\mathbf{X}}
\newcommand{\bZ}{\mathbf{Z}}
\newcommand{\by}{\mathbf{y}}
\newcommand{\bx}{\mathbf{x}}
\newcommand{\bz}{\mathbf{z}}
\newcommand{\bbeta}{\boldsymbol{\beta}}
\newcommand{\bhbeta}{\boldsymbol{\hat{\beta}}}

\newcommand{\bSig}{\boldsymbol{\Sigma}}

\newcommand{\R}{\mathbb{R}}

\newcommand{\sig}{\sigma}
\newcommand{\lone}{\mathit{l}_1}
\newcommand{\ltwo}{\mathit{l}_2}
\newcommand{\E}{\mathbb{E}}
\newcommand{\var}{\mbox{var}}
\newcommand{\cov}{\mbox{cov}}

\newcommand{\p}{\mathbb{P}}
\newcommand{\sgn}{\mbox{sign}}
\newcommand{\bzero}{\mathbf{0}}

\newcommand{\spc}{\hat{\rho}_n}
\newcommand{\ppc}{\rho_n}

\newcommand{\bC}{\mathbf{C}}

\newcommand{\bI}{\mathbf{I}}
\newcommand{\pin}{\pi_n}
\newcommand{\cCj}{\cC_{j}}
\newcommand{\cCja}{\cC_{j|\cA}}
\newcommand{\Pij}{\Pi_{j}}
\newcommand{\tXj}{\mathbf{\tilde{X}}_{j}}

\newcommand{\aj}{a_j}
\newcommand{\ajy}{a_{jy}}
\newcommand{\xjs}{X_j^*}
\newcommand{\xjo}{X_j^\circ}
\newcommand{\xjb}{X_j^\bullet}
\newcommand{\yjs}{\by_j^*}
\newcommand{\yjo}{\by_j^\circ}
\newcommand{\cjs}{c_j^*}
\newcommand{\lamj}{\lambda_j}
\newcommand{\Lamj}{\Lambda_j}
\newcommand{\bu}{\mathbf{u}}
\newcommand{\bv}{\mathbf{v}}
\newcommand{\bw}{\mathbf{w}}

\title{High-dimensional variable selection via tilting}
\author{
Haeran Cho and Piotr Fryzlewicz
\thanks{Department of Statistics, Columbia House, London School of Economics, Houghton Street, London, WC2A 2AE, UK.
\protect \ E-mail: {\tt \{h.cho1, p.fryzlewicz\}@lse.ac.uk}
}
}

\date{}

\begin{document}

\maketitle

\begin{abstract}

This paper considers variable selection in linear regression models where
the number of covariates is possibly much larger than the number of observations.
High dimensionality of the data brings in many complications, such as 
(possibly spurious) high correlations among the variables,
which result in marginal correlation being unreliable as a measure of association 
between the variables and the response.
We propose a new way of measuring the contribution of each variable to the response
which takes into account high correlations among the variables in a data-driven way.
The proposed {\em tilting} procedure provides an adaptive choice between the use of 
marginal correlation and {\em tilted correlation} for each variable,
where the choice is made depending on the values of the 
hard-thresholded sample correlation of the design matrix.
We study the conditions under which this measure can successfully discriminate 
between the relevant and the irrelevant variables and thus be used as a tool for variable selection.
Finally, an iterative variable screening algorithm is constructed 
to exploit the theoretical properties of tilted correlation,
and its good practical performance is demonstrated in a comparative simulation study.

\end{abstract}

\vspace{10pt} \textbf{keywords:} variable selection, correlation, high-dimensional linear regression

\section{Introduction}
\label{sec:intro}

Inferring the relationship between the response and the explanatory variables in linear models is an extremely important
and widely studied statistical problem, from the point of view of both practical applications and theory. In this work, 
we consider the following linear model:
\begin{eqnarray}
\by=\bX\bbeta+\bep, \label{lp}
\end{eqnarray}
where $\by=(y_1, \ldots, y_n)^T \in \R^n$ is an $n$-vector of the response,  $\bX=\left(X_1, \ldots, X_p\right)$ is 
an $n \times p$ design matrix and  $\bep=(\ep_1, \ldots, \ep_n)^T \in \R^n$ is an $n$-vector of i.i.d. random errors.

Recent technological advances have led to the explosion of data across many scientific disciplines, where the 
dimensionality of the data $p$ can be very large; examples can be found in genomics, functional MRI, tomography 
and finance, to name but a few. In such settings, difficulties arise in estimating the coefficient vector $\bbeta$.
Over the last two decades, substantial progress has been made in tackling this problem under the assumption that 
only a small number of variables actually contribute to the response, i.e., $\cS=\{1\le j \le p: \  \beta_j \ne 0\}$ 
is of cardinality $|\cS|\ll p$. By identifying $\cS$, we can improve both model interpretability and estimation
accuracy.

There exists a long list of literature devoted to the high-dimensional variable selection problem  
and an exhaustive survey can be found in \citet{fan2009}. 
The Lasso \citep{tibshirani1996} belongs to a class of penalised least squares estimators 
where the penalty is on the $\lone$-norm of $\bbeta$, 
which leads to a sparse solution by setting certain coefficients to be exactly zero. 
It has enjoyed considerable attention and substantial efforts in studying the consistency of the methodology and its extension can be found e.g. in \citet{meinshausen2006}, \citet{zhang2008}, \citet{zhao2006}, \citet{zou2006}, \citet{meinshausen2008}.

\citet{efron2004} proposed the Least Angle Regression (LARS) algorithm, 
which can be modified to compute the Lasso solution path for a range of penalty parameters.
The main criterion for determining which variables should enter the model in the progression of the
LARS algorithm is the screening of the marginal correlations between each variable and the current residual.
That is, denoting the current residual by $\bz$, the Lasso solution path is computed by taking a step 
of a suitably chosen size in the equiangular direction  between those variables which achieve the maximum $|X_j^T\bz|$ at 
each iteration. The Sure Independence Screening (SIS) proposed in \citet{fan2008} is a dimension 
reduction procedure, which screens the marginal correlations $X_j^T\by$ to choose which variables 
should remain in the model.

While the aforementioned methods show good theoretical properties as well as performing well in practice, 
we note that they heavily rely on marginal correlation to measure the strength of association between $X_j$ and $\by$.
\citet{fan2008} observed that, even when $X_1, \ldots, X_p$ were generated as i.i.d. Gaussian variables,
there might exist spurious correlations among the variables with growing dimensionality $p$. 
In general, when there are non-negligible correlations among the variables, whether spurious or not,
an irrelevant variable ($X_j, \ j\not\in\cS$) can have large marginal correlation with $\by$ 
due to its association with the relevant variables ($X_j, \ j\in\cS$), 
which implies that marginal correlation can be misleading, especially if $p$ is large.

There have been some efforts to introduce new measures of association between each variable and the
response in order to deal with the issue of high correlations among the variables.
\citet{buhlmann2009} proposed the PC-simple algorithm, which uses partial correlation 
in order to infer the association between each variable and the response conditional on other variables.
Also, we note that ``greedy'' algorithms such as the traditional forward selection 
(see e.g. Chapter 8.5 of \citet{weisberg1980}) or the forward regression \citep{wang2009} 
have an interpretation in this context due to their greediness 
(in the sense that the locally optimal choice is made at each iteration),
unlike less greedy algorithms generating a solution path, e.g. LARS. 
At each iteration, both forward selection and forward 
regression algorithms update the current residual $\bz$ by taking the greediest step towards the 
variables included in the current model, i.e., $\bz$ is obtained by projecting $\by$ onto the orthogonal 
complement of the current model space and this greedy progression can be seen as taking into account 
the correlations between those variables which are in the current model and those which are not. 
\citet{radchenko2011} proposed the forward-Lasso adaptive shrinkage (FLASH) which includes
the Lasso and forward selection as special cases at two extreme ends.
FLASH iteratively adds one variable at a time and adjusts each step size by introducing a 
new parameter so that their procedure is greedier than the Lasso, yet not as greedy as the forward selection. 
The regression framework proposed in \citet{witten2009} accounts for correlations among the variables 
using the so-called ``scout'' procedure, which obtains a shrunken estimate of the inverse covariance matrix of $\bX$
by maximising a penalised likelihood and then applies it to the estimation of $\bbeta$.
A more detailed description of the aforementioned methods, in comparison with our proposed methodology, 
is provided later in Section \ref{sec:relation}.

In this paper, we propose a new way of measuring the contribution of each variable to the response, 
which also accounts for the correlation structure among variables.
It is accomplished by ``tilting'' each column $X_j$ (so that it becomes $\xjs$) such that 
the impact of other variables $X_k, \ k\ne j$ on the ``tilted'' correlation between $\xjs$ and $\by$
is reduced and thus the relationship between the $j$th covariate and the response
can be identified more accurately. One key ingredient of this methodology, which sets it apart from other
approaches listed above, is the adaptive choice of the set $\cCj$ of variables $X_k$ whose
impact on $X_j$ is to be removed. Informally speaking, we note that 
$\cCj$ cannot include ``too many'' variables, as this would distort the 
association between the $j$th covariate and the response due to the large 
dimensionality $p$. However, we also observe that those $X_k$'s 
which have low marginal correlations with $X_j$ do not individually cause distortion in measuring 
this association anyway, so they can safely be omitted from the set $\cCj$.
Therefore, it appears natural to include in $\cCj$ only those variables $X_k$
whose correlations with $X_j$ exceed a certain threshold in magnitude,
and this hard thresholding step is an important element of our methodology.

Other key steps in our methodology are: projection of each variable onto a subspace chosen 
in the hard-thresholding step; and rescaling of such projected variables.  
We show that under certain conditions the tilted correlation can successfully discriminate between
relevant and irrelevant variables and thus can be applied as a tool for variable selection.
We also propose an iterative algorithm based on tilting and present its unique features in 
relation to the existing methods discussed above. 

The remainder of the paper is organised as follows.
In Section \ref{sec:tilt}, we introduce the tilting procedure and study 
the theoretical properties of tilted correlation in various scenarios.
Then, in Section \ref{sec:application}, we propose the TCS algorithm, 
which iteratively screens the tilted correlations to identify relevant variables, 
and compare it in detail to other existing methods. Section \ref{sec:sim} reports the 
outcome of extensive comparative simulation studies and 
the performance of TCS algorithm is further demonstrated in Section \ref{sec:boston}
on a real world dataset predicting real estate prices.
Section \ref{sec:conclusion} concludes the paper and the proofs of theoretical results are in the Appendix.

\section{Tilting: motivation, definition and properties}
\label{sec:tilt}

\subsection{Notation and model description}
\label{sec:notation}

For an $n$-vector $\bu\in\R^n$, we define the $\lone$ and $\ltwo$-norms as 
$\Vert\bu\Vert_1=\sum_{j}|u_j|$ and $\Vert\bu\Vert_2=\sqrt{\sum_{j}u_j^2}$, 
and the latter is frequently referred to as the norm.
Each column of $\bX$ is assumed to have a unit norm, 
and thus the sample correlation matrix of $\bX$ is defined as $\bC=\bX^T\bX=(c_{j,k})_{j,k=1}^p$.
We assume that $\ep_i, \ i=1, \ldots, n$ are i.i.d. random noise following a normal 
distribution $\cN(0, \sig^2/n)$ with $\sig^2<\infty$, where the $n^{-1}$ in the noise
variance is required due to our normalisation of the columns of $\bX$.
We denote the $i$th row of $\bX$ as $\bx_i=(X_{i,1}, \ldots, X_{i, p})$.
Let $\cD$ denote a subset of the index set $\cJ=\{1, \ldots, p\}$. 
Then $\bX_\cD$ denotes an $n\times |\cD|$-submatrix of $\bX$ with $X_j, \ j\in\cD$ as its columns 
for any $n\times p$ matrix $\bX$.
In a similar manner, $\bbeta_\cD$ denotes a $|\cD|$-subvector of a $p$-vector $\bbeta$ 
with $\beta_j, \ j\in\cD$ as its elements.
For a given submatrix $\bX_\cD$, we denote the projection matrix onto the column space of $\bX_\cD$ by $\Pi_\cD$.
Finally, $C$ and $C'$ are used to denote generic positive constants.

\subsection{Tilting: motivation and definition}
\label{sec:motv:def}

In this section, we introduce the procedure of tilting a variable 
and define the {\em tilted correlation} between each variable and the response. 
We first list typical difficulties encountered in high-dimensional problems, 
which were originally pointed out in \citet{fan2008}.
\begin{itemize}
\item[(a)] Irrelevant variables which are highly correlated with the relevant ones 
can have high priority to be selected in marginal correlation screening. 
\item[(b)] A relevant variable can be marginally uncorrelated but jointly correlated with the response.
\item[(c)] Collinearity can exist among the variables, 
i.e., $|c_{j, k}|=|X_j^TX_k|$ for $j\ne k$ can be close to 1.
\end{itemize}

We note that the marginal correlation between each variable $X_j$ and $\by$ has the following decomposition,
\begin{eqnarray}
X_j^T\by=X_j^T\left(\sum_{k=1}^p\beta_kX_k+\bep\right)=
\beta_j+\underline{\underline{\sum_{k\in\cS\setminus\{j\}}\beta_kX_j^TX_k}}+X_j^T\bep,
\label{marginal:corr}
\end{eqnarray}
which shows that the issues (a) and (b) arise from the underlined summand in (\ref{marginal:corr}).
The main idea behind tilting is to transform each $X_j$ in such a way that the corresponding
underlined summand for the transformed $X_j$ is zero or negligible, while not distorting
the contribution of the $j$th covariate to the response. By examining the form of the
underlined summand and viewing it as a ``bias'' term, it is apparent that its components
are particularly large for those $k$'s for which the corresponding term $X_j^TX_k$ is large.
If we were to transform $X_j$ by projecting it on the space orthogonal to those $X_k$'s,
a corresponding bias term for a thus-transformed $X_j$ would be significantly reduced.

For each $X_j$, denote the set of such $X_k$'s by $\cCj$. Without prior knowledge of $\cS$, 
one way of selecting $\cCj$ for each $X_j$ is to identify those variables $X_k, \ k\ne j$ which have 
non-negligible correlations with $X_j$. A careful choice of $\cCj$ is especially important when the 
dimensionality $p$ is high; when $\cCj$ is chosen to include too many variables, 
any vector in $\R^n$ may be well approximated by $X_k, \ k\in\cCj$,
which would result in the association between the transformed $X_j$ and $\by$
failing to reflect the true contribution of the $j$th covariate to the response.
Intuitively, those $X_k$'s having small sample correlations with $X_j$ do not significantly
contribute to the underlined bias term, and thus can be safely omitted from the set
$\cCj$. Below, we propose a procedure for selecting $\cCj$ adaptively for each $j$,
depending on the sample correlation structure of $\bX$.

We first find $\pin\in(0, 1)$ which will act as a threshold on 
each off-diagonal entry $c_{j, k}, \ j\ne k$ of the sample correlation matrix $\bC$
of $\bX$, identifying whether the sample correlation between $X_j$ and $X_k$ is non-negligible.
Then, the subset $\cCj$ is identified as
$\cCj=\{k\ne j: \  |X_j^TX_k|=|c_{j, k}|>\pin\}$ separately for each variable $X_j$.
We note that although the subset $\cCj$ is obviously different for each $j$, the
thresholding procedure for selecting it is always the same.
Our procedure for selecting $\pi_n$ itself is described in
Section \ref{sec:choice:thr}.
Tilting a variable $X_j$ is defined as the procedure of projecting $X_j$ onto 
the orthogonal complement of the space spanned by $X_k, \ k\in\cCj$,
which reduces to zero the impact of those $X_k$'s on the association 
between the projected version of $X_j$ and $\by$.

Hard-thresholding was previously adopted for the estimation of a high-dimensional covariance matrix,
although we emphasise that this was not in the context of variable selection. 
In \citet{bickel2008}, an estimator obtained by hard-thresholding the sample covariance matrix 
was shown to be consistent with the choice of $C\sqrt{\log{p}/n}$ as the threshold, 
provided the covariance matrix was appropriately sparse and the dimensionality $p$ satisfied $\log{p}/n\to 0$.
A similar result was reported in \citet{el2008} with the threshold of magnitude $Cn^{-\gamma}$ for some $\gamma\in(0, 1/2)$.
Our theoretical choice of threshold $\pin$ is described in Section \ref{sec:prop:tilt},
where we also briefly compare it to the aforementioned thresholds. 
In practice, we choose $\pin$ by controlling the false discovery rate, as presented in 
Section \ref{sec:choice:thr}.

Let $\tXj$ denote a submatrix of $\bX$ with $X_k, \ k\in\cCj$ as its columns, 
and $\Pij$ the projection matrix onto the space spanned by $X_k, \ k\in\cCj$, 
i.e., $\Pij\equiv\tXj(\tXj^T\tXj)^{-1}\tXj^T$.
The tilted variable $\xjs$ for each $X_j$ is defined as $\xjs\equiv(\bI_n-\Pij)X_j$. 
Then the correlation between the tilted variable $\xjs$ and $X_k, \ k\in\cCj$ is reduced to zero, 
and therefore such $X_k$'s no longer have any impact on $X_j^{*T}\by$.
However, $X_j^{*T}\by$ cannot directly be used as a measure of association between $X_j$ and $\by$, 
since the norm of the tilted variable $\xjs$, provided $\cCj$ is non-empty, satisfies 
$\Vert\xjs\Vert_2=X_j^T(\bI_n-\Pij)X_j<X_j^TX_j=1$.
Therefore, we need to rescale $X_j^{*T}\by$ so as to make it a reliable criterion for 
gauging the contribution of each $X_j$ to $\by$.

Let $\aj$ and $\ajy$ denote the squared proportion of $X_j$ and $\by$ (respectively) represented by $X_k, \ k\in\cCj$, i.e.,
$\aj\equiv\Vert\Pij X_j\Vert_2^2/\Vert X_j\Vert_2^2$ and $\ajy\equiv\Vert\Pij\by\Vert_2^2/\Vert\by\Vert_2^2$.
We denote the tilted correlation between $X_j$ and $\by$ with respect to a rescaling factor $s_j$ by
$\cjs(s_j) \equiv s_j^{-1}\cdot X_j^{*T}\by$, and propose two rescaling rules below.
\begin{description}
\item[Rescaling 1.] 
Decompose $X_j^{*T}\by$ as
\begin{eqnarray}
X_j^{*T}\by&=&X_j^T(\bI_n-\Pij)\by
=X_j^T\left\{\sum_{k=1}^p\beta_k(\bI_n-\Pij)X_k+(\bI_n-\Pij)\bep\right\} 
\nonumber \\
&=&\beta_jX_j^T(\bI_n-\Pij)X_j+\sum_{k\in\cS\setminus\cCj, k\ne j}\beta_kX_j^T(\bI_n-\Pij)X_k+X_j^T(\bI_n-\Pij)\bep.
\label{decom:tilt:corr}
\end{eqnarray}
Provided the second and third summands in (\ref{decom:tilt:corr}) are negligible in comparison with the first, 
rescaling the inner product $X_j^{*T}\by$ by $1-\aj=X_j^T(\bI_n-\Pij)X_j$ 
can ``isolate'' $\beta_j$, which amounts to the contribution of $X_j$ to $\by$, in the sense that
$X_j^{*T}\by / (1-\aj)$ can be represented as $\beta_j$ plus a ``small'' term
(our theoretical results later make this statement more precise).
Motivated by this, we use the rescaling factor of $\lamj\equiv(1-\aj)$ to define
a rescaled version of $\xjs$ as $\xjb\equiv(1-\aj)^{-1}\cdot\xjs$ 
and the corresponding tilted correlation as 
$\cjs(\lamj)=(1-\aj)^{-1}\cdot X_j^{*T}\by=X_j^{\bullet T}\by$.

\item[Rescaling 2.] 
Since $\bI_n-\Pij$ is also a projection matrix, 
we note that $X_j^{*T}\by$ is equal to the inner product between $\xjs=(\bI_n-\Pij)X_j$ and $\yjs=(\bI_n-\Pij)\by$,
with their norms satisfying $\Vert\xjs\Vert_2=\sqrt{1-\aj}$ and $\Vert\yjs\Vert_2=\sqrt{1-\ajy}\cdot\Vert\by\Vert_2$.
By rescaling $\xjs$ and $\yjs$ by $\sqrt{1-\aj}$ and $\sqrt{1-\ajy}$ respectively,
we obtain vectors $\xjo\equiv(1-\aj)^{-1/2}\cdot\xjs$ and $\yjo\equiv(1-\ajy)^{-1/2}\cdot\yjs$, whose
norms satisfy $\|\xjo\|_2 = \|X_j\|_2$ and $\|\yjo\|_2 = \|\by\|_2$.
Therefore, with the rescaling factor set equal to $\Lamj\equiv\{(1-\aj)(1-\ajy)\}^{1/2}$, 
we define the tilted correlation as $\cjs(\Lamj)=\{(1-\aj)(1-\ajy)\}^{-1/2}\cdot X_j^{*T}\by=X_j^{\circ T}\yjo$.
\end{description}

We note that, with the rescaling factor $\lamj$ (rescaling 1), 
the tilted correlation $\cjs(\lamj)$ coincides with the ordinary least squares estimate of $\beta_j$
when regressing $\by$ onto $X_k, \ k\in\cCj\cup\{j\}$.
When rescaled by $\Lamj$ (rescaling 2), the tilted correlation coincides with 
the sample partial correlation between $X_j$ and $\by$ given $X_k, \ k\in\cCj$ (denoted by $\spc(j, \by|\cCj)$), 
up to a constant multiplicative factor $\Vert\by\Vert_2$,
i.e., $\cjs(\Lamj)=\Vert\by\Vert_2\cdot\spc(j, \by|\cCj)$.
Although partial correlation is also used in the PC-simple algorithm \citep{buhlmann2009},
we emphasise that a crucial difference between tilting and PC-simple is that 
tilting makes an adaptive choice of the conditioning subset $\cCj$ for each $X_j$,
as described earlier in this section. For a detailed discussion of this point, see 
Section \ref{sec:relation}. In what follows, whenever the tilted correlation is denoted
by $\cjs$ without specifying the rescaling factor $s_j$, the relevant statement is valid 
for either of the rescaling factors $\lamj$ and $\Lamj$.

Finally, we note that if the set $\cCj$ turns out to be empty for a certain index $j$, 
then for such $X_j$, our tilted correlation with either rescaling factor 
would reduce to standard marginal correlation, which in this case is expected to work well 
(in measuring the association between the $j$th covariate and the response) 
due to the fact that no other variables $X_k$ are significantly correlated with $X_j$.
In summary, our proposed tilting procedure enables an adaptive choice
between the use of marginal correlation and tilted correlation for each variable $X_j$,
depending on the sample correlation structure of $\bX$.

In the following section, we study some properties of tilted correlation and show
that the corresponding properties do not always hold for marginal correlation. This  
prepares the ground for the algorithm proposed in Section \ref{sec:algorithm} 
which adopts tilted correlation for variable screening.

\subsection{Properties of the tilted correlation}
\label{sec:prop:tilt}

In studying the theoretical properties of tilted correlation, 
we make the following assumptions on the linear model in (\ref{lp}).
\begin{itemize}
\item[(A1)] The number of non-zero coefficients $|\cS|$ satisfies $|\cS|=O(n^{\delta})$ for $\delta\in[0, 1/2)$.
\item[(A2)] The number of variables satisfies $\log p=O(n^{\theta})$ with 
$\theta\in[0, 1-2\gamma)$ for $\gamma\in(\delta, 1/2)$.
\item[(A3)] With the same $\gamma$ as in (A2), the threshold is chosen as $\pin=C_1n^{-\gamma}$ for some $C_1>0$.
We assume that there exists $C>0$ such that $\cCj=\{k\ne j: \  |c_{j, k}|>\pin\}$ is of cardinality 
$|\cCj|\le Cn^\xi$ uniformly over all $j$, where $\xi\in[0, 2(\gamma-\delta))$.
\item[(A4)] Non-zero coefficients satisfy $\max_{j\in\cS}|\beta_j|<M$ for $M\in(0, \infty)$ and
$n^\mu\min_{j\in\cS}|\beta_j|\to\infty$ for $\mu\in[0, \gamma-\delta-\xi/2)$.
\item[(A5)] There exists $\alpha\in(0, 1)$ satisfying $1-X_j^T\Pij X_j=1-\aj > \alpha$ for all $j$.
\item[(A6)] For those $j$ whose corresponding $\cCj$ satisfies $\cS\nsubseteq\cCj$, we have
\[\
n^\kappa\cdot \frac{\Vert(\bI_n-\Pij)\bX_\cS\bbeta_\cS\Vert_2^2}{\Vert\bX_\cS\bbeta_\cS\Vert_2^2}\to\infty,
\]
for $\kappa$ satisfying $\kappa/2+\mu\in[0, \gamma-\delta-\xi/2)$.
\end{itemize}

In (A1) and (A2), we let the sparsity $|\cS|$ and dimensionality $p$ of the linear model grow with the sample size $n$.
Intuitively, if some non-zero coefficients tend to zero too rapidly, 
identifying them as relevant variables is difficult. 
Therefore (A4) imposes a lower bound on the magnitudes of the non-zero coefficients,
which still allows the minimum non-zero coefficient to decay to 0 as $n$ grows.
It also imposes an upper bound, which is needed to ensure that the ratio
between the largest and smallest coefficients in absolute value does not
grow too quickly with $n$.

We now clarify the rest of assumptions which are imposed on the correlation structure of $\bX$, 
and compare them to related conditions in existing literature. 
It is common practice in high-dimensional variable selection literature 
to study the performance of proposed methods under some conditions on $\bX$. 
For the Lasso, it was shown that the irrepresentable condition \citep{zhao2006}, 
also referred to as the neighbourhood stability condition \citep{meinshausen2006} on $\bX$ 
was sufficient and almost necessary for consistent variable selection. 
This condition required that 
\begin{eqnarray*}
\max_{j\notin\cS}\left\vert\sgn(\bbeta_\cS)^T(\bX_\cS^T\bX_\cS)^{-1}\bX_\cS^TX_j\right\vert<1,
\end{eqnarray*}
which can roughly be interpreted as saying that the portion of the irrelevant variable 
$X_j, \ j\notin\cS$, represented by relevant variables $\bX_\cS$ is bounded from above by 1.
\citet{zhang2008} showed the variable selection consistency of Lasso under the sparse Riesz condition.
It requires the existence of $C>0$ for which the eigenvalues of $\bX_\cD^T\bX_\cD$ are bounded 
uniformly over any $\cD\subset\cJ$ with $|\cD|\le C|\cS|$.
\citet{candes2007} showed the consistency of the Dantzig selector under the uniform uncertainty principle (UUP), 
which also similarly restricts the behaviour of the sparse eigenvalues of $\bX_\cD^T\bX_\cD$. 

We note that the assumption (A3) is not directly comparable to the above conditions
in the sense that it requires the number of highly correlated variables for each variable not to exceed
a certain polynomial rate in $n$. This bound is needed in order to guarantee the existence of the 
projection matrix $\Pij$, as well as to prevent tilted correlations from being distorted by high dimensionality
as explained in Section \ref{sec:motv:def}. 
We now give an example of when (A3) is satisfied. 
Suppose for instance that each observation $\bx_i, \ i=1, \ldots, n$ is independently 
generated from a multivariate normal distribution $\cN_p(\bzero, \Sigma)$ with 
$\Sigma_{j, k}=\varphi^{|j-k|}$ for some $\varphi\in(-1, 1)$.
Then using Lemma 1 in \citet{kalisch2007}, we have that
\begin{eqnarray}
\p\left(\max_{j\ne k}\left\vert c_{j, k}-\Sigma_{j, k}\right\vert
\le C_2n^{-\gamma}\right)\ge 1-\frac{Cnp(p-1)}{2}\cdot\exp\left(-\frac{C_2(n-4)n^{-2\gamma}}{2}\right),
\label{dist:sample:pop}
\end{eqnarray}
for some $C_2\in(0, C_1)$ and $C>0$. 
The right-hand side of (\ref{dist:sample:pop}) tends to 1, provided 
$\log\,p = O(n^\theta)$ with $\theta\in[0, 1/2-\gamma)$. 
Then (A3) holds with probability tending to 1 since $|c_{j, k}| \le |\varphi|^{|j-k|}+C_2n^{-\gamma}<\pin$ for $|j-k|\gg\log n$ ($|a_n| \gg |b_n|$ means $|a_n b_n^{-1}| \to \infty$).
The choice of $\pin=C_1n^{-\gamma}$ is in agreement with \citet{bickel2008} and \citet{el2008}
in the sense that their threshold is also greater than $n^{-1/2}$.
However, as we describe in Section \ref{sec:choice:thr}, our procedure requires
a data-dependent, rather than a fixed threshold, and we propose to choose it
by controlling the false discovery rate.

(A5) is required to rule out strong collinearity among the variables. 
From the fact that $1 - \aj = \det\left(\bX_{\cCj\cup\{j\}}^T\bX_{\cCj\cup\{j\}}\right)/\det\left(\tXj^T\tXj\right)$, 
we can find a connection between (A5) and the condition requiring 
strict positive definiteness of the population covariance matrix of $\bX$,
which is often found in the variable selection literature including \citet{fan2001}, \citet{buhlmann2009} and \citet{zou2006}.

Further, we show in Appendix \ref{append:one} that assumptions (A5) and (A6) are satisfied
under a certain mild assumption on $\bX$ and $\bep$, also used e.g. in \cite{wang2009}.

As far as variable selection is concerned, 
if the absolute values of tilted correlations for $j\in\cS$ are markedly larger than 
those for $j\notin\cS$, we can use the tilted correlations for the purpose of variable screening. 
Before studying the properties of the tilted correlation in details, 
we provide a simple example to throw light on the situations where tilted correlation screening is successful while marginal correlation is not. 
The following set-up is consistent with Condition \ref{cond:three} in Section \ref{sec:scen:one}:
$p=3$, $\cS=\{1, 2\}$, noise is not present, $|c_{1, 3}|$ and $|c_{2, 3}|$ exceed the threshold.
Then, even when $c_{1, 2}, c_{1, 3}, c_{2, 3}$ and the non-zero coefficients $\beta_1, \beta_2$ 
are chosen so that the marginal correlation screening fails 
(i.e., $|X_3^T\by|>\max(|X_1^T\by|, |X_2^T\by|)$), it is still the case that
$|(X_3^*)^T\by|=0$ and thus tilted correlation screening can avoid picking up $X_3$ as relevant.

In the following Sections \ref{sec:scen:one}--\ref{sec:scen:three}, 
we introduce different scenarios under which the tilted correlation screening 
(with either rescaling factor) achieves separation between relevant and irrelevant variables.

\subsubsection{Scenario 1}
\label{sec:scen:one}

In the first scenario, we assume the following condition on $\bX$. 
\begin{cond}
There exists $C>0$ such that $\left\vert(\Pij X_j)^TX_k\right\vert \le Cn^{-\gamma}$
for all $j \in \cJ$ and $k\in\cS\setminus\cCj,\  k\ne j$.
\label{cond:one}
\end{cond}
This condition implies that when $X_j$ is projected onto the space spanned by $X_l, \ l\in\cCj$,
any $X_k\in\cS$ which are not close to $X_j$ (in the sense that $k\notin\cCj$) 
remain not ``too close'' to the projected $X_j$ ($\Pij X_j$).
In Appendix \ref{appendix:one:ex}, it is shown that Condition \ref{cond:one} holds asymptotically
when each column $X_j$ is generated independently as a random vector on a sphere of radius 1,
which is the surface of the Euclidean ball $B_2^n=\left\{\bx\in\R^n: \ \sum_{i=1}^nx_i^2\le 1\right\}$.
The following theorem states that, under Condition \ref{cond:one}, 
the tilted correlations of the relevant variables dominate those of the irrelevant variables.
\begin{thm}
Under assumptions (A1)--(A6), if Condition \ref{cond:one} holds, then $\p(\cE_1)\to 1$ where
\begin{eqnarray}
\cE_1=\left\{\frac{|c^*_k(s_k)|}{\min_{j\in\cS}|\cjs(s_j)|}\to 0
\mbox{ for all } k\notin\cS\right\},
\label{separation}
\end{eqnarray}
regardless of the choice of the rescaling factor (that is, with $s_j=\lamj$ or $s_j=\Lamj$).
On the event $\cE_1$, the following holds.
\begin{itemize}
\item $n^\mu\cdot\cjs\to 0$ for $j\notin\cS$.
\item $n^\mu\cdot|\cjs|\to\infty$ for $j\in\cS$. 
\item With the rescaling 1, $\cjs(\lamj)/\beta_j\to 1$ when $\beta_j\ne 0$. 
\end{itemize}
\label{thm:one}
\end{thm}
As noted in the Introduction, in high-dimensional problems, the maximum sample correlation
of the columns of $\bX$ can be non-negligible, even if the columns are
generated as independent. Therefore marginal correlations $X_j^T\by$ for $j\in\cS$ cannot 
be expected to have the same dominance over those for $j\notin\cS$ as in (\ref{separation}).

\subsubsection{Scenario 2}
\label{sec:scen:two}

Let $\cK$ denote a subset of $\cJ$ such that $X_k, \ k\in\cK$ are either relevant ($k\in\cS$) 
or highly correlated with at least one of relevant variables ($k\in\cup_{j\in\cS}\cCj$). 
That is, $\cK=\cS\cup\left\{\cup_{j\in\cS}\cCj\right\}$,
and we impose the following condition on the sample correlation structure of $\bX_\cK$.
\begin{cond}
For each $j\in\cS$, if $k\in\cK \setminus \{\cCj \cup \{j\}\,\}$, then $\cC_k\cap\cCj=\emptyset$.
\label{cond:two}
\end{cond}
In other words, this condition implies that for each relevant variable $X_j$, 
if $X_k, \ k\in\cK$ is not highly correlated with $X_j$, there does not exist an $X_l, \ l\ne j, k$, 
which achieves sample correlations greater than the threshold $\pin$ with both $X_j$ and $X_k$ simultaneously.

Suppose that the sample correlation matrix of $\bX_\cK$ is ``approximately bandable'', 
i.e., $|c_{j,k}|>\pin$ for any $j, k\in\cK$ satisfying $|j-k| \le B$ and $|c_{j,k}|<\pin$ otherwise, 
with the band width $B$ satisfying $B|\cS|^2/p\to 0$.
Then, if $\cS$ is selected randomly from $\cJ$ with each $j\in\cJ$ 
having equal probability to be included in $\cS$,
Condition \ref{cond:two} holds with probability bounded from below by
\[
\left(1-\frac{4B}{p-1}\right)\cdot\left(1-\frac{8B}{p-2}\right)\cdots \left(1-\frac{4(|\cS|-1)B}{p-|\cS|+1}\right)\ge \left(1-\frac{4|\cS|B}{p-|\cS|+1}\right)^{|\cS|-1} \to 1.
\]
Another example satisfying Condition \ref{cond:two} is when each column of $\bX_\cK$ 
is generated as a linear combination of common factors in such a way that 
every off-diagonal element of the sample correlation matrix of $\bX_\cK$ exceeds the threshold $\pin$.

Under this condition, we can derive a similar result as in Scenario 1, 
with the dominance of the tilted correlations for relevant variables restricted within $\cK$.
\begin{thm}
Under (A1)--(A6), if Condition \ref{cond:two} holds, then $\p(\cE_2)\to 1$ where
\begin{eqnarray*}
\cE_2=\left\{\frac{|c^*_k(s_k)|}{\min_{j\in\cS}|\cjs(s_j)|}\to 0
\mbox{ for all } k\in\cK\setminus\cS\right\},
\end{eqnarray*}
regardless of the choice of the rescaling factor (that is, with $s_j=\lamj$ or $s_j=\Lamj$).
On the event $\cE_2$, the following holds.
\begin{itemize}
\item $n^\mu\cdot\cjs\to 0$ for $j\notin\cS$.
\item $n^\mu\cdot|\cjs|\to\infty$ for $j\in\cS$. 
\item With the rescaling 1, $\cjs(\lamj)/\beta_j\to 1$ when $\beta_j\ne 0$. 
\end{itemize}
\label{thm:two}
\end{thm}

\subsubsection{Scenario 3}
\label{sec:scen:three}

Finally, we consider a case when $\bX$ satisfies a condition weaker than Condition \ref{cond:two}. 
\begin{cond}
\begin{itemize}
\item[(C1)]
For each $j\in\cS$, if $k\in\cK \setminus \{\cCj \cup \cS \,\}$, then $\cC_k\cap\cCj=\emptyset$.
\item[(C2)] The marginal correlation between $\xjs=(\bI_n-\Pij)X_j$ for $j\in\cS$ 
and $\E\by=\bX_\cS\bbeta_\cS$ satisfies \linebreak
$n^\mu\cdot\inf_{j\in\cS}\left\vert X_j^{*T}\bX_\cS\bbeta_\cS\right\vert \to \infty$.
\end{itemize}
\label{cond:three}
\end{cond}
It is clear that Condition \ref{cond:two} is stronger than (C1), 
as the latter does not impose any restriction between $\cCj$ and $\cC_k$ if both $j, k\in\cS$.
\citet{buhlmann2009} placed a similar lower bound as that in (C2) on the 
population partial correlation $\ppc(j, \by|\cD)$ of relevant variables $X_j, \ j\in\cS$,
for \emph{any} subset $\cD\subset\cJ\setminus\{j\}$ satisfying $|\cD|\le|\cS|$.
Combined with the assumptions (A4)--(A5), (C2) rules out an ill-posed case where 
the parameters $\beta_j, \ j\in\cS$ take values which cancel out the ``tilted covariance'' 
among the relevant variables (this statement is explained more precisely 
in the proof of Theorem \ref{thm:three}).
It is shown in Appendix \ref{appendix:three} that Condition \ref{cond:three} is satisfied 
if Condition \ref{cond:two} holds and thus Condition \ref{cond:three} is indeed weaker than Condition \ref{cond:two}.
With Condition \ref{cond:three}, we can show similar results to those in Theorem \ref{thm:two}. 
\begin{thm}
Under (A1)--(A6), if Condition \ref{cond:three} holds, then $\p(\cE_3)\to 1$ where
\begin{eqnarray*}
\cE_3=\left\{\frac{|c^*_k(s_k)|}{\min_{j\in\cS}|\cjs(s_j)|}\to 0
\mbox{ for all } k\in\cK\setminus\cS\right\},
\end{eqnarray*}
regardless of the choice of the rescaling factor (that is, with $s_j=\lamj$ or $s_j=\Lamj$).
On the event $\cE_3$, the following holds.
\begin{itemize}
\item $n^\mu\cdot\cjs\to 0$ for $j\notin\cS$.
\item $n^\mu\cdot|\cjs|\to\infty$ for $j\in\cS$. 
\end{itemize}
\label{thm:three}
\end{thm}
In contrast to Scenario 2, tilted correlations $\cjs(\lamj)$ no longer necessarily converge to $\beta_j$ as 
$n\to\infty$ in this scenario.

In the next section, we use the theoretical properties of tilted correlations derived in this section
to construct a variable screening algorithm.

\section{Application of tilting}
\label{sec:application}

Recalling issues (a)--(c) listed at the beginning of Section \ref{sec:tilt}
which are typically encountered in high-dimensional problems, 
it is clear that tilting is specifically designed to tackle the occurrence of (a) and (b).
First turning to (a), for an irrelevant variable $X_j$ which attains high marginal
correlation with $\by$ due to its high correlations with relevant variables 
$X_k, \ k\in\cCj\cap\cS$, the impact of those high correlations is reduced to 0 in 
the tilted correlation of $X_j$ and $\by$, and thus tilted correlation provides a more 
accurate measure of its association with $\by$, as demonstrated in our theoretical
results of the previous section. Similar arguments apply to (b),
where tilting is capable of fixing {\em low} marginal correlations between
{\em relevant variables} and $\by$.
(As for (c), it is common practice to impose assumptions which rule out strong 
collinearity among variables, and we have also followed this route.)
In what follows, we present an algorithm, specifically constructed to exploit our 
theoretical study in Section \ref{sec:prop:tilt} by iteratively applying the tilting 
procedure.

\subsection{Tilted correlation screening algorithm}
\label{sec:algorithm}

In Scenario 3, under a relatively weaker condition than those in Scenarios 1--2, 
it is shown that the tilted correlations of relevant variables dominate 
those of irrelevant variables within $\cK=\cS\cup\left(\cup_{j\in\cS}\cCj\right)$.
Even though $\cK$ is unknown in practice, we can exploit the theoretical results 
by iteratively screening both marginal correlations and 
tilted correlations within a specifically chosen subset of variables. 

When every off-diagonal entry of the sample correlation matrix is small, 
marginal correlation screening can be used as a reliable way of measuring the strength of association 
between each $X_j$ and $\by$,
and indeed, $\cjs$ for the variable $X_j$ with an empty $\cCj$ is equal to the marginal correlation $X_j^T\by$,
with either choice of the rescaling factor $s_j$.
Therefore if a variable $X_j$ with $\cCj=\emptyset$ achieves the maximum marginal correlation, 
such $X_j$ is likely to be relevant.
On the other hand, if $\cCj\ne\emptyset$, then high marginal correlation between $X_j$ and $\by$
may have resulted from the high correlations of $X_j$ with $X_k, \ k\in\cCj\cap\cS$, even when $j\notin\cS$.
In this case, by screening the tilted correlations of $X_k, \ k\in\cCj\cup\{j\}$, 
we can choose the variable attaining the maximum $|c^*_k|$ as a relevant variable.
In either case, one variable is selected and added to the \emph{active set} $\cA$ 
which represents the currently chosen model.
The next step is to update the linear model by projecting it onto 
the orthogonal complement of the current model space $\bX_\cA$, i.e., 
\begin{eqnarray}
(\bI_n-\Pi_\cA)\by=(\bI_n-\Pi_\cA)\bX\bbeta+(\bI_n-\Pi_\cA)\bep.
\label{update:model}
\end{eqnarray}
With the updated response and design matrix, we iteratively continue the above screening procedure. 
Below we present the algorithm which is referred to as the tilted correlation screening (TCS) algorithm
throughout the paper.

\begin{itemize}
\item[Step 0] Start with an empty active set $\cA=\emptyset$, current residual $\bz=\by$, 
and current design matrix $\bZ=\bX$.

\item[Step 1]
Find the variable which achieves the maximum marginal correlation with $\bz$ and 
let $k=\arg\max_{j\notin\cA}|Z_j^T\bz|$.
Identify $\cC_k=\{j\notin\cA, j\ne k: \  |Z_k^TZ_j|>\pin\}$ and
if $\cC_k=\emptyset$, let $k^*=k$ and go to Step 3.

\item[Step 2] 
If $\cC_k\ne\emptyset$, screen the tilted correlations $\cjs$ between $Z_j$ and $\bz$ for $j\in\cC_k\cup\{k\}$ 
and find $k^*=\arg\max_{j\in\cC_k\cup\{k\}}|\cjs|$.

\item[Step 3] 
Add $k^*$ to $\cA$ and update the current residual and the current design matrix
$\bz\leftarrow(\bI_n-\Pi_\cA)\by$ and $\bZ\leftarrow(\bI_n-\Pi_\cA)\bX$,
respectively. Further, rescale each column $j \not\in \cA$ of $\bZ$ to have norm one.

\item[Step 4] Repeat Steps $1$--$3$ until the cardinality of active set $|\cA|$ reaches a pre-specified $m<n$.
\end{itemize}

We note that Theorems \ref{thm:one}--\ref{thm:three} do not guarantee the selection consistency 
of the TCS algorithm itself. However, they do demonstrate a certain `separation' property of the
tilted correlation (as a measure of association). Steps 1--2 of the above algorithm exploit this
property in the sense that they attempt to ``operate'' within the set $\cK$
(which is unknown without the knowledge of $\cS$), 
since we either directly choose a variable indexed $k$ which is believed to lie in the set $\cS$ or
screen its corresponding set $\cC_k$ (recall that $\cK=\cS\cup\left\{\cup_{j\in\cS}\cCj\right\}$).

In Step 4, we need to specify $m$ which acts as a stopping index in the TCS algorithm.
The TCS algorithm iteratively builds a solution path of the active set $\cA_{(1)}\subset \cdots \subset \cA_{(m)}=\cA$,
and the final model $\hat{\cS}$ can be chosen as either one of the submodels $\cA_{(i)}$ or a subset of $\cA$.
We discuss the selection of the final model $\hat{\cS}$ in Section \ref{sec:final:model}.
In the simulation study, we used $m=\lfloor n/2 \rfloor$, which was an empirical choice made in order 
to ensure that the projections performed in the algorithm were numerically stable, 
while a sufficiently large number of variables were selected in the final model, if necessary.
In practice however, if the TCS algorithm combined with the chosen model selection criterion
returned $m$ variables (i.e. if it reached the maximum permitted number of active variables), we
would advise re-running the TCS algorithm with the limit of $m$ slightly raised,
until the number of final active variables was less than the current value of $m$.

During the application of the TCS algorithm, the linear regression model (\ref{lp}) is updated in Step 3 
by projecting both $\by$ and $\bX$ onto the orthogonal complement of the current model space spanned by $\bX_\cA$. 
Therefore, with a non-empty active set $\cA$, it is interesting to observe that the tilted correlation $\cjs$ measures 
the association between $X_j$ and $\by$ conditional on both the current model $X_k, \ k\in\cA$ and
the following subset of variables adaptively chosen for each $j\notin\cA$,
\begin{eqnarray}
\cCja=\{k\notin\cA, k\ne j: \  \spc(j, k|\cA)>\pin\},
\label{update:ccj}
\end{eqnarray} 
where $\spc(j, k|\cA)$ denotes the sample partial correlation between $X_j$ and $X_k$ conditional on $\bX_\cA$.

Finally, we discuss the computational cost of the TCS algorithm. 
When $p \gg n$, the computational complexity of the algorithm is dominated by 
the computation of the threshold at Step 1,
which is $O(np+np^2+p^2\log\,p+p^2)=O(np^2)$.
Since the procedure is repeated $m$ times, with $m$ set to satisfy $m=O(n)$, 
the computational complexity of the entire algorithm is $O(n^2p^2)$,
which is $n$ times the cost of computing a $p\times p$ sample covariance matrix.

\subsection{Final model selection}
\label{sec:final:model}

Once the size of the active set reaches a pre-specified value $m$, 
the final model $\hat{\cS}$ needs to be chosen from $\cA$.
In this section, we present two methods which can be applied in our framework.
One of the most commonly used methods for model selection is cross-validation (CV), in which 
the observations would be divided into a training set and a test set
such that the models returned after each iteration (i.e. $\cA_{(1)}\subset \cdots \subset \cA_{(m)}=\cA$) could
be tested using an appropriate error measure. However, we expect that for a CV-based method to work well, it 
would have to be computationally intensive: for example, a leave-one-out CV or a leave-half-out CV with 
averaging over different test and training sets.

One less computationally demanding option is to use e.g. an
extended version of the Bayesian information criterion (BIC) proposed in
\citet{bogdan2004} and \citet{chen2008} as
\begin{eqnarray}
\mbox{BIC}(\cA)=\log\left\{\frac{1}{n}\Vert(\bI_n-\Pi_\cA)\by\Vert_2^2\right\}+
\frac{|\cA|}{n}(\log n+2\log p).
\label{bic}
\end{eqnarray}
This new BIC takes into account high dimensionality of the data by adding a penalty term dependent on $p$.
Since the TCS algorithm generates a solution path which consists of $m$ sub-models
$\cA_{(1)}\subset \cdots \subset \cA_{(m)}=\cA$, 
we can choose our final model as $\hat{\cS}=\cA_{(m^*)}$ where 
$m^*=\arg\min_{1\le i \le m}\mbox{BIC}(\cA_{(i)})$.

\citet{chen2008} showed the consistency of this new BIC under stronger conditions 
than those imposed in (A1), (A2) and (A4):
the level of sparsity was $|\cS|=O(1)$, the dimensionality was $p=O(n^C)$ for $C>0$, 
and non-zero coefficients satisfied $\min_{j\in\cS}|\beta_j|>C'$ for $C'>0$.
Then, under the asymptotical identifiability condition introduced in \citet{chen2008}, 
(see (\ref{asym:iden}) in Appendix \ref{append:one}), 
the modified BIC as defined in (\ref{bic}) was shown to be consistent in the sense that 
\[
\p\left(\min_{|\cD|\le m, \ \cD\ne\cS}\mbox{BIC}(\cD)>\mbox{BIC}(\cS)\right)\to 1 \mbox{ for } m\ge|\cS|,
\]
i.e., the probability of selecting any model other than $\cS$ converged to zero.  
It was also noted that the original BIC was likely to fail when $p>n^{1/2}$.
At the price of replacing $\log n/n$ with $n^{-\kappa}$ in (\ref{asym:iden}), 
the consistency of the new BIC (\ref{bic}) can be shown 
with the level of sparsity growing with $n$ as in (A1) and the dimensionality increasing exponentially with $n$ as in (A2).
The proof of this statement follows the exact line of proof in \citet{chen2008} and so we omit
the details. 

\subsection{Relation to existing literature}
\label{sec:relation}

We first note that our use of the term ``tilting'' is 
different from the use of the same term in \citet{hall2009},
where it applies to distance-based classification and
denotes an entirely different procedure.

In the Introduction, we briefly discuss a list of existing variable selection
techniques in which care is taken of the correlations among the variables
in measuring the association between each variable and the response.
Having now a complete picture of the TCS algorithm, we provide a more detailed comparison 
between our methodology and the aforementioned methods.

\citet{buhlmann2009} proposed the PC-simple algorithm, which iteratively removes variables 
having small association with the response.
Sample partial correlations $\spc(j, \by|\cD)$ are used as the measure of association
between $X_j$ and $\by$,
where $\cD$ is \emph{any} subset of the active set $\cA$ (those variables still remaining in the model excluding $X_j$) 
with its cardinality $|\cD|$ equal to the number of iterations taken so far. 
Behind the use of partial correlations lies the concept of partial faithfulness 
which implies that, at the population level, if $\ppc(j, \by|\cD)=0$ for some $\cD\subset\cJ\setminus\{j\}$, 
then $\ppc(j, \by|\cJ\setminus\{j\})=0$.
Their PC-simple algorithm starts with $\cA=\cJ$ and iteratively repeats 
the following:
(i) screening sample partial correlations $\spc(j, \by|\cD)$ for all $j\in\cA$ 
and for all $\cD$ satisfying the cardinality condition, 
(ii) applying Fisher's Z-transform to test the null hypotheses $H_0: \ppc(j, \by|\cD)=0$,
(iii) removing irrelevant variables from $\cA$,
until $|\cA|$ falls below the number of iterations taken so far.
Recalling the definition of the rescaling factor $\Lamj$, 
we can see the connection between $\cjs(\Lamj)$ and $\spc(j, \by|\cD)$, 
as both are (up to a multiplicative factor $\Vert\by\Vert_2$) partial correlations between $X_j$ and $\by$ 
conditional on a certain subset of variables.
However, a significant difference comes from the fact that 
the PC-simple algorithm takes all $\cD\subset\cA\setminus\{j\}$ with fixed $|\cD|$ at each iteration, 
whereas our TCS algorithm adaptively selects $\cCj$ (or $\cCja$ when $\cA\ne\emptyset$) for each $j$.
Also, while $\lamj$ is also a valid rescaling factor in our tilted correlation methodology, 
partial correlations are by definition computed using $\Lamj$ only.

As for the forward regression \citep[FR]{wang2009} and the forward selection (FS),
although the initial stage of the two techniques is simple marginal correlation screening, 
their progression has a new interpretation given a non-empty active set ($\cA\ne\emptyset$).
Both algorithms obtain the current residual $\bz$ by projecting the response $\by$ 
onto the orthogonal complement of the current model space, i.e., $\bz=(\bI_n-\Pi_\cA)\by$.
Therefore they also measure the association between each $X_j, \ j\notin\cA$ and $\by$ 
conditional on the current model space spanned by $\bX_\cA$ and thus take into account
the correlations between $X_j, \ j\notin\cA$ and $X_j, \ j\in\cA$. 
The difference between FR and FS comes from the fact that 
FR updates not only the current residual $\bz$ but also the current design matrix as 
$\bZ=(\bI_n-\Pi_\cA)\bX$ (as in Step 3 of the TCS algorithm).
Therefore FR eventually screens the rescaled version of $X_j^T(\bI_n-\Pi_\cA)\by$ 
with the rescaling factor defined similarly to $\lamj$, replacing $\cCj$ with $\cA$, 
i.e., $X_j^T(\bI_n-\Pi_\cA)X_j=1-X_j^T\Pi_\cA X_j$.
On the other hand, there is no rescaling step in FS and it screens 
the terms $X_j^T(\bI_n-\Pi_\cA)\by, \ j\notin\cA$, themselves.

By contrast, we note that while both FR and FS apply straight marginal correlation
at each stage of their progression, our method, if and as necessary, uses the tilted correlation,
which provides an adaptive choice between the marginal correlation and conditional
correlation, depending on the correlation structure of the current design matrix.
Indeed, in the extreme case where $\pin=1$ is used, we have $\cCj=\emptyset$ and therefore the TCS 
algorithm becomes identical to FR. 
Another crucial difference is as already mentioned above in the context of the PC-simple algorithm:
the tilting algorithm employs an adaptive choice of the conditioning set, unlike
FR and FS.

In conclusion, the TCS algorithm, the PC-simple algorithm, FR and FS share the common ingredient of 
measuring the contribution of each variable $X_j$ to $\by$ conditional on certain other variables;
however, there are also important differences between them, and Table \ref{table:comp} summarises this 
comparison. We emphasise yet again that the TCS algorithm is distinguished from the rest 
in its adaptive choice of the conditioning subset via hard-thresholding of the sample correlations among 
the variables. Also, we note that the theoretical results of Section \ref{sec:prop:tilt} hold for 
\emph{both} rescaling methods, while the other algorithms use only one of them 
(FR, PC-simple) or none (FS).

\begin{table}
\caption{Comparison of variable selection methods.}
\label{table:comp}
\centering
\begin{tabular}{ c | c | c | c | c }
\hline
\hline
 & TCS algorithm & PC-simple & FR & FS 
\\
\hline 
Step 0 & $\cA=\emptyset$ & $\cA=\cJ$ & $\cA=\emptyset$ & $\cA=\emptyset$ 
\\
\hline
\multirow{2}{*}{action} &
one & multiple & one & one \\ 
& selected & removed & selected & selected 
\\
\hline 
\multirow{3}{*}{
conditioning set $\cD$} & 
$\cA\cup\cCja$ & remaining & current & current  \\
& $=\cA\cup\{k\notin\cA, k\ne j:$ & variables, & model & model \\
& $|\spc(j, k|\cA)|>\pin\}$ & $|\cD|$ fixed & $\cA$ & $\cA$
\\
\hline
rescaling & $\lamj$ or $\Lamj$ & $\Lamj$ & $\lamj$ & none \\
\hline
\hline
\end{tabular}
\end{table}

Finally, we note the relationship between the TCS algorithm and 
the covariance-regularised regression method proposed in \citet{witten2009}.
A key difference between the two is that
the TCS algorithm works with the sample marginal correlations among the variables
whereas in the scout procedure, it is the conditional correlations among the variables 
(i.e., $\ppc(j, k|\cJ \setminus \{j, k\})\ne 0$) that are subject to regularisation.
Also, the scout procedure achieves such regularisation 
by maximising a penalised likelihood function rather than hard-thresholding,
and the thus-obtained estimate of the covariance structure of $\bX$ is applied to estimate $\bbeta$, 
again by solving an optimisation problem. 
By contrast, the tilted correlation method uses the outcome from thresholding 
the sample correlation structure to compute the tilted correlations and select the variable with maximum tilted correlation in an iterative algorithm, and therefore does not involve any optimisation problems.

\subsection{Choice of threshold}
\label{sec:choice:thr}

In this section, we discuss the practical choice of the unknown threshold $\pin$ from the sample correlation matrix $\bC$. 
Due to the lack of information on the correlation structure of $\bX$ in general 
and the possibility of spurious sample correlation among the variables,
a deterministic choice of $\pin$ is not expected to perform well universally and we need a data-driven way of selecting a threshold.
\citet{bickel2008} proposed a cross-validation method for this purpose, 
while \citet{el2008} conjectured the usefulness of a procedure based on controlling the false discovery rate (FDR). 
Since our aim is different from the accurate estimation of the correlation matrix itself, 
we propose a threshold selection procedure which is a modified version of the approach taken in the latter paper. 
In the following, we assume that $\bX$ is a realisation of a random matrix with each row generated as
$\bx_i\sim_{\mbox{\scriptsize{i.i.d.}}}(\bzero, \Sigma)$,
where each diagonal element of $\Sigma$ equals one.

The procedure is a multiple hypothesis testing procedure and thus requires $p$-values of 
the $d=p(p-1)/2$ hypotheses $H_0: \ |\Sigma_{j, k}|=0$ defined for all $j<k$. 
We propose to compute the $p$-values as follows.
First, an $n$-vector with i.i.d. Gaussian entries is repeatedly generated $p$ times,
and sample correlations $\{r_{l, m}: \ 1 \le l < m \le p\}$ among those vectors are obtained as a reference.
Then, the p-value for each null hypothesis $H_0: \ |\Sigma_{j, k}|=0$ is defined as 
$P_{j, k}=d^{-1}\cdot\left\vert\left\{r_{l, m}: \ 1 \le l < m \le p, \ |r_{l, m}|\ge|c_{j, k}|\right\}\right\vert$. 
The next step is to apply the testing technique proposed in \citet{benjamini1995} to control the false discovery rate.
Denoting $P_{(1)}\le \ldots \le P_{(d)}$ as the ordered $p$-values, 
we find the largest $i$ for which $P_{(i)}\le i/d\cdot\nu^*$ and reject all $H_{(j)}, \ j=1, \ldots, i$.
Then $\hat{\pi}_{thr}$ is chosen as the absolute value of the correlation corresponding to $P_{(i)}$.
FDR is controlled at level $\nu^*$ and we use $\nu^*=p^{-1/2}$ as suggested in \citet{el2008}.

An extensive simulation study described below confirms good practical performance of the above threshold selection procedure. 
We also checked the sensitivity of our algorithm to the choice of threshold by applying a grid of thresholds in model (C) below. 
Apart from the threshold $\hat{\pi}_{thr}$ selected as above, we ran versions of our algorithm where $\hat{\pi}_{thr}$ was multiplied by the constant factors of $0.75, 0.9, 1.1, 1.25$ each time it was used. 
Performance of our algorithm was similar across the different thresholds, 
which provides evidence for robustness of our procedure to the choice of threshold within reason.

\section{Simulation study}
\label{sec:sim}

In this section, we compare the performance of the TCS algorithm on simulated data 
with that of other related methods discussed in the Introduction and Section \ref{sec:relation}, 
which are the PC-simple algorithm, FR, FS, iterative SIS (ISIS) and FLASH (for ease of implementation, 
we adopt the ``global'' approach for FLASH), as well as Lasso for completeness.
Furthermore, some non-convex penalised least squares (PLS) estimation techniques 
are included in the comparison study,
such as the SCAD \citep{fan2001} and the minimax concave (MC+) penalty \citep{zhang2010c}.
Sub-optimality of the Lasso in terms of model selection
has been noted in recent literature (see e.g. \citet{zhang2008} and \citet{zou2008}),
and non-convex penalties are proposed as a greedier alternative to achieve better variable selection.
In the following simulation study, the SCAD estimator is produced using the local linear approximation \citep{zou2008} and the MC+ penalised criterion is optimised using the SparseNet \citep{mazumder2009}.

The TCS algorithm is applied using both rescaling methods (denoted by TCS1 and TCS2, respectively),
with the maximum cardinality of the active set $\cA$ (Step 4) set at $m=\lfloor n/2\rfloor$, 
a value also used for FR.
The extended BIC is adopted (see Section \ref{sec:final:model}) to select the final model 
for the one-at-a-time algorithms, i.e. TCS1, TCS2, FR and FS.
For the thus-selected final models, the coefficient values are estimated using least squares. 
We note that, when the aim is to construct a well-performing predictive model, a shrinkage method 
can be applied to the least squares estimate. 
However, since our focus is on the variable selection aspect of the different techniques,
we use the plain (i.e. unshrunk) least squares estimates. 

As for the rest of the methods, we select the tuning parameters for each method as follows:
the data is divided into the training and validation sets such that
the training observations are used to compute the solution paths over a range of tuning parameters,
and those which give the smallest mean squared error between the response and the predictions on the validation data are selected.

Finally, we note that FS and the Lasso are implemented using the R package \texttt{lars},
and the ISIS and the SCAD by the package \texttt{SIS}. 

\subsection{Simulation models}
\label{sec:sim:models}

Our simulation models were generated as below. For models (A)--(C) and (F), the
procedure for generating the sparse coefficient vectors $\bbeta$ is outlined
below the itemised list which follows.

\begin{description}
\item[(A) Factor model with 2 factors:] 
Let $\phi_1$ and $\phi_2$ be two independent standard normal variables. 
Each variable $X_j$, $j = 1, \ldots, p$, is generated as $X_j=f_{j, 1}\phi_1+f_{j, 2}\phi_2+\eta_j$, 
where $f_{j, 1}, f_{j, 2}, \eta_j$ are also generated independently from a standard normal distribution.
The model is taken from \citet{meinshausen2008}.

\item[(B) Factor model with 10 factors:] 
Identical to (A) but with 10 instead of 2 factors.

\item[(C) Factor model with 20 factors:] 
Identical to (A) but with 20 instead of 2 factors.

\item[(D) Taken from \citet{fan2008} Section 4.2.2:] 
\begin{eqnarray*}
\by=\beta X_1+\beta X_2+\beta X_3-3\beta \sqrt{\varphi}X_4+\ep,
\end{eqnarray*} 
where $\ep\sim\cN_n(0, \bI_n)$ and $(X_{i, 1}, \ldots, X_{i, p})^T$ are generated from 
a multivariate normal distribution $\cN_n(\bzero, \Sigma)$ independently for $i=1, \ldots, n$.
The population covariance matrix $\Sigma=\left(\Sigma_{j, k}\right)_{j, k=1}^p$ satisfies
$\Sigma_{j, j}=1$ and $\Sigma_{j, k}=\varphi, j\ne k$, except $\Sigma_{4,k}=\Sigma_{j,4}=\sqrt{\varphi}$, 
such that $X_4$ is marginally uncorrelated with $\by$ at the population level.
In the original model of \citet{fan2008}, $\beta=5$ and $\varphi=0.5$ were used, 
but we chose $\beta=2.5$ and $\varphi=0.5, 0.95$ to investigate the performance of the
variable selection methods in more challenging situations.

\item[(E) Taken from \citet{fan2008} Section 4.2.3:] 
\begin{eqnarray*}
\by=\beta X_1+\beta X_2+\beta X_3-3\beta \sqrt{\varphi}X_4+0.25\beta X_5+\ep,
\end{eqnarray*} 
with the population covariance matrix of $\bX$ as in (D) 
except $\Sigma_{5,k}=\Sigma_{j,5}=0$, such that 
$X_5$ is uncorrelated with any $X_j, \ j\ne 5$, and relevant. However, it has only a very small 
contribution to $\by$.

\item[(F) Leukemia data analysis:] 
\citet{golub1999} analysed the Leukaemia dataset from high density Affymetrix oligonucloeotide arrays
(available on \url{http://www.broadinstitute.org/cgi-bin/cancer/datasets.cgi}),
which has 72 observations and 7129 genes (i.e. variables).
In \citet{fan2008}, the dataset was used to investigate the performance of Sure Independence Screening in 
a feature selection problem. Here, instead of using the actual response from the dataset, 
we used the design matrix to create simulated models as follows.
Each column $X_j$ of the design matrix was normalised to $\Vert X_j\Vert_2^2=n$, 
and out of 7129 such columns, $p$ were randomly selected to generate an $n\times p$-matrix $\bX$.
Then we generated a sparse $p$-vector $\bbeta$ and the response $\by$ as in (\ref{lp}).
In this manner, the knowledge of $\cS$ could be used to assess the performance of the competing 
variable selection techniques. A similar approach was taken in \citet{meinshausen2008} to generate 
simulation models from real datasets.

\end{description}

With the exception of (D)--(E), we generated the simulated data as below. 
Sparse coefficient vectors $\bbeta$ were generated by randomly sampling 
the indices of $\cS$ from $1, \ldots, p$, with $|\cS|=10$. 
The non-zero coefficient vector $\bbeta_\cS$ was drawn from a zero-mean normal distribution
such that $\bC_{\cS, \cS}\bbeta_\cS\sim\cN_{|\cS|}(\bzero, n^{-1}\bI_{|\cS|})$,
where $\bC_{\cS, \cS}$ denotes the sample correlation matrix of $\bX_\cS$.
In this manner, $\arg\max_{j\in\cJ}|X_j^T(\bX_\cS\bbeta_\bX)|$ may not always be attained by $j\in\cS$,
which makes the correct identification of relevant variables more challenging.  
The noise level $\sig$ was chosen to set $R^2=\var(\bx_i^T\bbeta)/\var(y_i)$ at $0.3$, $0.6$, or $0.9$,
adopting a similar approach to that taken in \citet{wang2009}.
In models (A)--(E), the number of observations was $n=100$ 
while the dimensionality $p$ varied from 500 to 2000 (except (D)--(E) where it was fixed at 1000), and 
finally, 100 replicates were generated for each set-up.

\subsection{Simulation results}
\label{sec:sim:results}

\begin{figure}
\label{fig:roc:a}
\centering
\begin{tabular}{ccc}
\epsfig{file=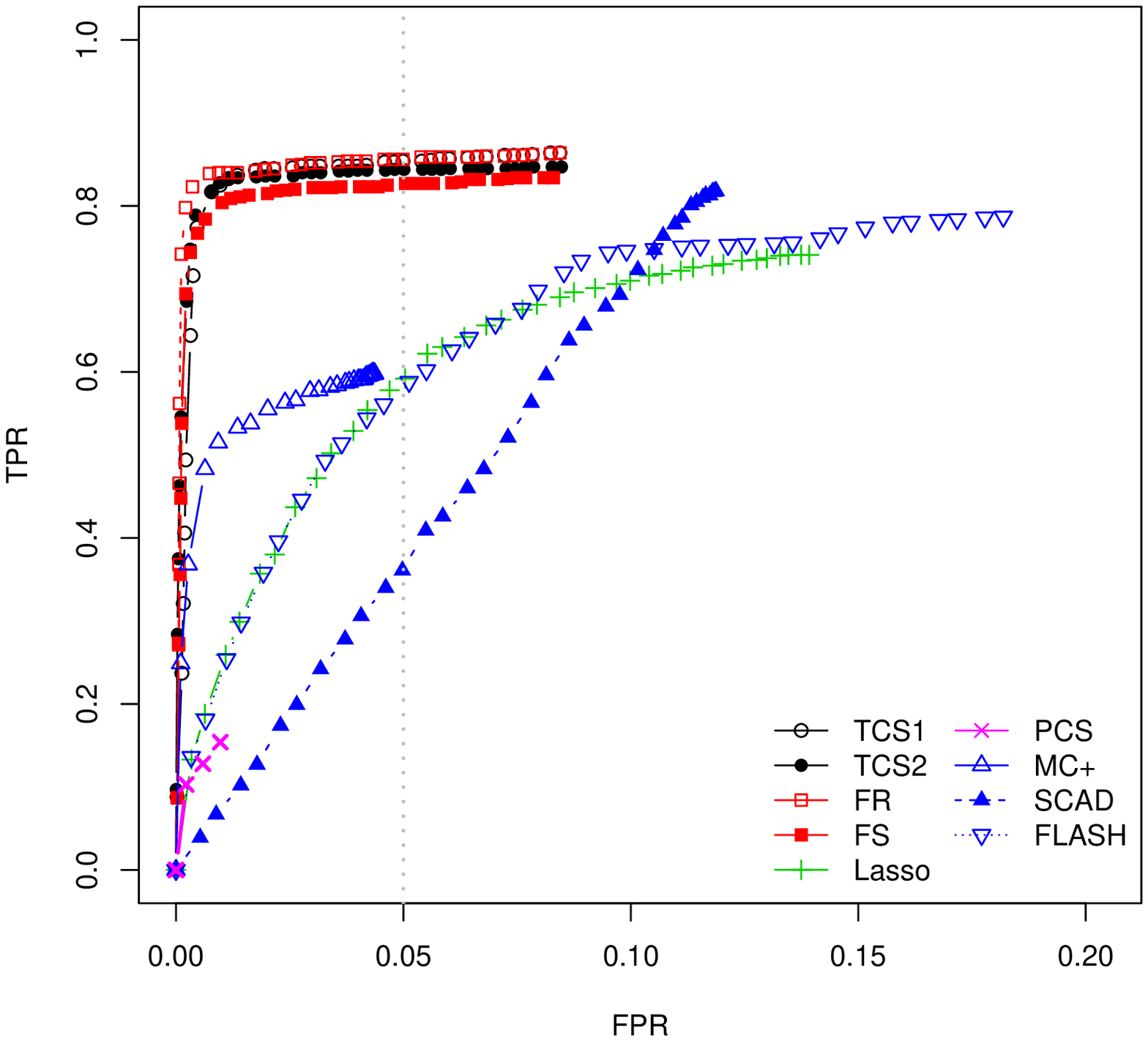, width=0.33\linewidth,clip=} & 
\epsfig{file=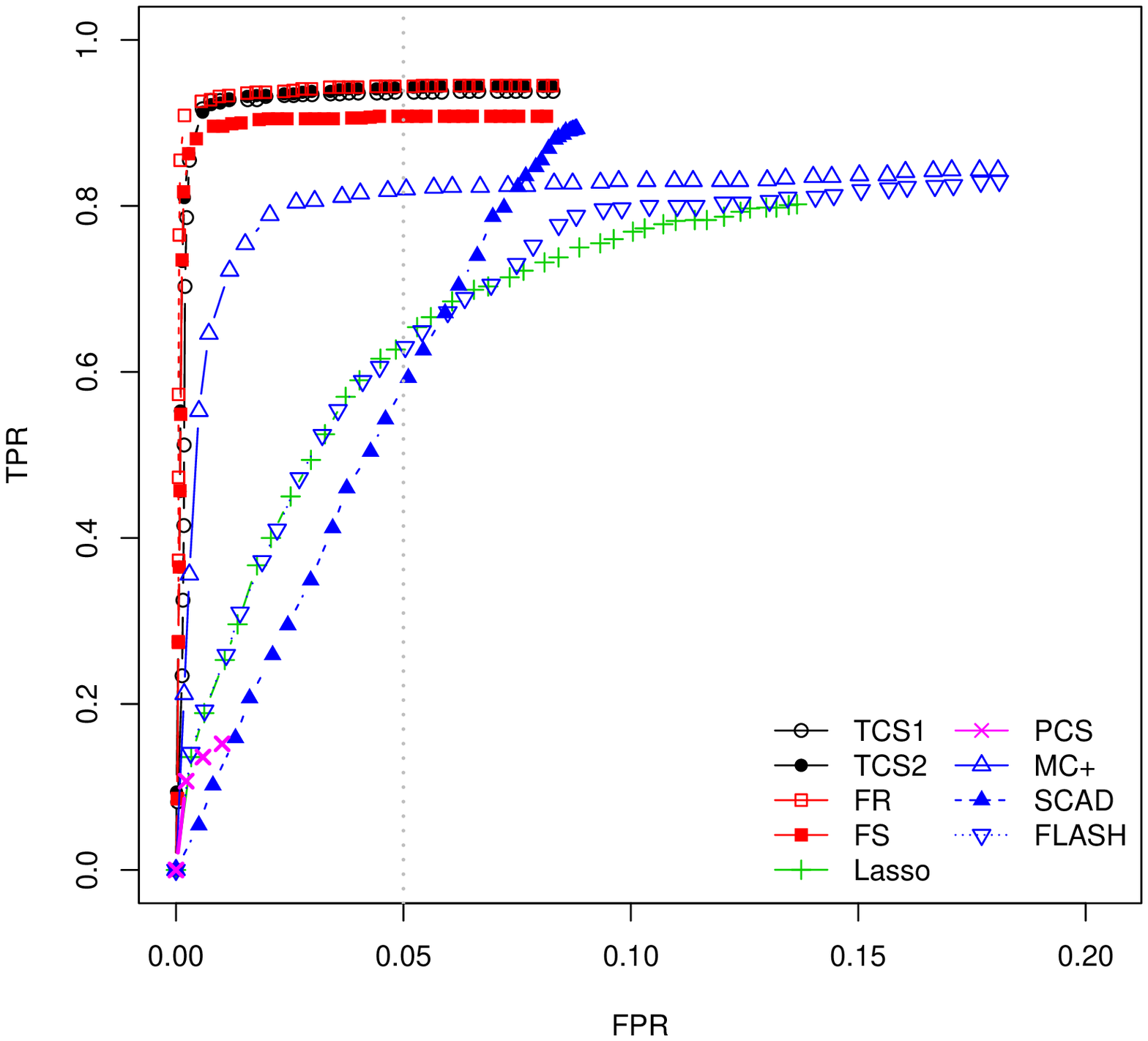, width=0.33\linewidth,clip=} & 
\epsfig{file=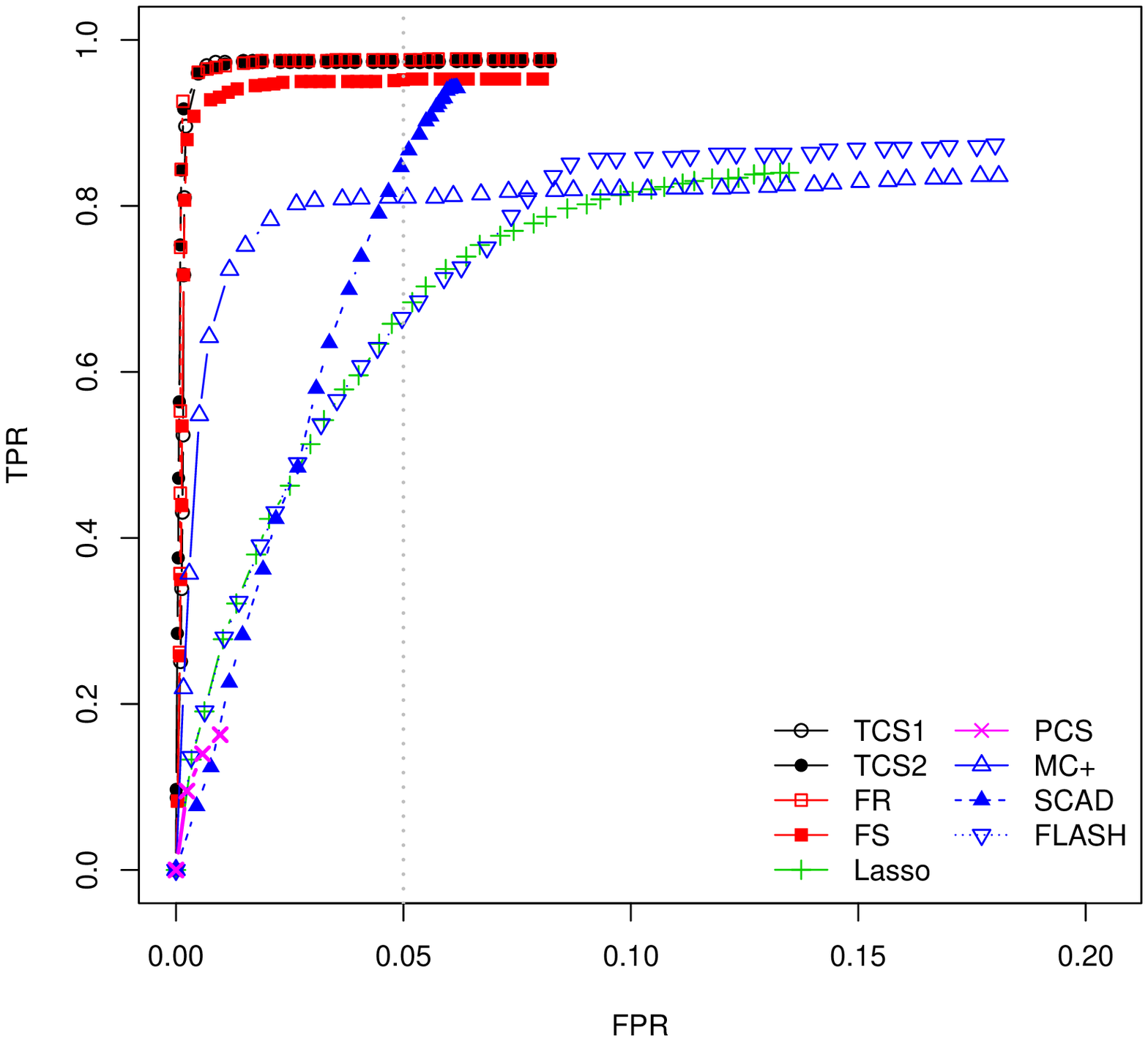, width=0.33\linewidth,clip=} 
\\
\epsfig{file=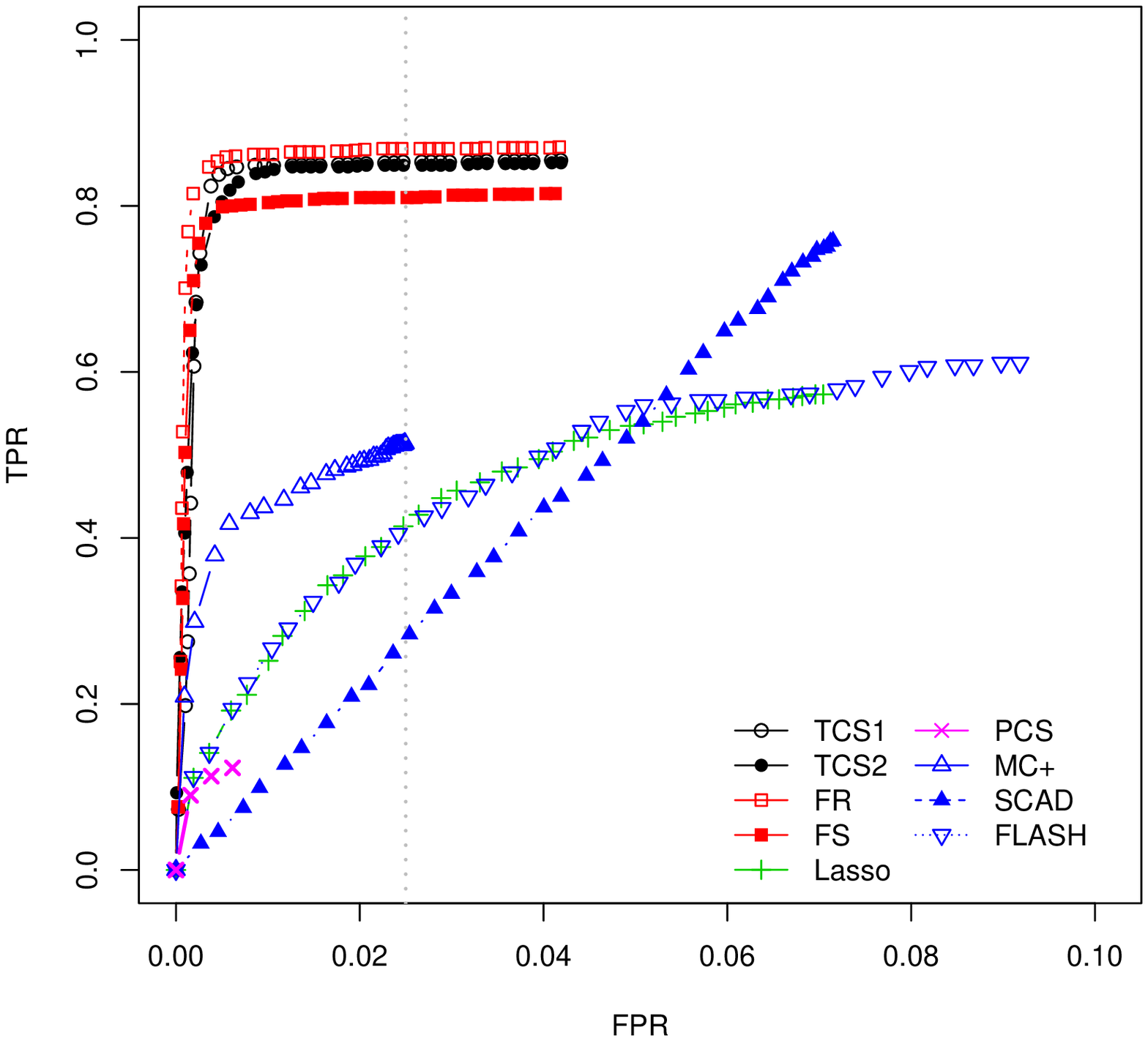, width=0.33\linewidth,clip=} & 
\epsfig{file=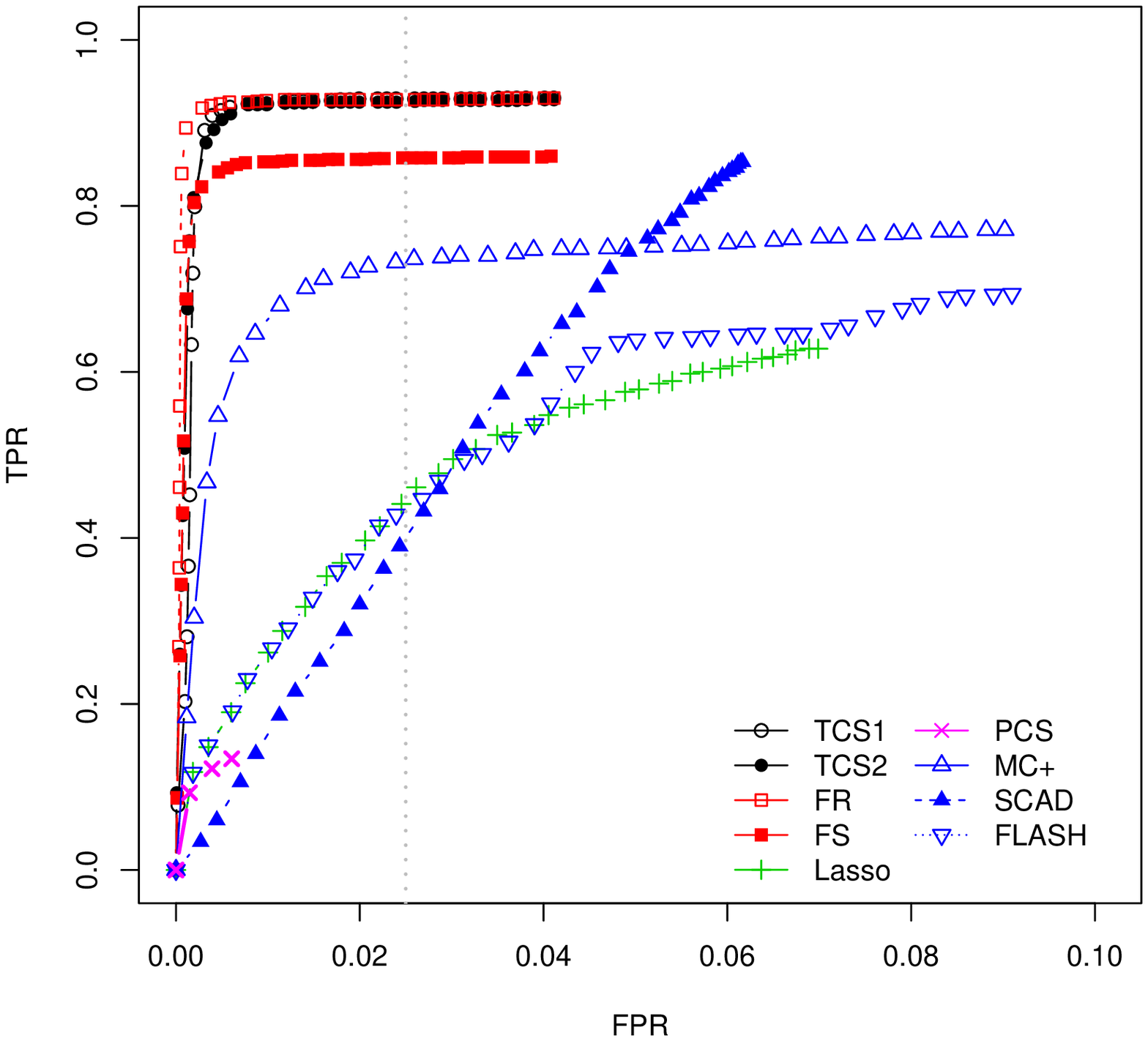, width=0.33\linewidth,clip=} & 
\epsfig{file=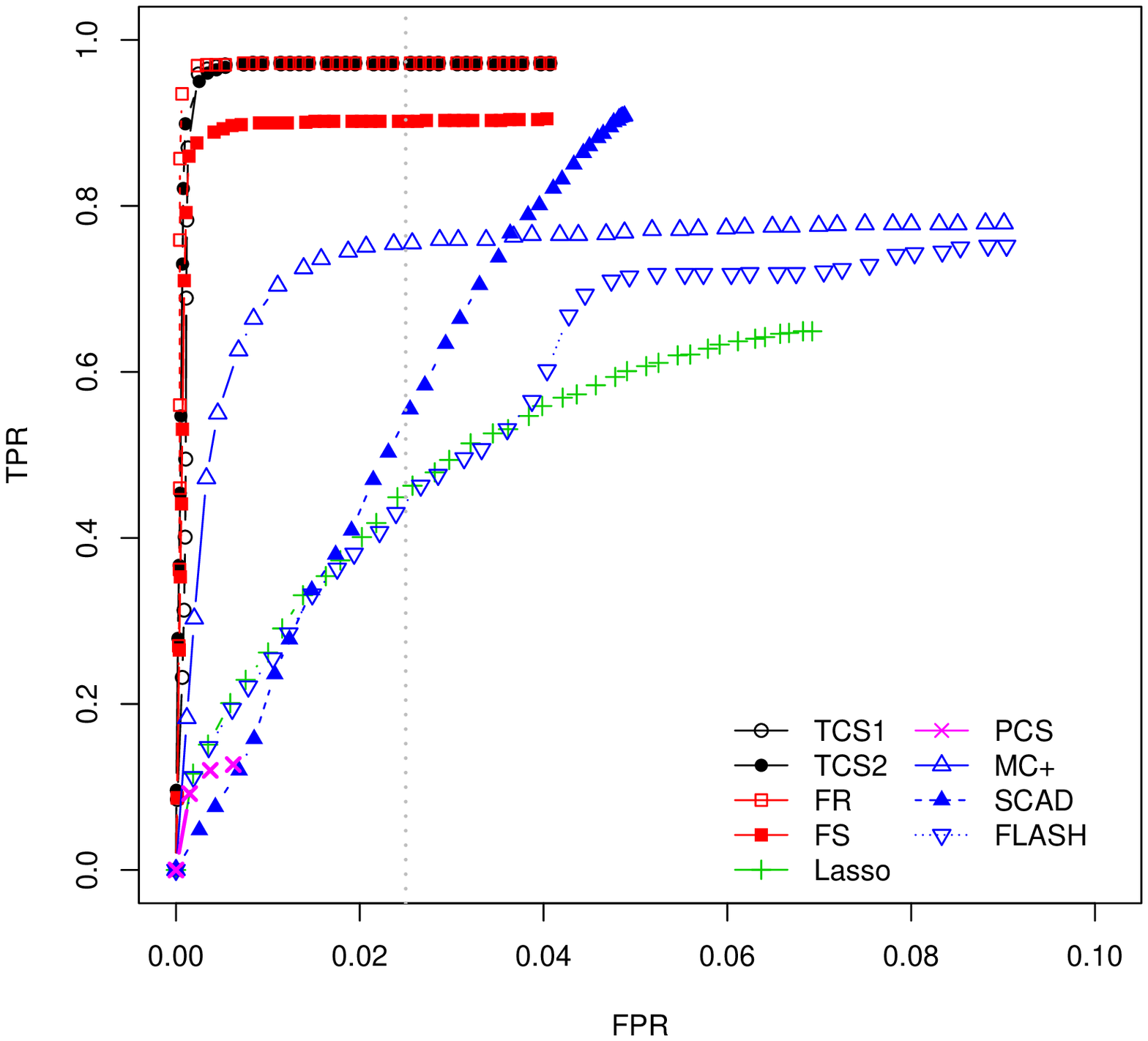, width=0.33\linewidth,clip=} 
\\
\epsfig{file=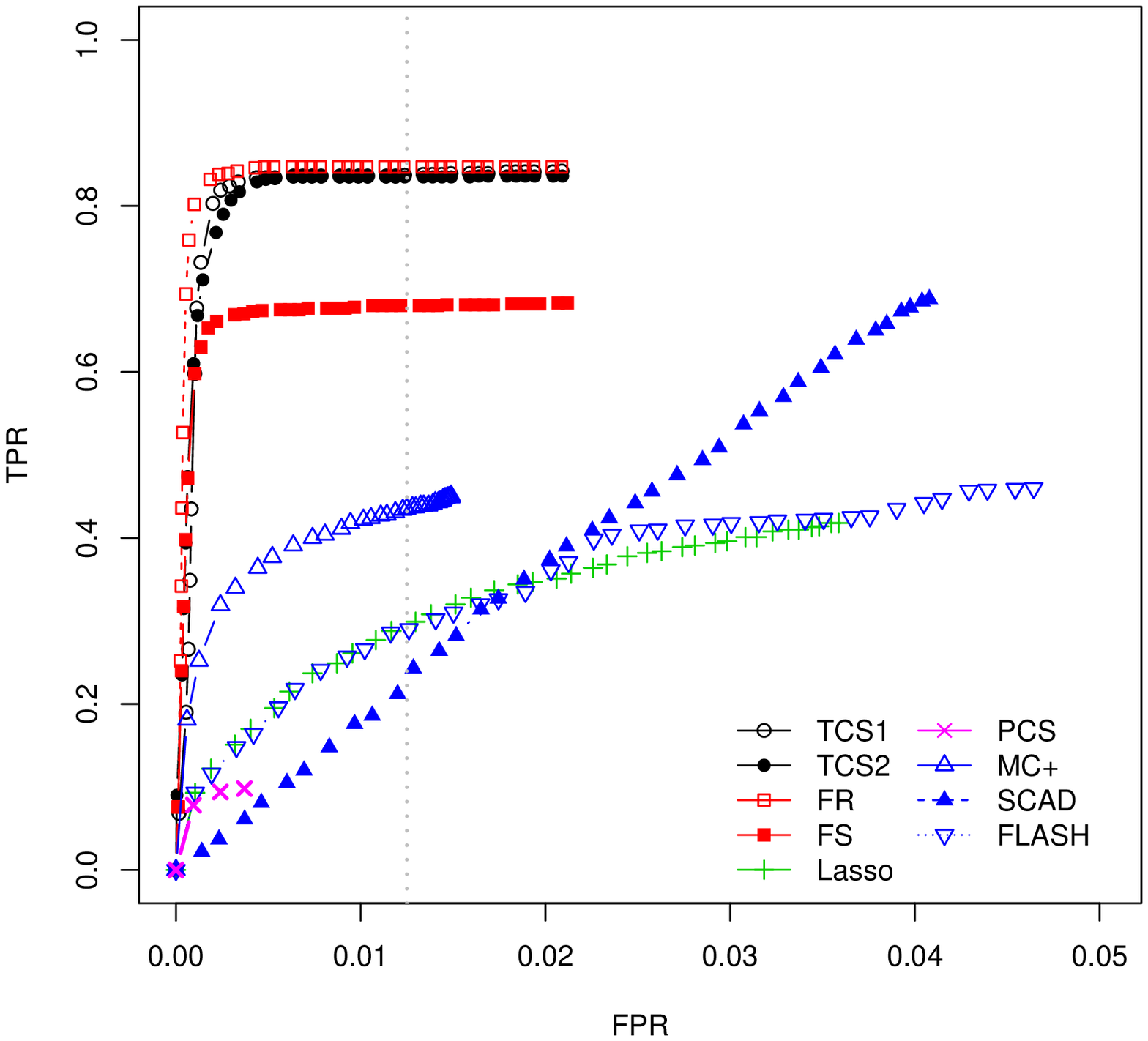, width=0.33\linewidth,clip=} & 
\epsfig{file=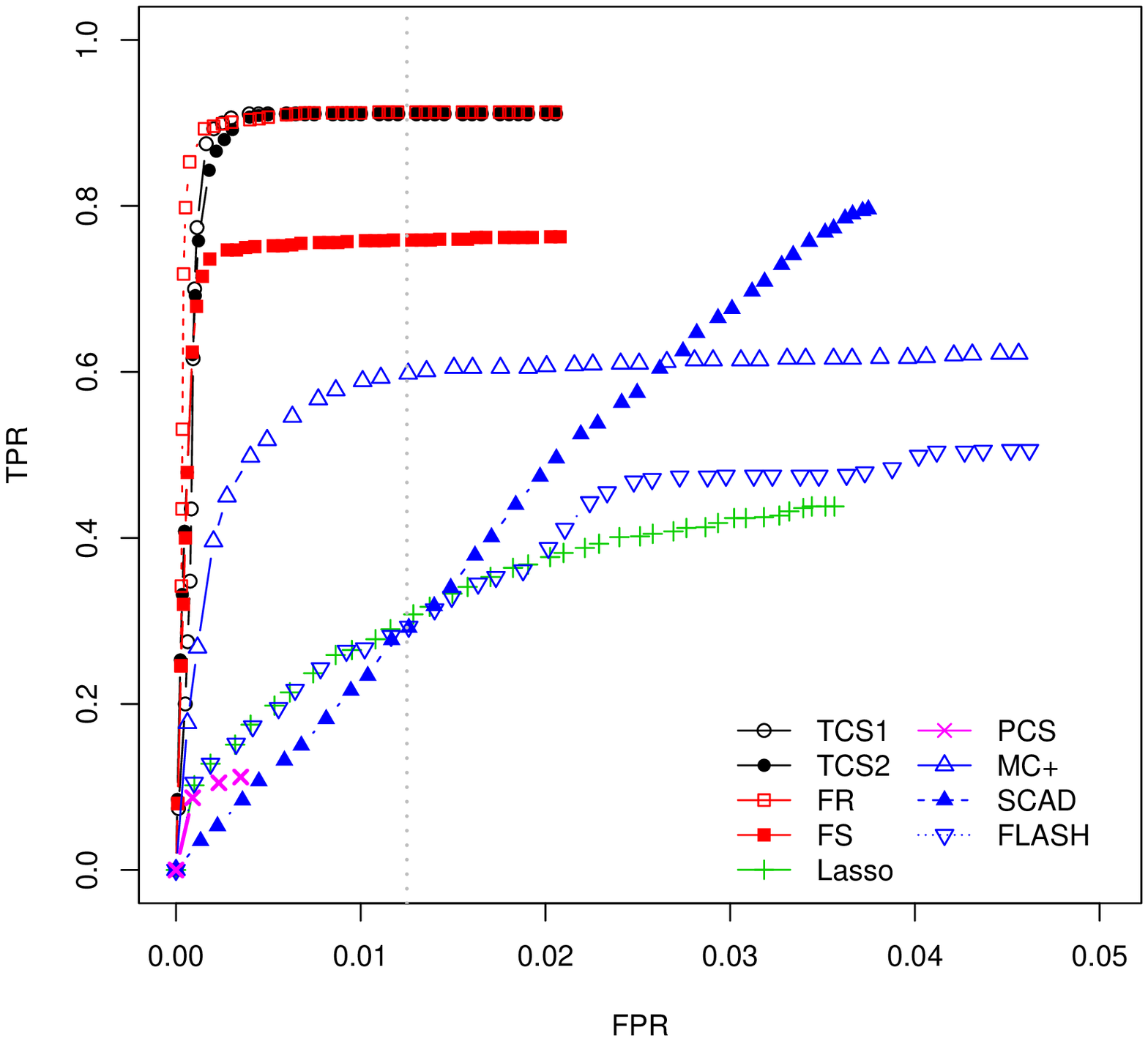, width=0.33\linewidth,clip=} & 
\epsfig{file=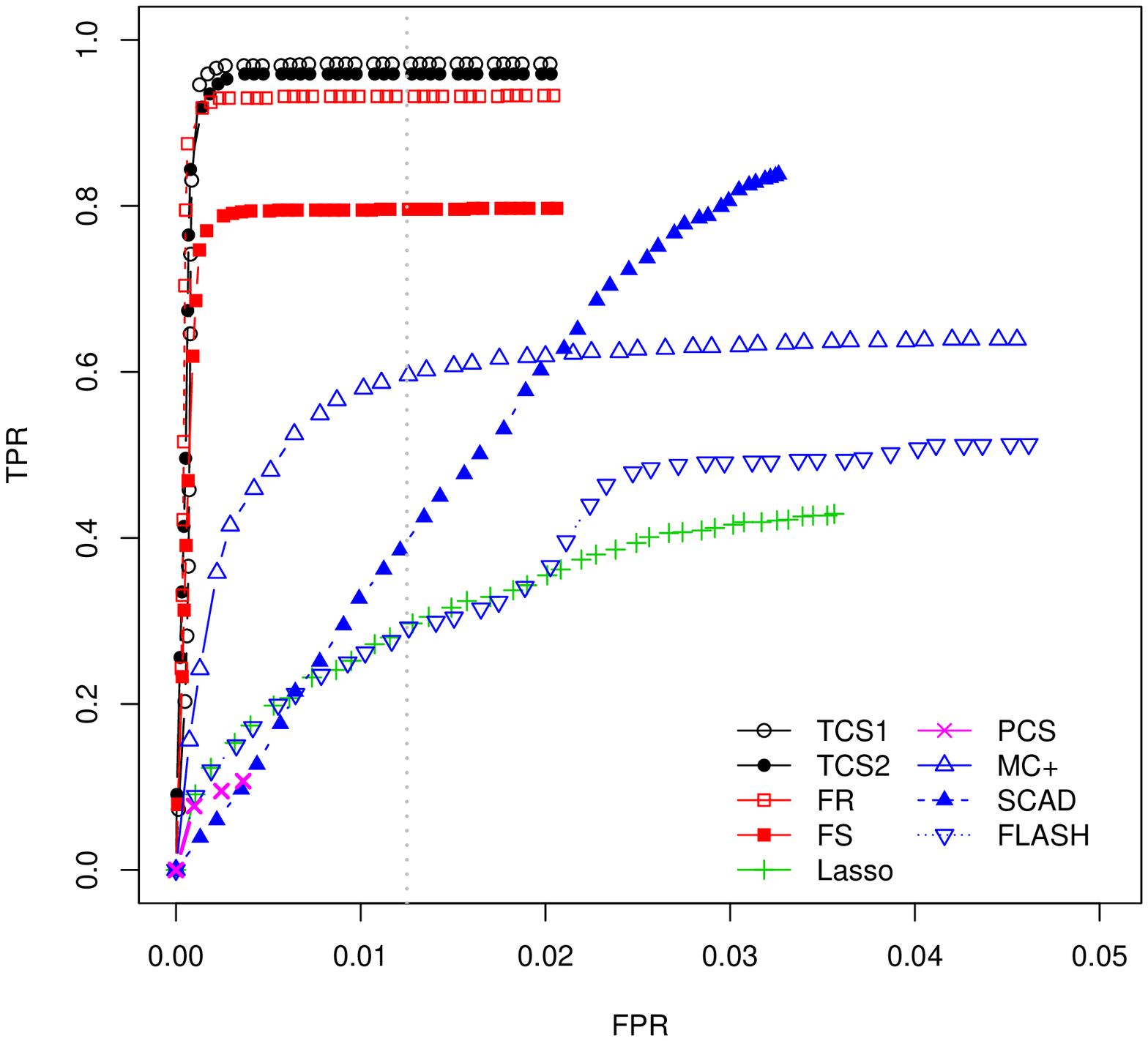, width=0.33\linewidth,clip=} 
\end{tabular}
\caption{\small{ROC curves for the simulation model (A) with $n=100$: 
TCS1 (black empty circle), TCS2 (black filled circle), FR (red empty square), FS (red filled square),
Lasso (green crossed circle) PC-simple algorithm (magenta two triangles), MC+ (blue empty triangle),
SCAD (blue filled triangle) and FLASH (blue reversed triangle); FPR$=2.5|\cS|/p$ (vertical dotted);
first row: $p=500$, second row: $p=1000$, third row: $p=2000$; 
first column: $R^2=0.3$, second column: $R^2=0.6$, third column: $R^2=0.9$.}}
\end{figure}

\begin{figure}
\label{fig:roc:c}
\centering
\begin{tabular}{ccc}
\epsfig{file=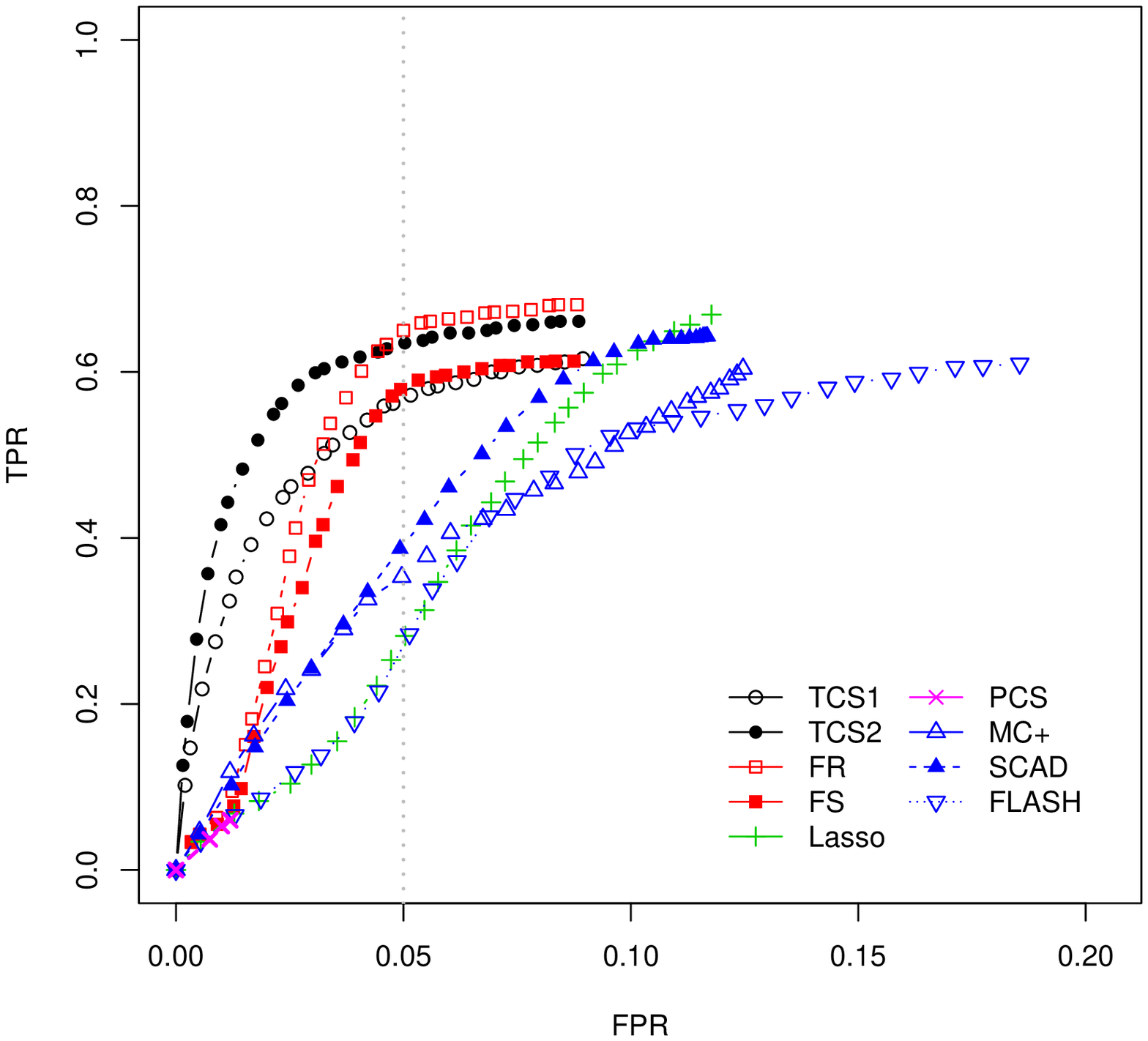, width=0.33\linewidth,clip=} & 
\epsfig{file=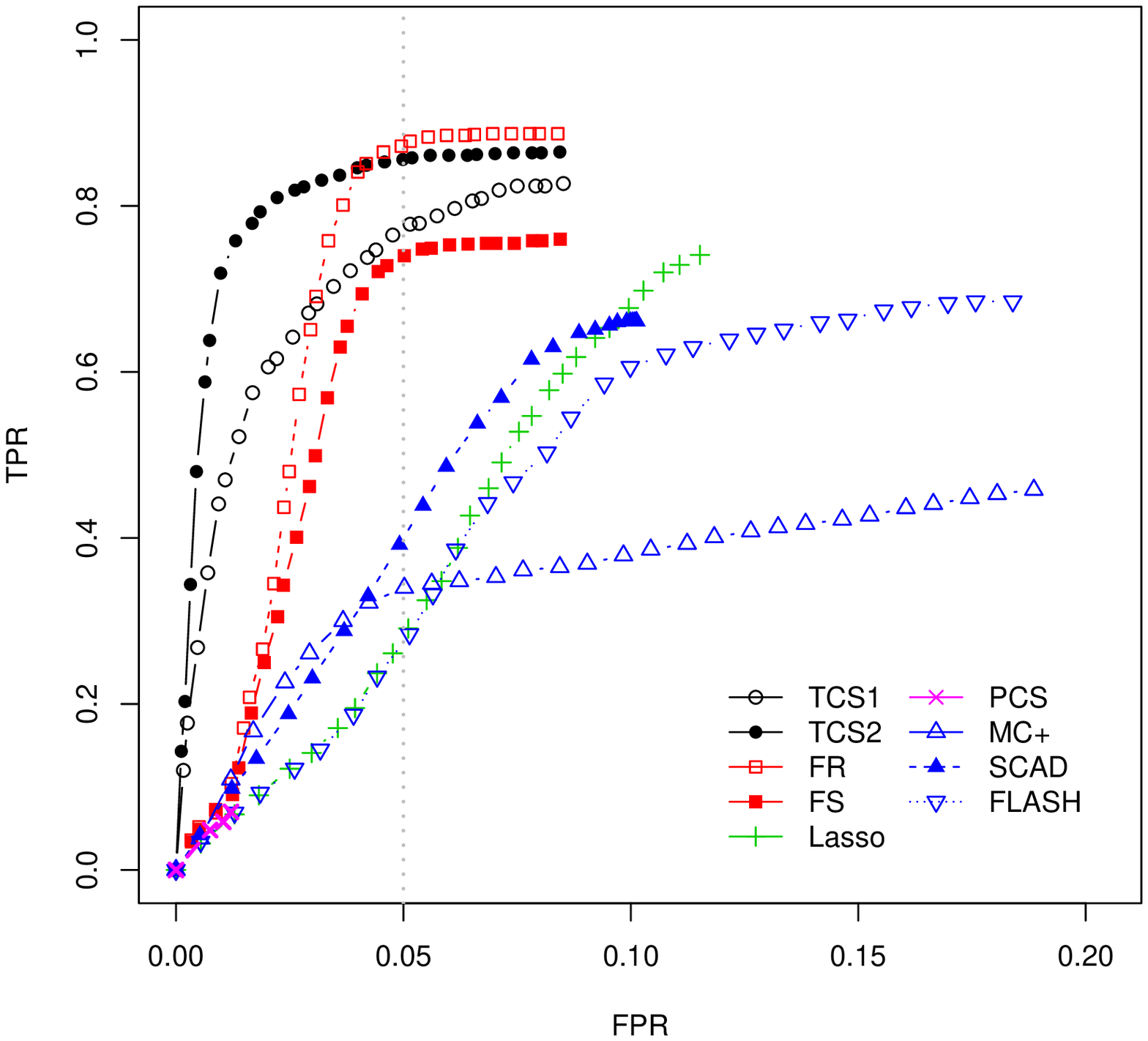, width=0.33\linewidth,clip=} & 
\epsfig{file=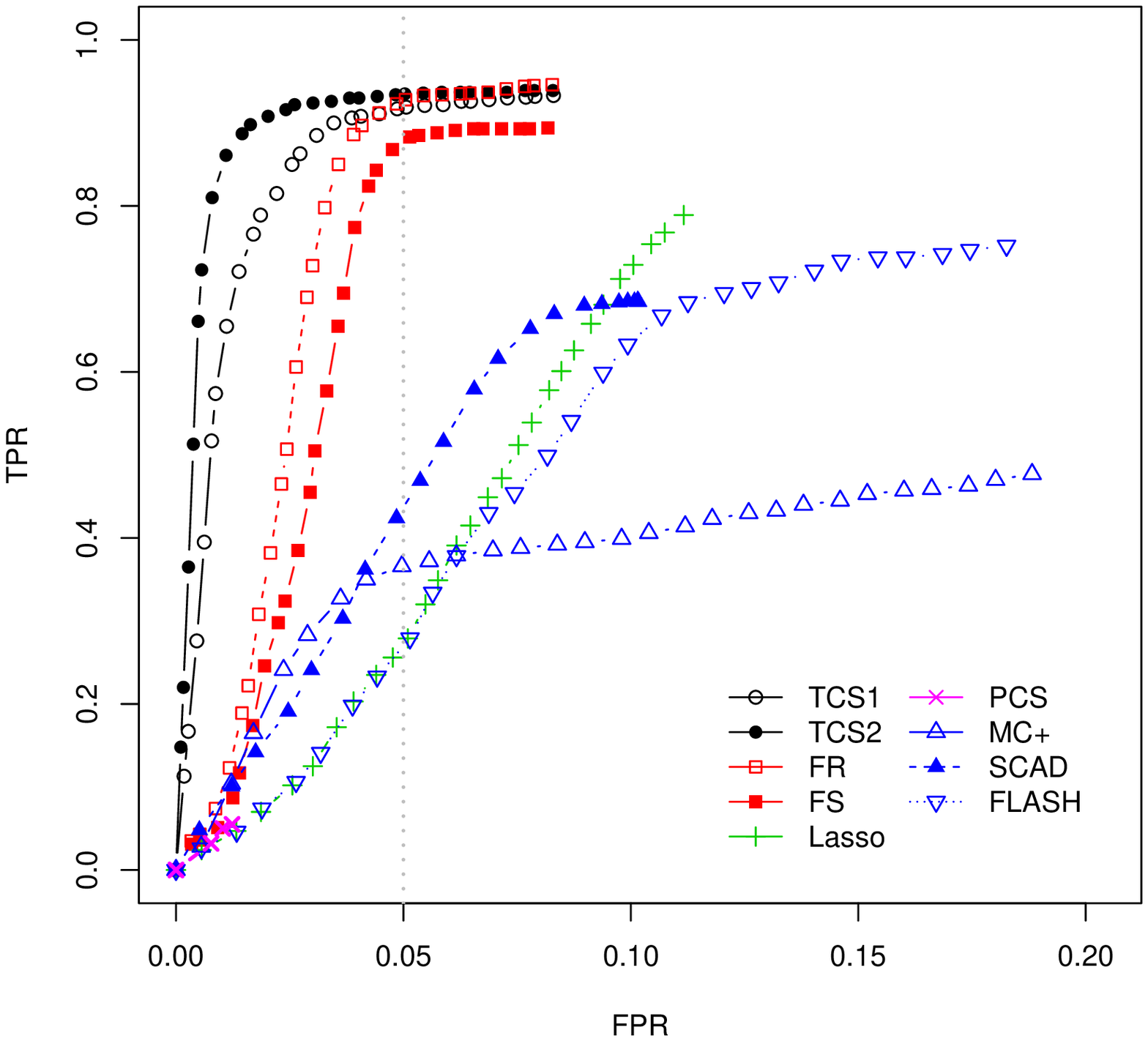, width=0.33\linewidth,clip=} 
\\
\epsfig{file=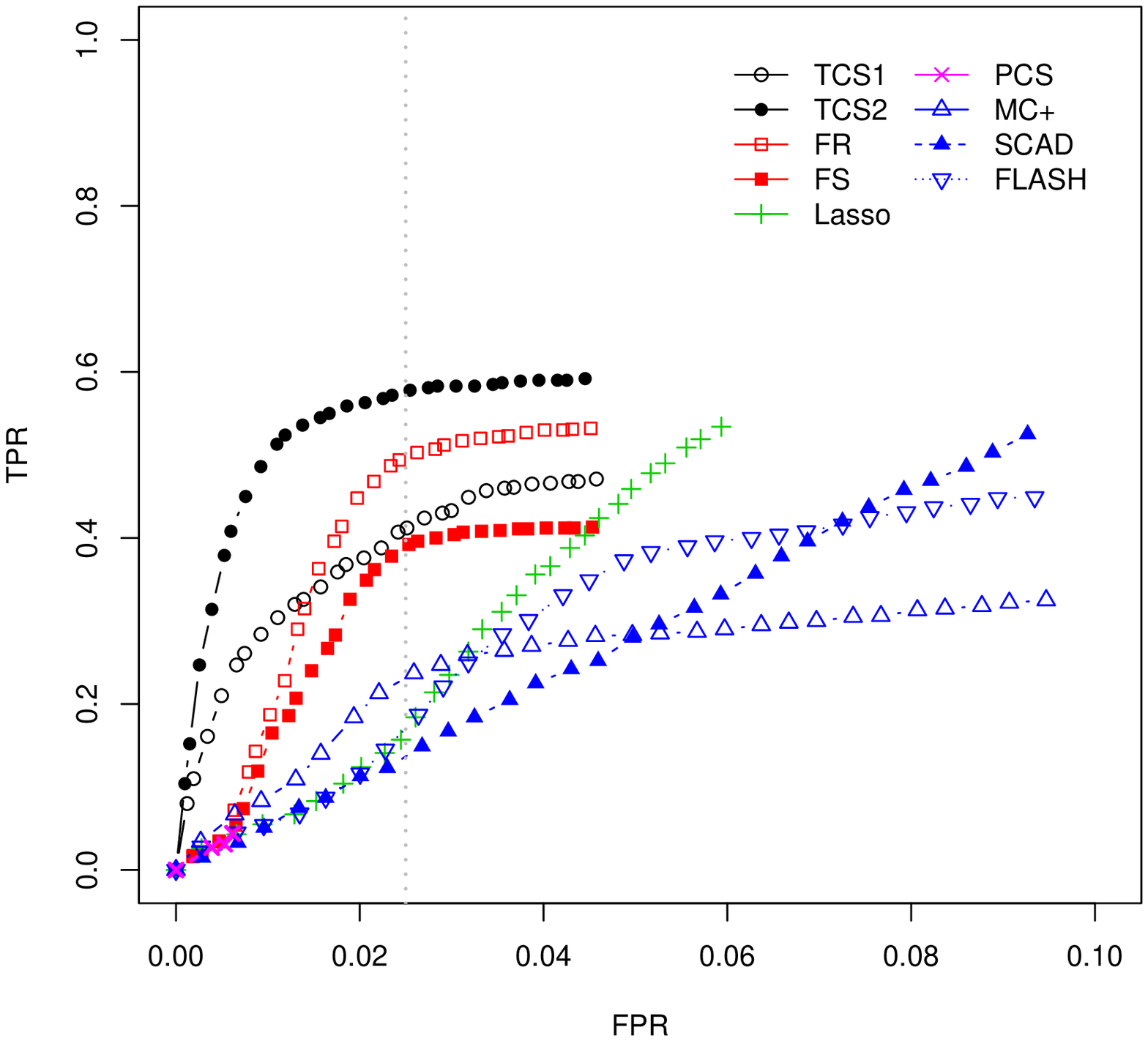, width=0.33\linewidth,clip=} & 
\epsfig{file=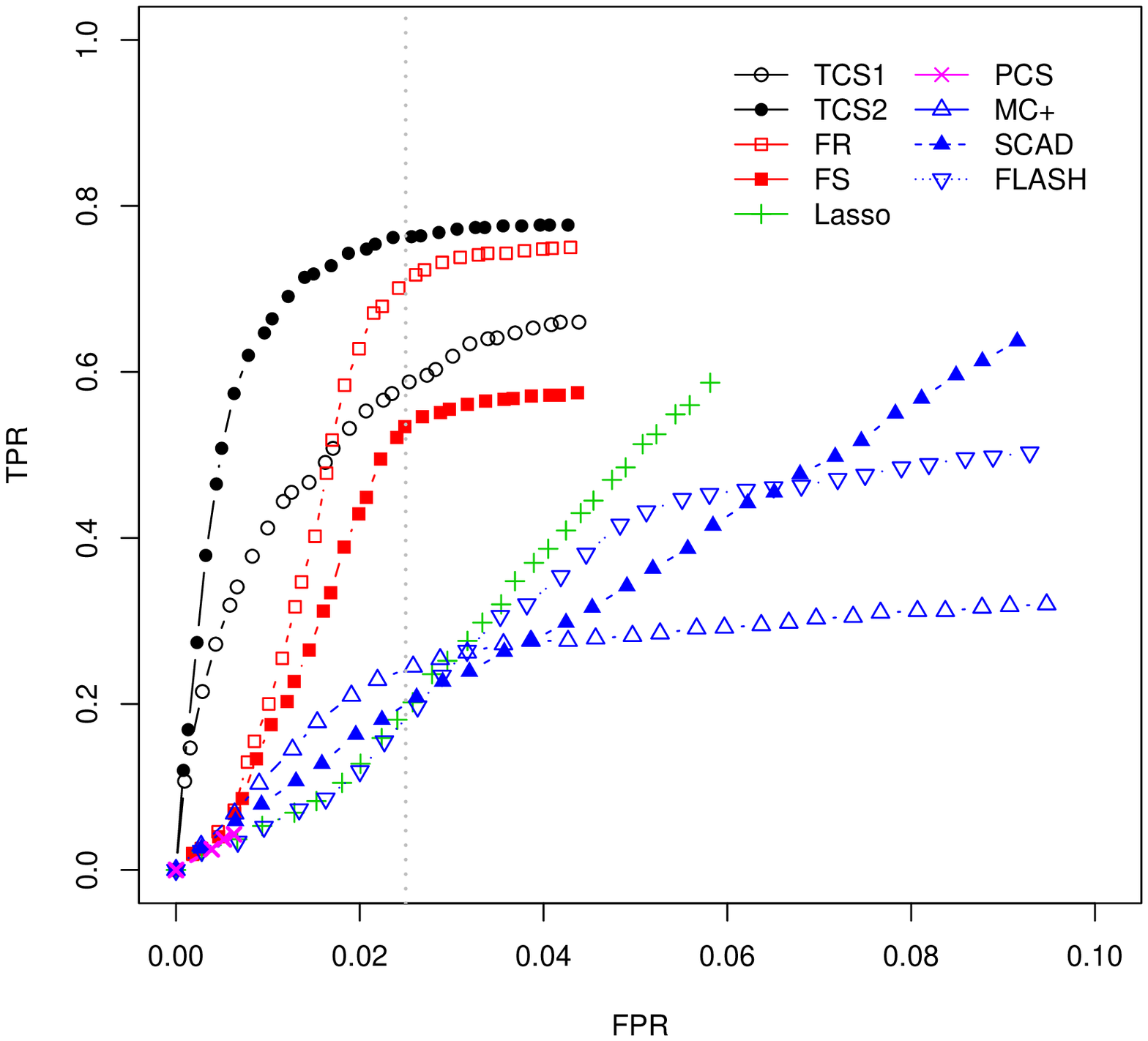, width=0.33\linewidth,clip=} & 
\epsfig{file=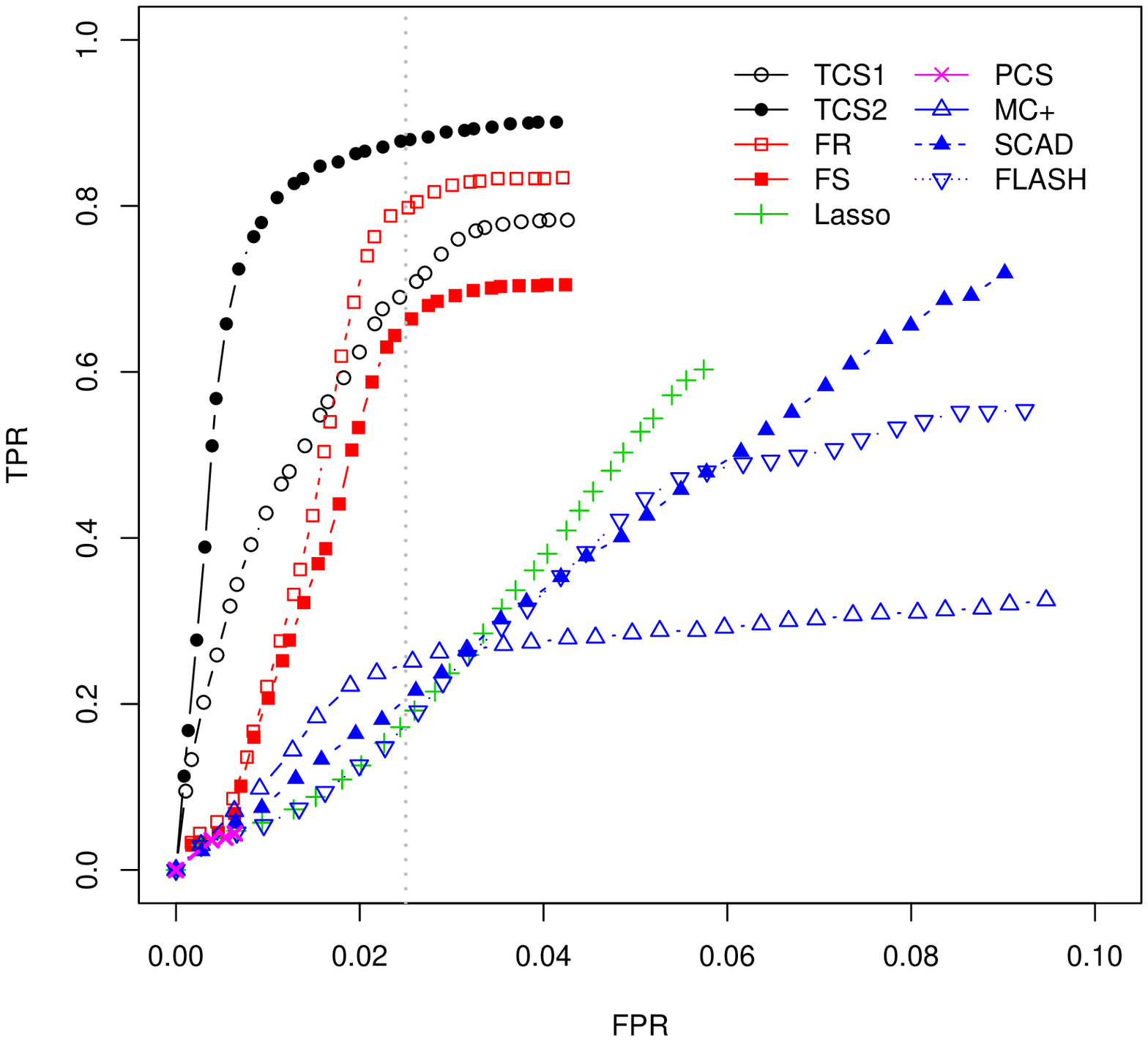, width=0.33\linewidth,clip=} 
\\
\epsfig{file=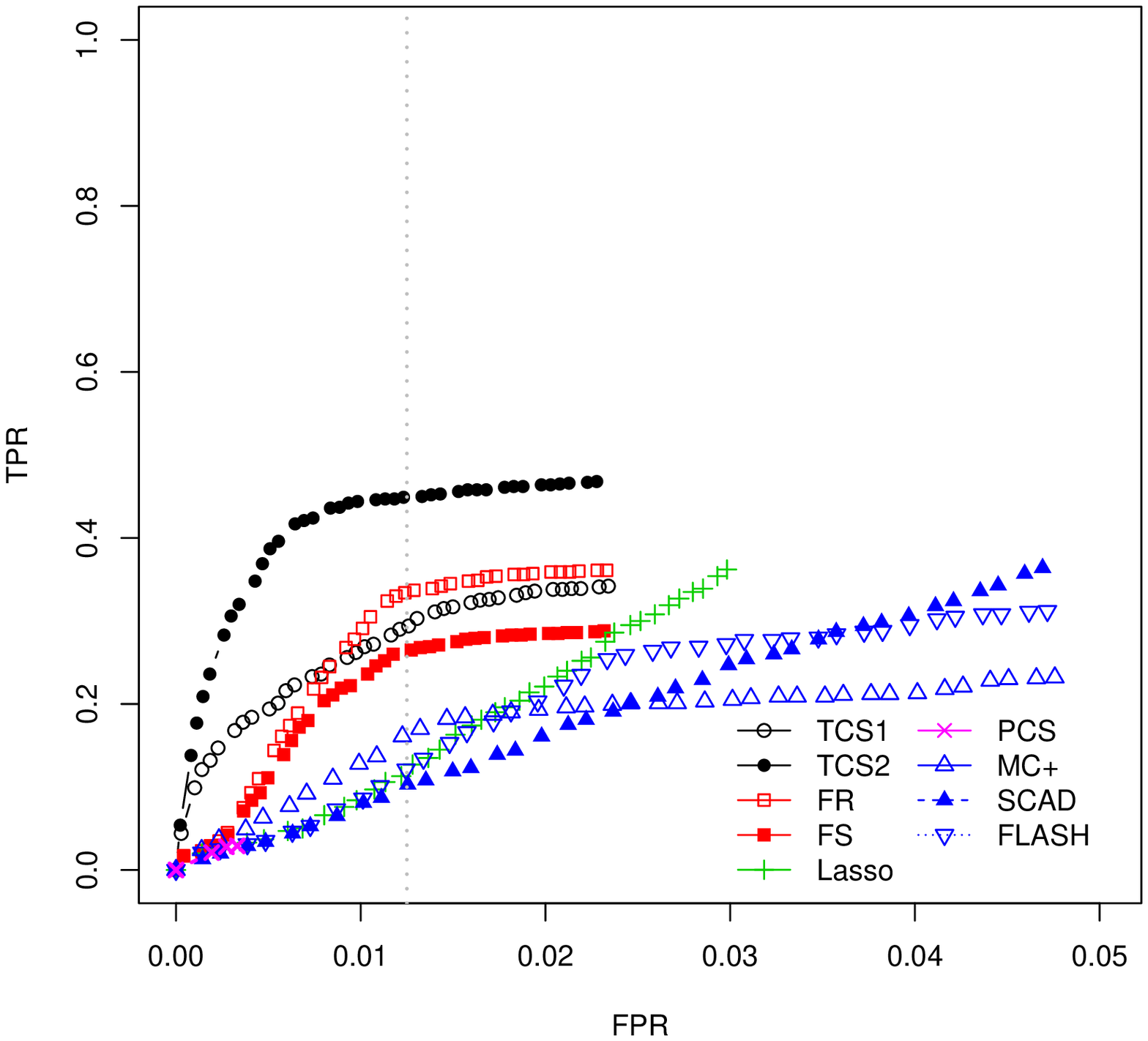, width=0.33\linewidth,clip=} & 
\epsfig{file=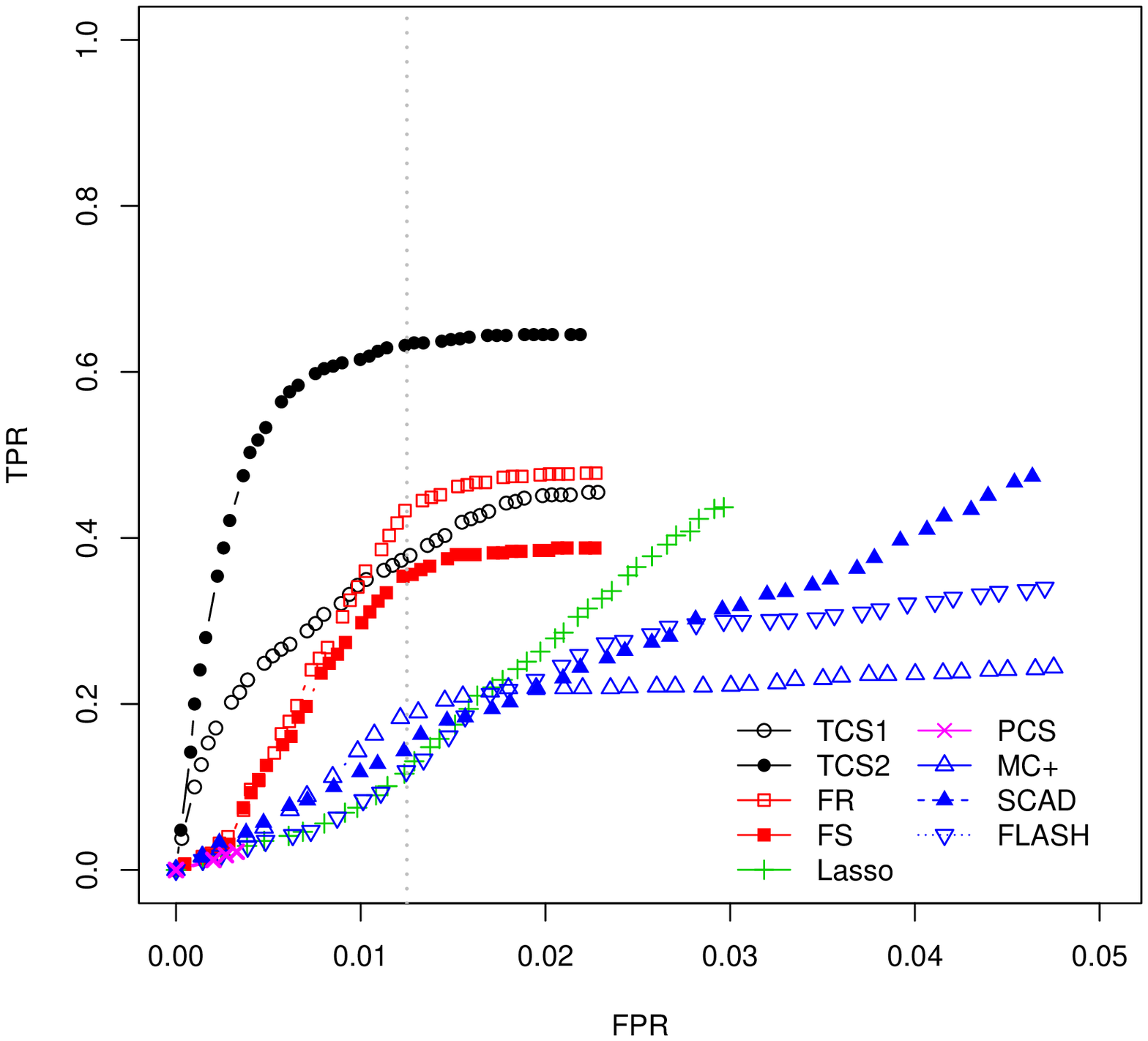, width=0.33\linewidth,clip=} & 
\epsfig{file=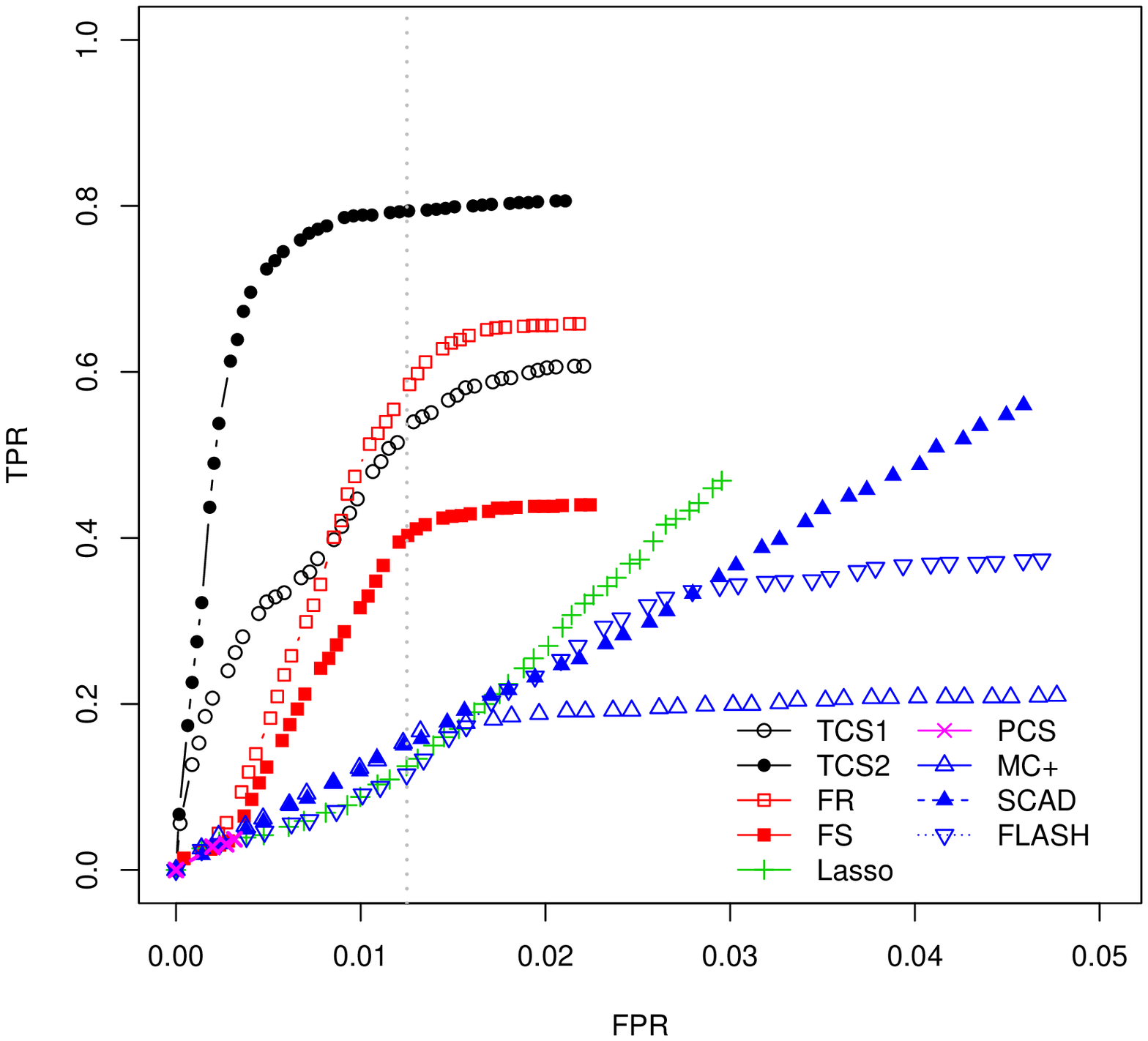, width=0.33\linewidth,clip=} 
\end{tabular}
\caption{\small{ROC curves for the simulation model (C) with $n=100$.}}
\end{figure}

\begin{figure}
\label{fig:roc:de}
\centering
\begin{tabular}{cc}
\epsfig{file=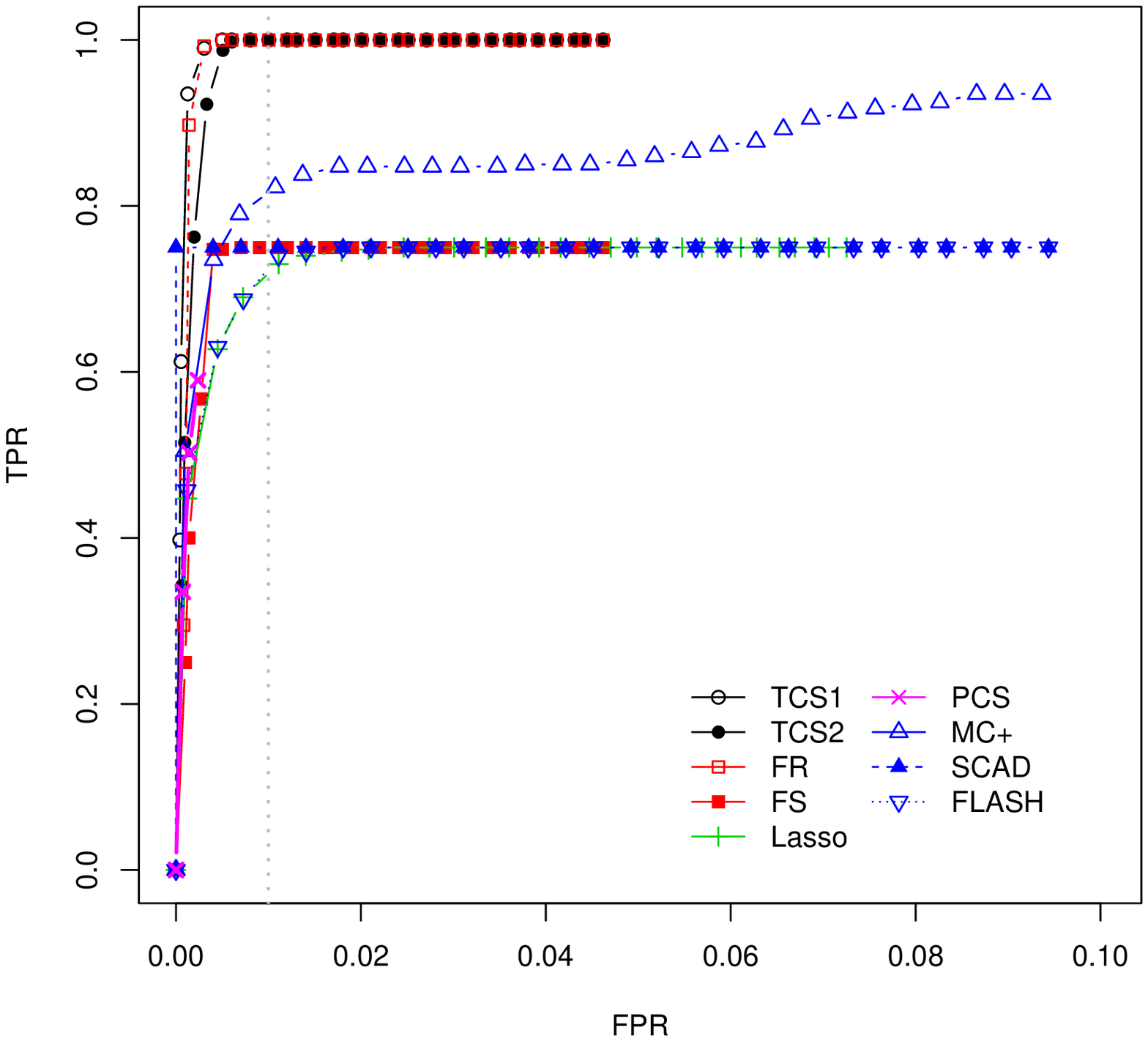, width=0.45\linewidth,clip=} &  
\epsfig{file=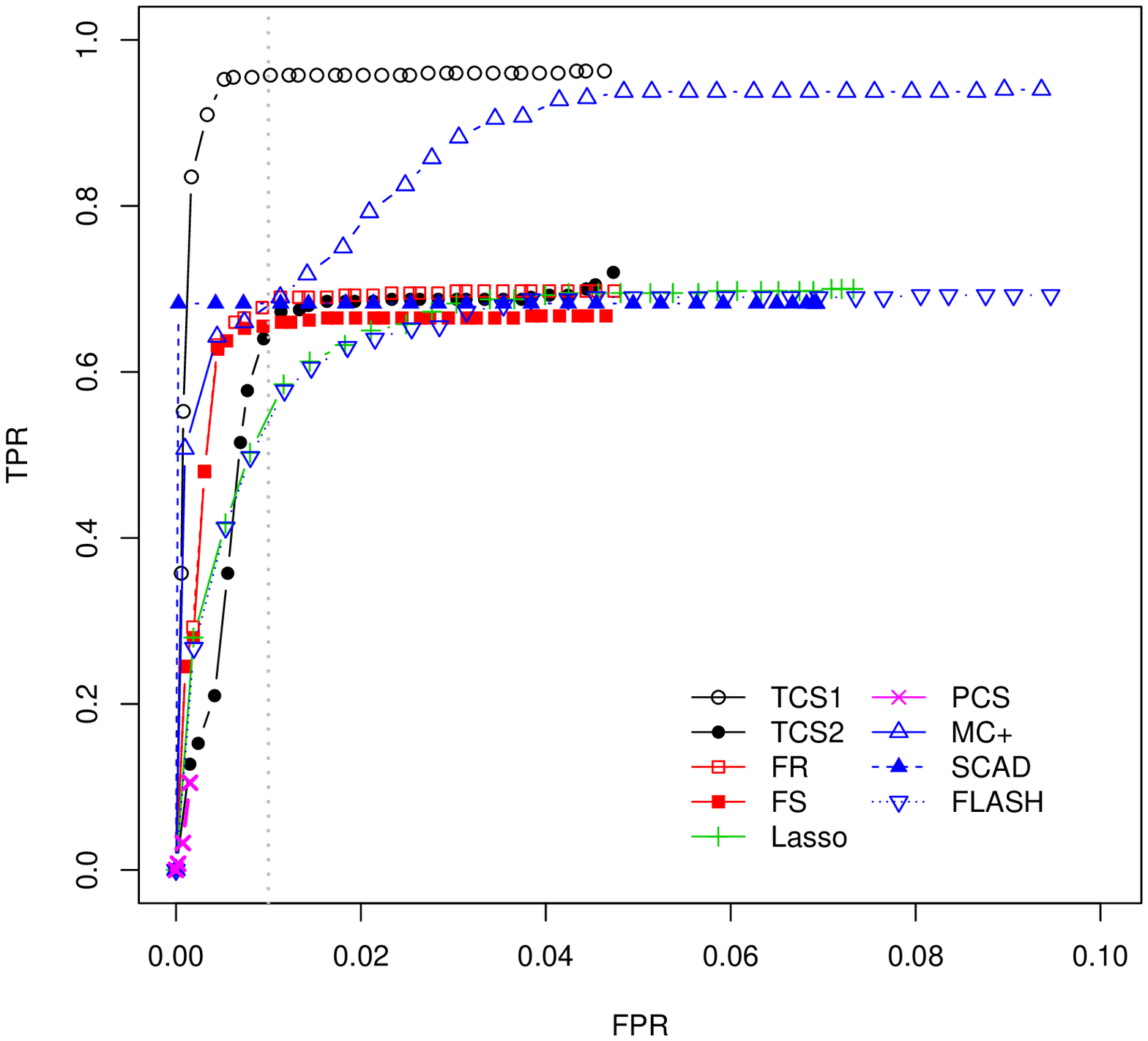, width=0.45\linewidth,clip=} 
\\
\epsfig{file=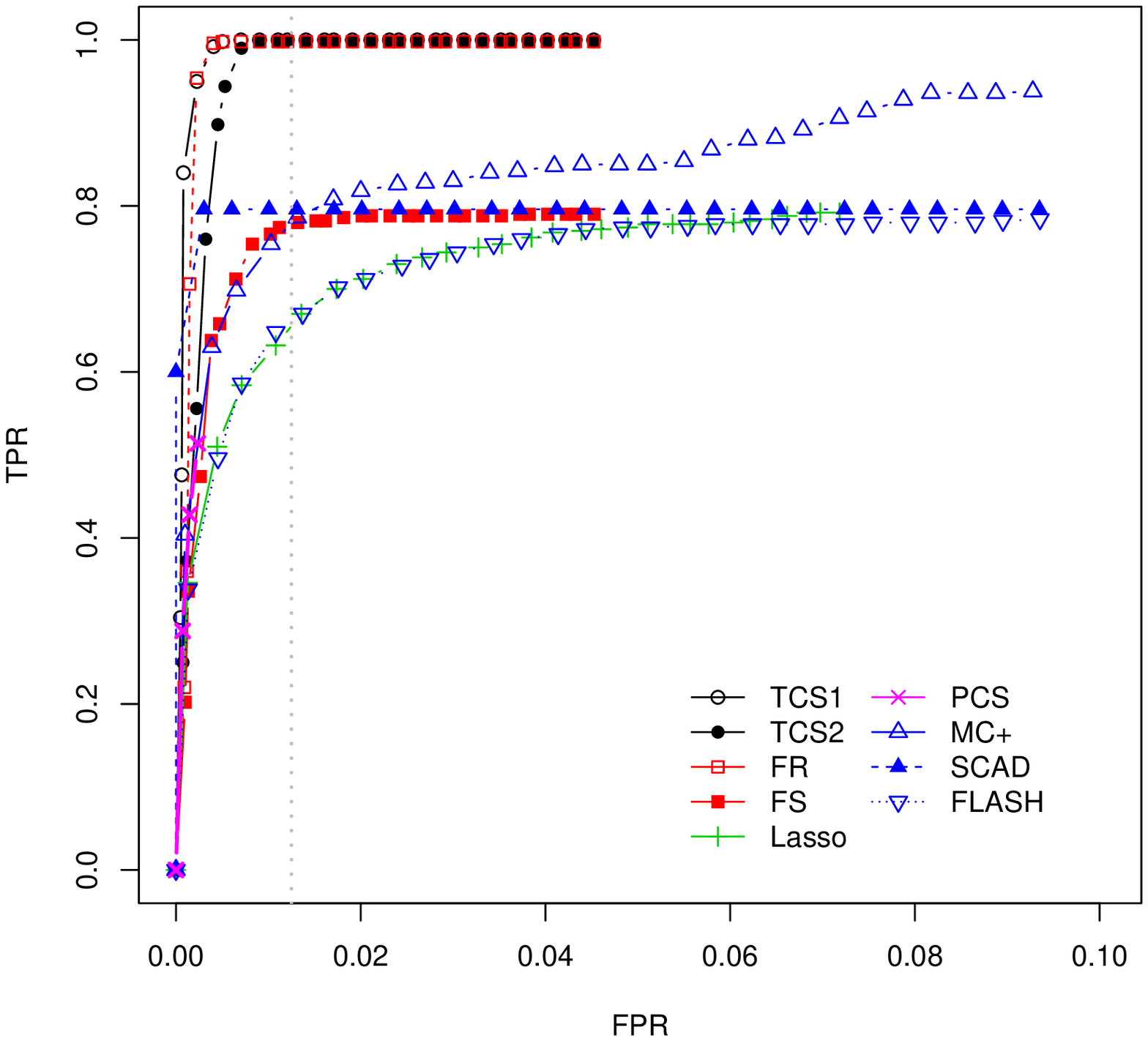, width=0.45\linewidth,clip=} &  
\epsfig{file=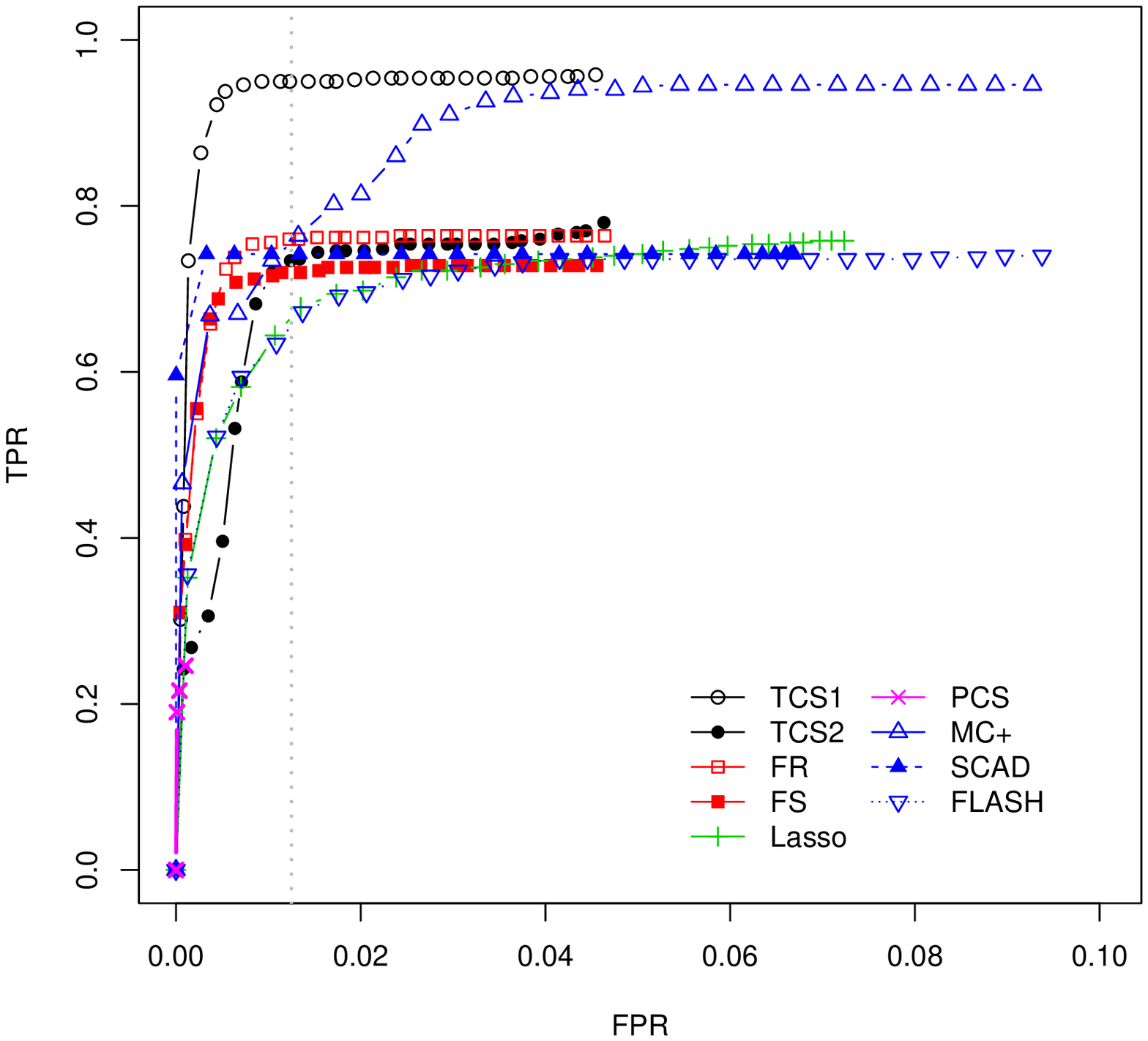, width=0.45\linewidth,clip=} 
\end{tabular}
\caption{\small{ROC curves for the simulation models (D) (first row) and (E) (second row) with $n=100$; 
first column: $\varphi=0.5$, second column: $\varphi=0.95$.}}
\end{figure}

\begin{figure}
\centering
\begin{tabular}{ccc}
\epsfig{file=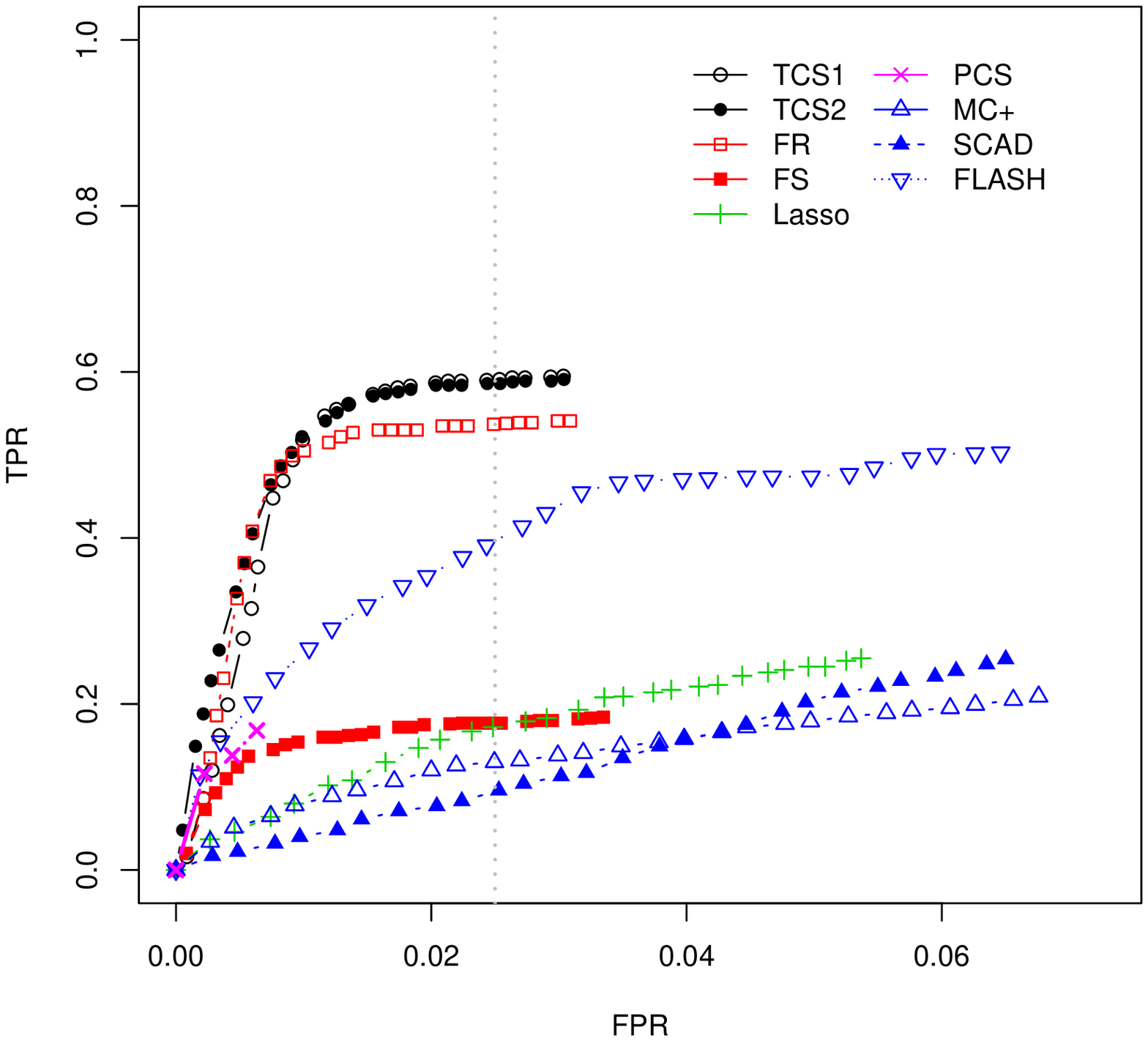, width=0.33\linewidth,clip=} &  
\epsfig{file=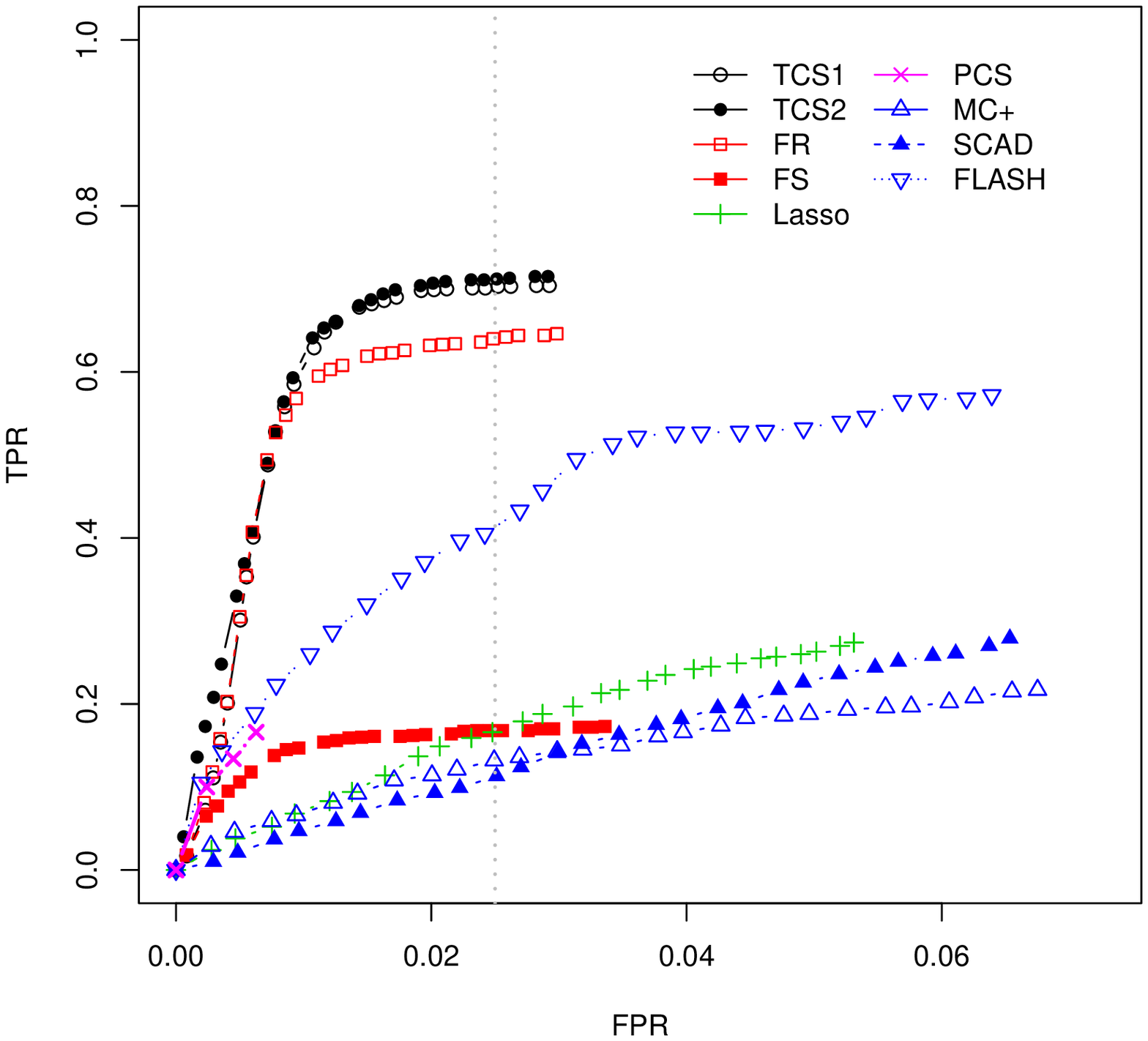, width=0.33\linewidth,clip=} & 
\epsfig{file=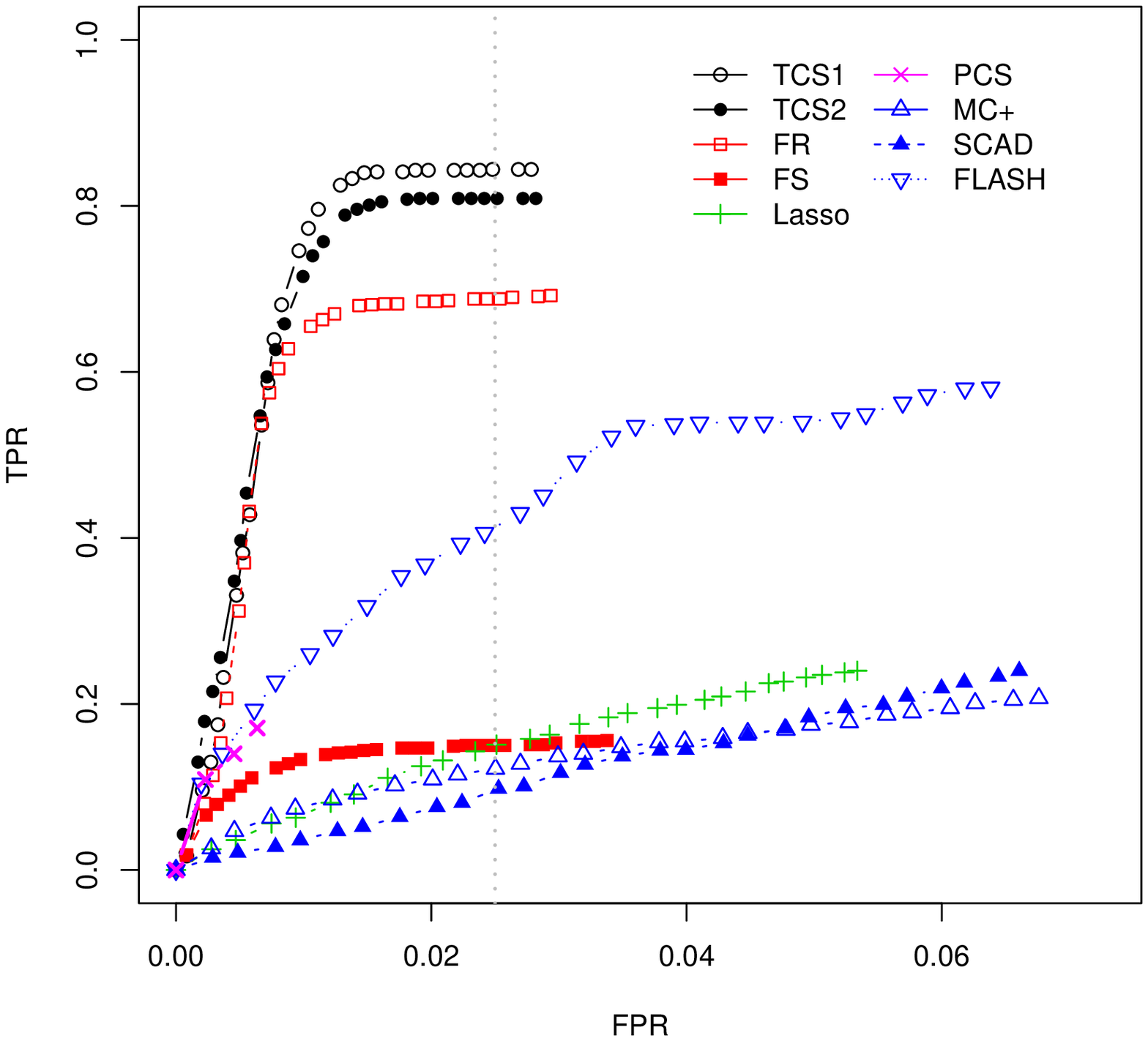, width=0.33\linewidth,clip=} 
\\
\epsfig{file=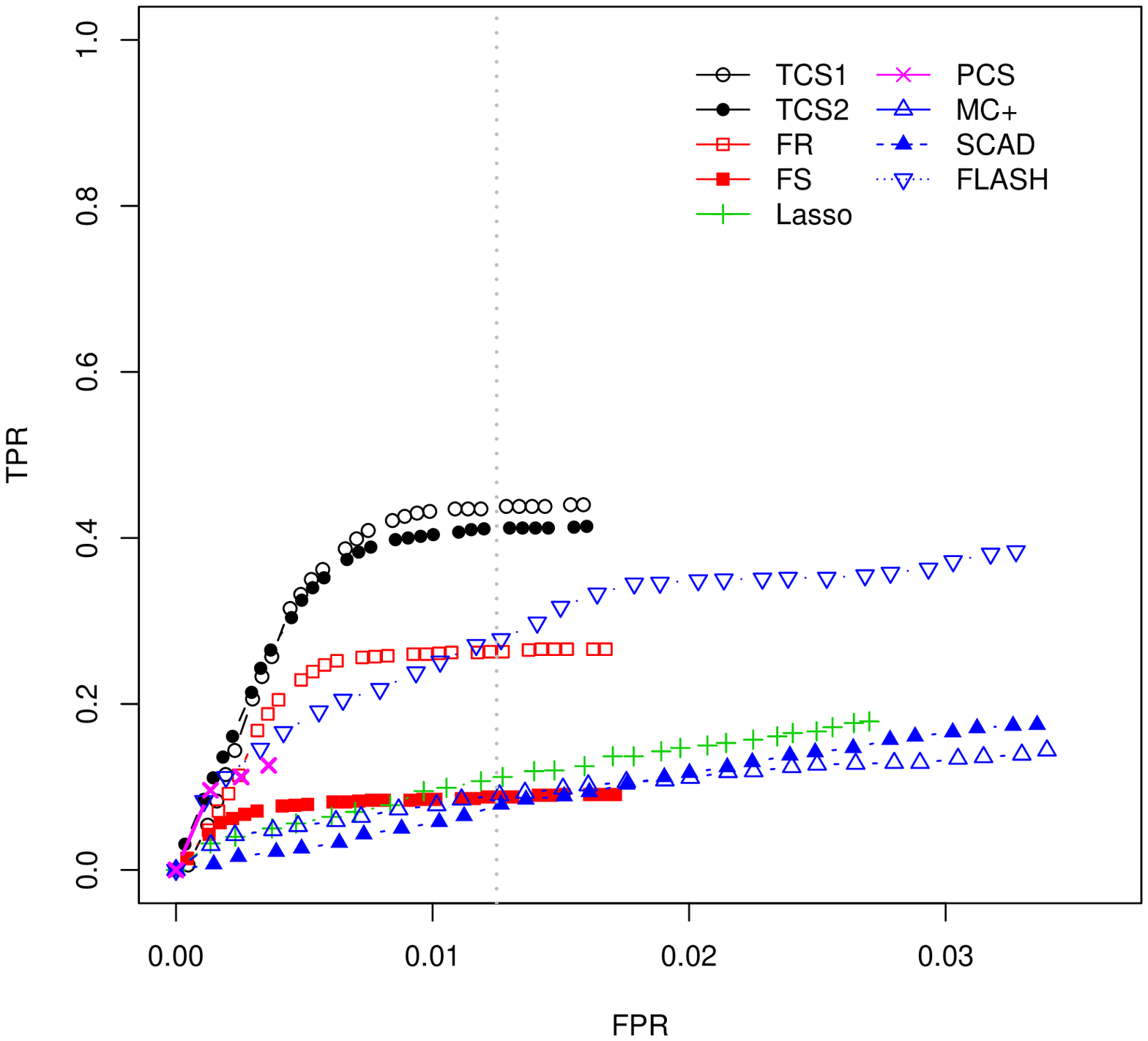, width=0.33\linewidth,clip=} &  
\epsfig{file=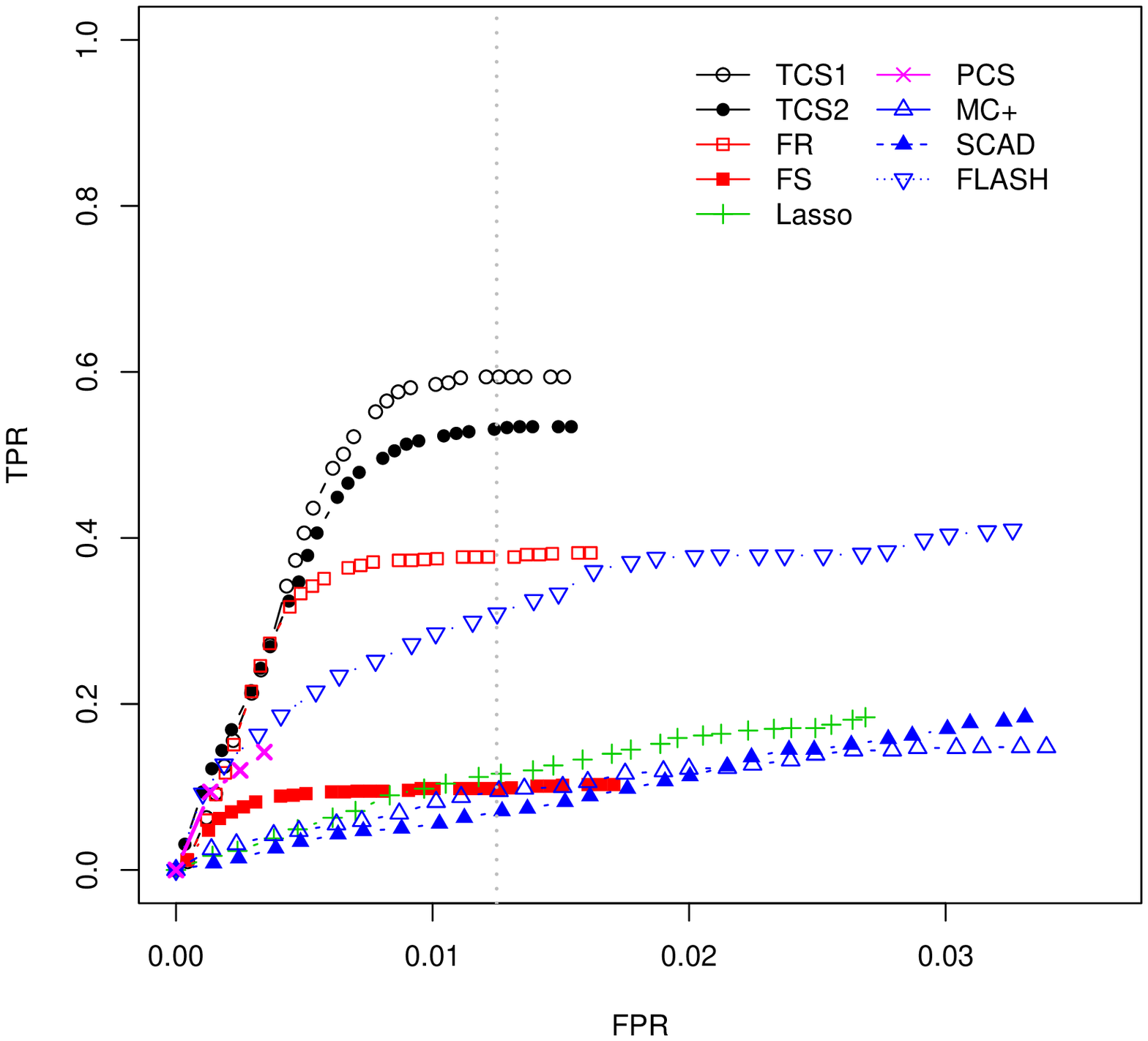, width=0.33\linewidth,clip=} & 
\epsfig{file=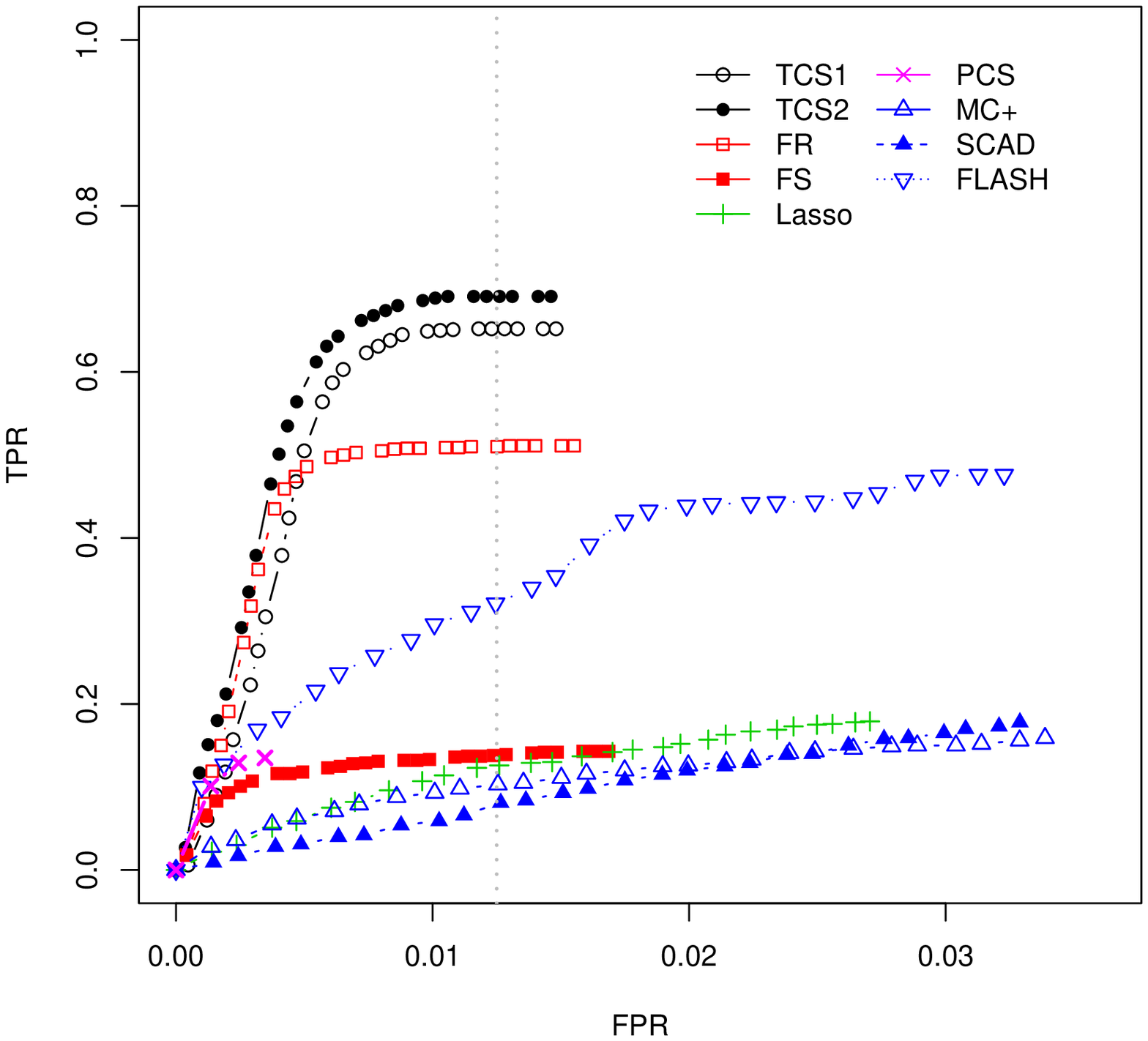, width=0.33\linewidth,clip=} 
\end{tabular}
\caption{\small{ROC curves for the simulation model (F) with $n=72$;
first row: $p=1000$, second row: $p=2000$; 
first column: $R^2=0.3$, second column: $R^2=0.6$, third column: $R^2=0.9$.}}
\label{fig:roc:f}
\end{figure}

For each method and simulation setting, we report the following error measures which are often adopted to evaluate the performance of variable selection:
the number of False Positives (FP, the number of irrelevant variables incorrectly identified as relevant), 
the number of False Negatives (FN, the number of relevant variables incorrectly identified as irrelevant)
and the L2 distance $\Vert \bbeta-\bhbeta \Vert_2^2$; all averaged over 100 simulated data sets.
The summary of the simulation results can be found in Tables \ref{table:sim:a}--\ref{table:sim:f}.
We also present the receiver operating characteristic (ROC) curves,
which plot the true positive rate (TPR) against the false positive rate (FPR),
in Figures \ref{fig:roc:a}--\ref{fig:roc:f}.
Note that the simulation results from model (B) are discussed in the text only 
and the corresponding figure and table are omitted for brevity.
The steep slope of an ROC implies that relevant variables have been selected without including too many irrelevant ones.
Vertical lines are plotted as a guideline to indicate when the FPR reaches $2.5|\cS|/p$.
Since the existing R implementation of ISIS (package \verb+SIS+) returns the final selection
of variables only, rather than an entire path, we did not produce the ROC curves for that
method.

\begin{table}
\caption{\footnotesize{Simulation results for model (A) with $|\cS|=10$. Results in bold font mean the
value of FP+FN is the lowest or within 10\% of the lowest; the same for L2. The value of
0 means less than $5 \times 10^{-4}$}.}
\label{table:sim:a}
\centering
\footnotesize{
\begin{tabular}{c|c|c|c|c|c|c|c|c|c|c|c|c}
\hline
\hline
 $p$	& $R^2$ &	& TCS1 &	TCS2 &	FR &	FS &	Lasso &	ISIS &	PCS &	MC+ &	SCAD & FLASH	\\
\hline
500 &	0.3 &	FP &	1.2 &	0.55 &	3.8 &	1.04 &	44.93 &	1.06 &	4.59 &	5.33 &	57.28 &	5.66	\\		   
&	&	FN &	2.47 &	2.52 &	1.82 &	2.2 &	2.93 &	9.18 &	8.45 &	4.31 &	1.8 &	2.9	\\		   
&	&	FP+FN &	3.67 &	\textbf{3.07} &	5.62 &	\textbf{3.24} &	47.86 &	10.24 &	13.04 &	9.64 &	59.08 &	8.56	\\		   
&	&	L2 &	\textbf{0.012} &	\textbf{0.012} &	\textbf{0.012} &	\textbf{0.013} &	0.264 &	1.006 &	0.914 &	0.134 &	0.096 &	0.081	\\	\hline	   
&	0.6 &	FP &	1.05 &	0.74 &	4.49 &	1.07 &	47.92 &	1.09 &	4.76 &	3.25 &	40.76 &	6.45	\\		   
&	&	FN &	1.07 &	1.12 &	0.87 &	1.16 &	2.24 &	9.29 &	8.45 &	1.96 &	1.06 &	1.94	\\		   
&	&	FP+FN &	2.12 &	\textbf{1.86 }&	5.36 &	2.23 &	50.16 &	10.38 &	13.21 &	5.21 &	41.82 &	8.39	\\		   
&	&	L2 &	\textbf{0.002 }&	\textbf{0.002} &	0.003 &	0.055 &	0.242 &	1.021 &	0.812 &	0.042 &	0.04 &	0.106	\\	\hline	   
&	0.9 &	FP &	0.92 &	0.57 &	2.64 &	1.17 &	47.97 &	1.06 &	4.52 &	1.57 &	27.48 &	7.15	\\		   
&	&	FN &	0.43 &	0.41 &	0.32 &	0.62 &	1.75 &	9.21 &	8.37 &	2.03 &	0.58 &	1.49	\\		   
&	&	FP+FN &	1.35 &	\textbf{0.98} &	2.96 &	1.79 &	49.72 &	10.27 &	12.89 &	3.6 &	28.06 &	8.64	\\		   
&	&	L2 &	\textbf{0} &	\textbf{0} &	0.001 &	0.074 &	0.292 &	1.075 &	0.982 &	0.085 &	0.02 &	0.205	\\	\hline\hline	   
1000 &	0.3 &	FP &	1.79 &	1.38 &	22.28 &	1.61 &	44.56 &	1.38 &	5.53 &	6.6 &	69.77 &	10.05	\\		   
&	&	FN &	2.18 &	2.54 &	1.41 &	2.31 &	4.73 &	9.48 &	8.73 &	5.21 &	2.34 &	4.76	\\		   
&	&	FP+FN &	\textbf{3.97} &	\textbf{3.92} &	23.69 &	\textbf{3.92} &	49.29 &	10.86 &	14.26 &	11.81 &	72.11 &	14.81	\\		   
&	&	L2 &	\textbf{0.01} &	0.027 &	0.035 &	0.039 &	0.463 &	1.073 &	0.787 &	0.219 &	0.159 &	0.318	\\	\hline	   
&	0.6 &	FP &	1.67 &	1.35 &	24.91 &	1.3 &	46 &	1.19 &	5.55 &	4.39 &	55.93 &	7.76	\\		   
&	&	FN &	1.13 &	1.17 &	0.78 &	1.65 &	4.16 &	9.33 &	8.63 &	2.79 &	1.51 &	3.33	\\		   
&	&	FP+FN &	2.8 &	\textbf{2.52} &	25.69 &	2.95 &	50.16 &	10.52 &	14.18 &	7.18 &	57.44 &	11.09	\\		   
&	&	L2 &	\textbf{0.002} &	\textbf{0.003} &	0.009 &	0.126 &	0.498 &	1.016 &	0.868 &	0.13 &	0.117 &	0.32	\\	\hline	   
&	0.9 &	FP &	1.21 &	0.8 &	25.75 &	1.11 &	47.38 &	1.23 &	5.45 &	1.84 &	43.93 &	7.42	\\		   
&	&	FN &	0.43 &	0.45 &	0.3 &	1.1 &	3.86 &	9.38 &	8.69 &	2.58 &	0.94 &	2.61	\\		   
&	&	FP+FN &	1.64 &	\textbf{1.25 }&	26.05 &	2.21 &	51.24 &	10.61 &	14.14 &	4.42 &	44.87 &	10.03	\\		   
&	&	L2 &	\textbf{0} &	\textbf{0} &	0.002 &	0.088 &	0.405 &	0.916 &	0.803 &	0.078 &	0.063 &	0.192	\\	\hline\hline	   
2000 &	0.3 &	FP &	1.77 &	1.65 &	41.53 &	1.64 &	38.27 &	1.48 &	6.71 &	10.53 &	80.9 &	9.07	\\		   
&	&	FN &	2.33 &	2.36 &	1.53 &	3.48 &	6.48 &	9.59 &	8.98 &	5.79 &	3.07 &	6.22	\\		   
&	&	FP+FN &	\textbf{4.1} &	\textbf{4.01 }&	43.06 &	5.12 &	44.75 &	11.07 &	15.69 &	16.32 &	83.97 &	15.29	\\		   
&	&	L2 &	\textbf{0.013 }&	0.016 &	0.047 &	0.116 &	0.603 &	0.99 &	0.804 &	0.311 &	0.199 &	0.467	\\	\hline	   
&	0.6 &	FP &	1.89 &	1.89 &	40.87 &	1.39 &	41.32 &	1.35 &	6.37 &	6.1 &	66.65 &	7.82	\\		   
&	&	FN &	1.4 &	1.46 &	0.87 &	2.77 &	6.18 &	9.48 &	8.82 &	4.06 &	2.21 &	5.06	\\		   
&	&	FP+FN &	\textbf{3.29} &	\textbf{3.35} &	41.74 &	4.16 &	47.5 &	10.83 &	15.19 &	10.16 &	68.86 &	12.88	\\		   
&	&	L2 &	\textbf{0.004} &	\textbf{0.004} &	0.024 &	0.252 &	0.752 &	1.243 &	0.989 &	0.338 &	0.18 &	0.496	\\	\hline	   
&	0.9 &	FP &	1.61 &	1.32 &	39.5 &	1.45 &	39 &	1.35 &	6.87 &	19.99 &	59.73 &	6.96	\\		   
&	&	FN &	0.44 &	0.56 &	0.68 &	2.21 &	6.32 &	9.55 &	8.9 &	3.88 &	1.6 &	5.11	\\		   
&	&	FP+FN &	\textbf{2.05} &	\textbf{1.88} &	40.18 &	3.66 &	45.32 &	10.9 &	15.77 &	23.87 &	61.33 &	12.07	\\		   
&	&	L2 &	\textbf{0} &	0.005 &	0.314 &	0.285 &	0.711 &	1.126 &	0.978 &	0.367 &	0.147 &	0.577	\\	\hline\hline	 
\end{tabular}}
\end{table}

\begin{table}
\caption{\footnotesize{Simulation results for model (C) with $|\cS|=10$.
Results in bold font mean the
value of FP+FN is the lowest or within 10\% of the lowest; the same for L2.}}
\label{table:sim:c}
\centering
\footnotesize{
\begin{tabular}{c|c|c|c|c|c|c|c|c|c|c|c|c}
\hline
\hline
 $p$	& $R^2$ &	& TCS1 &	TCS2 &	FR &	FS &	Lasso &	ISIS &	PCS &	MC+ &	SCAD & FLASH	\\
\hline
500 &	0.3 &	FP &	4.21 &	3.57 &	9.56 &	8.44 &	43.82 &	1.81 &	5.73 &	38.84 &	42.23 &	19.97	\\		   
&	&	FN &	6.27 &	5.45 &	5.81 &	7.44 &	3.08 &	9.81 &	9.42 &	4.49 &	3.69 &	5.19	\\		   
&	&	FP+FN &	10.48 &	\textbf{9.02} &	15.37 &	15.88 &	46.9 &	11.62 &	15.15 &	43.33 &	45.92 &	25.16	\\		   
&	&	L2 &	0.207 &	\textbf{0.172} &	0.246 &	0.427 &	\textbf{0.166 }&	0.718 &	0.648 &	0.322 &	0.189 &	0.271	\\	\hline	   
&	0.6 &	FP &	6.57 &	4.44 &	15.67 &	15.61 &	45.36 &	1.83 &	5.78 &	64.69 &	38.82 &	19.07	\\		   
&	&	FN &	3.44 &	2.01 &	1.57 &	3.35 &	1.99 &	9.83 &	9.31 &	5.73 &	3.4 &	4.09	\\		   
&	&	FP+FN &	10.01 &	\textbf{6.45} &	17.24 &	18.96 &	47.35 &	11.66 &	15.09 &	70.42 &	42.22 &	23.16	\\		   
&	&	L2 &	0.066 &	0.024 &	\textbf{0.019} &	0.114 &	0.093 &	0.858 &	0.782 &	0.36 &	0.164 &	0.207	\\	\hline	   
&	0.9 &	FP &	6.89 &	3.49 &	16.22 &	17.58 &	48.62 &	1.79 &	5.9 &	58.78 &	39.17 &	18.66	\\		   
&	&	FN &	1.06 &	0.86 &	0.63 &	1.43 &	1.01 &	9.79 &	9.47 &	5.7 &	3.16 &	3.16	\\		   
&	&	FP+FN &	7.95 &	\textbf{4.35} &	16.85 &	19.01 &	49.63 &	11.58 &	15.37 &	64.48 &	42.33 &	21.82	\\		   
&	&	L2 &	0.011 &	\textbf{0.002} &	0.025 &	0.078 &	0.035 &	0.82 &	0.752 &	0.374 &	0.157 &	0.2	\\	\hline\hline	   
1000 &	0.3 &	FP &	2.29 &	3.45 &	8 &	6.73 &	45.22 &	1.92 &	5.86 &	109.1 &	114.8 &	19.63	\\		   
&	&	FN &	7.9 &	5.77 &	7.75 &	8.67 &	4.33 &	9.92 &	9.58 &	6.48 &	3.63 &	6.92	\\		   
&	&	FP+FN &	10.19 &	\textbf{9.22} &	15.75 &	15.4 &	49.55 &	11.84 &	15.44 &	115.6 &	118.4 &	26.55	\\		   
&	&	L2 &	0.558 &	\textbf{0.342} &	0.694 &	0.835 &	0.414 &	0.993 &	0.897 &	0.588 &	\textbf{0.343} &	0.554	\\	\hline	   
&	0.6 &	FP &	5.04 &	4.72 &	15.21 &	11.93 &	48.97 &	1.92 &	6.13 &	90.51 &	110.8 &	19.86	\\		   
&	&	FN &	5.79 &	3.6 &	4.31 &	6.41 &	3.27 &	9.92 &	9.6 &	6.74 &	2.51 &	5.97	\\		   
&	&	FP+FN &	10.83 &	\textbf{8.32} &	19.52 &	18.34 &	52.24 &	11.84 &	15.73 &	97.25 &	113.3 &	25.83	\\		   
&	&	L2 &	0.286 &	\textbf{0.138 }&	0.293 &	0.456 &	0.287 &	1.006 &	0.905 &	0.537 &	0.214 &	0.404	\\	\hline	   
&	0.9 &	FP &	9.15 &	5.44 &	20.3 &	15.99 &	52.41 &	1.8 &	6.23 &	78.06 &	100.4 &	20.67	\\		   
&	&	FN &	3.74 &	1.72 &	2.18 &	4.22 &	2.28 &	9.8 &	9.56 &	6.75 &	1.75 &	5.16	\\		   
&	&	FP+FN &	12.89 &	\textbf{7.16} &	22.48 &	20.21 &	54.69 &	11.6 &	15.79 &	84.81 &	102.1 &	25.83	\\		   
&	&	L2 &	0.258 &	\textbf{0.058} &	0.147 &	0.52 &	0.174 &	1.09 &	0.985 &	0.612 &	0.137 &	0.43	\\	\hline\hline	   
2000 &	0.3 &	FP &	1.75 &	2.25 &	5.12 &	4.97 &	47.13 &	1.89 &	6.4 &	133.6 &	129.4 &	19.9	\\		   
&	&	FN &	8.72 &	7.34 &	9.13 &	9.44 &	5.63 &	9.89 &	9.74 &	7.39 &	4.81 &	7.89	\\		   
&	&	FP+FN &	\textbf{10.47} &	\textbf{9.59} &	14.25 &	14.41 &	52.76 &	11.78 &	16.14 &	141 &	134.3 &	27.79	\\		   
&	&	L2 &	0.649 &	\textbf{0.446} &	0.855 &	0.894 &	0.499 &	0.951 &	0.87 &	0.669 &	\textbf{0.438} &	0.678	\\	\hline	   
&	0.6 &	FP &	3.4 &	4.76 &	11.64 &	6.85 &	49.4 &	1.94 &	6.31 &	187.3 &	125.4 &	20.29	\\		   
&	&	FN &	7.83 &	4.62 &	7.27 &	8.66 &	4.56 &	9.94 &	9.78 &	6.67 &	3.68 &	7.69	\\		   
&	&	FP+FN &	11.23 &	\textbf{9.38} &	18.91 &	15.51 &	53.96 &	11.88 &	16.09 &	194 &	129 &	27.98	\\		   
&	&	L2 &	0.512 &	\textbf{0.164} &	0.629 &	0.761 &	0.418 &	0.943 &	0.857 &	0.566 &	0.31 &	0.675	\\	\hline	   
&	0.9 &	FP &	7.02 &	4.93 &	19.17 &	10.77 &	52.8 &	1.91 &	6.16 &	149.3 &	117.3 &	20.81	\\		   
&	&	FN &	5.75 &	2.64 &	4.3 &	7.17 &	3.87 &	9.91 &	9.65 &	7.25 &	2.85 &	7.3	\\		   
&	&	FP+FN &	12.77 &	\textbf{7.57 }&	23.47 &	17.94 &	56.67 &	11.82 &	15.81 &	156.6 &	120.2 &	28.11	\\		   
&	&	L2 &	0.36 &	\textbf{0.104} &	0.292 &	0.516 &	0.284 &	0.796 &	0.708 &	0.552 &	0.196 &	0.56	\\	\hline\hline	 
\end{tabular}}
\end{table}

\begin{table}
\caption{\footnotesize{Simulation results for models (D)--(E) with $|\cS|=4$ and $5$.
Results in bold font mean the
value of FP+FN is the lowest or within 10\% of the lowest; the same for L2.}}
\label{table:sim:de}
\centering
\footnotesize{
\begin{tabular}{c|c|c|c|c|c|c|c|c|c|c|c}
\hline
\hline
$\varphi$ &	& TCS1 &	TCS2 &	FR &	FS &	Lasso &	ISIS &	PCS &	MC+ &	SCAD & FLASH	\\
\hline
0.5 &	FP &	0.71 &	2.4 &	22.41 &	27.86 &	58.73 &	1.21 &	2.33 &	27.94 &	111 &	26.18 	\\		   
&	FN &	0 &	0 &	0 &	1 &	1 &	3.21 &	1.65 &	0.6 &	1 &	1 	\\		   
&	FP+FN &	\textbf{0.71} &	2.4 &	22.41 &	28.86 &	59.73 &	4.42 &	3.98 &	28.54 &	112 &	27.18	\\		   
&	L2 &	\textbf{0.149} &	0.351 &	2.876 &	33.46 &	30.92 &	47.9 &	38.74 &	19.12 &	30.96 &	31.85 	\\	\hline	   
0.95 &	FP &	0.39 &	0.76 &	19.84 &	7.14 &	28.37 &	1.45 &	1.42 &	49.58 &	46.68 &	12.88 	\\		   
&	FN &	1.43 &	3.64 &	1.89 &	2.05 &	1.54 &	3.71 &	3.58 &	1.7 &	2.07 &	1.61 	\\		   
&	FP+FN &	\textbf{1.82} &	4.4 &	21.73 &	9.19 &	29.91 &	5.16 &	5 &	51.28 &	48.75 &	14.49	\\		   
&	L2 &	\textbf{26.71} &	71.17 &	76.23 &	70.87 &	65.82 &	73.73 &	71.61 &	67.07 &	69.23 &	67.21 	\\	\hline\hline	   
0.5 &	FP &	0.85 &	3.31 &	30.2 &	29.06 &	56.92 &	1.23 &	2.31 &	32.56 &	112.3 &	27.04 	\\		   
&	FN &	0.03 &	0.11 &	0.01 &	1.15 &	1.05 &	4.23 &	2.42 &	0.79 &	1.02 &	1.19 	\\		   
&	FP+FN &	\textbf{0.88} &	3.42 &	30.21 &	30.21 &	57.97 &	5.46 &	4.73 &	33.35 &	113.3 &	28.23	\\		   
&	L2 &	\textbf{0.177 }&	0.528 &	4.102 &	33.5 &	31.46 &	48.83 &	39.46 &	22.11 &	31.46 &	32.18 	\\	\hline	   
0.95 &	FP &	0.05 &	0.05 &	26.08 &	4.5 &	28.74 &	1.03 &	1.01 &	35.82 &	43.73 &	12.78 	\\		   
&	FN &	2.76 &	3.96 &	1.75 &	2.32 &	1.56 &	4.1 &	3.77 &	1.86 &	2.11 &	1.83 	\\		   
&	FP+FN &	\textbf{2.81} &	4.01 &	27.83 &	6.82 &	30.3 &	5.13 &	4.78 &	37.68 &	45.84 &	14.61	\\		   
&	L2 &	\textbf{49.89} &	71.56 &	81.1 &	69.81 &	65.9 &	76.37 &	71.88 &	66.78 &	68.76 &	67.28 	\\	\hline\hline	 
\end{tabular}}
\end{table}

\begin{table}
\caption{\footnotesize{Simulation results for model (F) with $|\cS|=10$. Results in bold font mean the
value of FP+FN is the lowest or within 10\% of the lowest; the same for L2.}}
\label{table:sim:f}
\centering\footnotesize{
\begin{tabular}{c|c|c|c|c|c|c|c|c|c|c|c|c}
\hline
\hline
 $p$	& $R^2$ &	& TCS1 &	TCS2 &	FR &	FS &	Lasso &	ISIS &	PCS &	MC+ &	SCAD & FLASH	\\
\hline
1000 &	0.3 &	FP &	2.27 &	2.08 &	13.68 &	1.65 &	23.69 &	0.87 &	6.09 &	130.6 &	23.61 &	7.97	\\		   
&	&	FN &	7.2 &	6.45 &	5.12 &	8.94 &	8.22 &	9.92 &	8.33 &	7.81 &	8.42 &	5.96	\\		   
&	&	FP+FN &	9.47 &	\textbf{8.53 }&	18.8 &	10.59 &	31.91 &	10.79 &	14.42 &	138.4 &	32.03 &	13.93	\\		   
&	&	L2 &	3.376 &	\textbf{2.579} &	3.549 &	6.487 &	6.33 &	7.577 &	5.144 &	6.654 &	6.346 &	\textbf{2.605}	\\	\hline	   
&	0.6 &	FP &	3.97 &	3.87 &	16.36 &	1.58 &	21.89 &	0.78 &	5.98 &	106.4 &	23.54 &	8.48	\\		   
&	&	FN &	4.65 &	4.11 &	4.07 &	9.1 &	8.24 &	9.89 &	8.37 &	7.88 &	8.46 &	5.22	\\		   
&	&	FP+FN &	\textbf{8.62} &	\textbf{7.98 }&	20.43 &	10.68 &	30.13 &	10.67 &	14.35 &	114.2 &	32 &	13.7	\\		   
&	&	L2 &	3.029 &	\textbf{2.515 }&	6.604 &	10.53 &	10.25 &	11.5 &	7.181 &	10.64 &	10.38 &	4.229	\\	\hline	   
&	0.9 &	FP &	5.97 &	5.17 &	14.54 &	1.77 &	20.29 &	0.83 &	6.1 &	115.2 &	20.72 &	7.73	\\		   
&	&	FN &	1.95 &	2.42 &	3.45 &	9.14 &	8.7 &	9.88 &	8.3 &	8.03 &	8.87 &	4.81	\\		   
&	&	FP+FN &	\textbf{7.92} &	\textbf{7.59} &	17.99 &	10.91 &	28.99 &	10.71 &	14.4 &	123.2 &	29.59 &	12.54	\\		   
&	&	L2 &	\textbf{0.573} &	2.055 &	5.81 &	9.555 &	9.501 &	10.65 &	8.428 &	9.736 &	9.51 &	5.428	\\	\hline\hline	   
2000 &	0.3 &	FP &	1.76 &	1.53 &	12.56 &	1.49 &	21.06 &	0.84 &	6.89 &	154.2 &	26.63 &	8.88	\\		   
&	&	FN &	8.66 &	8.25 &	7.73 &	9.48 &	8.89 &	9.9 &	8.75 &	8.37 &	8.86 &	7.06	\\		   
&	&	FP+FN &	\textbf{10.42} &	\textbf{9.78 }&	20.29 &	10.97 &	29.95 &	\textbf{10.74} &	15.64 &	162.6 &	35.49 &	15.94	\\		   
&	&	L2 &	4.774 &	\textbf{3.952} &	5.626 &	6.371 &	6.267 &	7.756 &	5.484 &	6.403 &	6.286 &	\textbf{4.27}	\\	\hline	   
&	0.6 &	FP &	3.18 &	2.51 &	16.9 &	1.62 &	20.89 &	0.85 &	6.45 &	250.1 &	29.89 &	8.46	\\		   
&	&	FN &	6.94 &	7.04 &	6.56 &	9.51 &	8.83 &	9.9 &	8.56 &	8.05 &	8.86 &	6.56	\\		   
&	&	FP+FN &	\textbf{10.12} &	\textbf{9.55} &	23.46 &	11.13 &	29.72 &	10.75 &	15.01 &	258.2 &	38.75 &	15.02	\\		   
&	&	L2 &	\textbf{2.424} &	2.9 &	5.74 &	6.891 &	6.901 &	8.071 &	6.072 &	7.013 &	6.902 &	4.79	\\	\hline	   
&	0.9 &	FP &	5.4 &	4.42 &	18.96 &	1.83 &	22.73 &	0.83 &	6.73 &	202.3 &	29.23 &	9.04	\\		   
&	&	FN &	4.29 &	3.98 &	5.17 &	9 &	8.72 &	9.92 &	8.64 &	8.25 &	8.99 &	5.86	\\		   
&	&	FP+FN &	9.69 &	\textbf{8.4 }&	24.13 &	10.83 &	31.45 &	10.75 &	15.37 &	210.6 &	38.22 &	14.9	\\		   
&	&	L2 &	\textbf{1.675 }&	\textbf{1.745} &	3.64 &	5.232 &	5.254 &	6.67 &	4.133 &	5.401 &	5.275 &	2.841	\\	\hline\hline	 
\end{tabular}}
\end{table}

Overall, compared with other methods, TCS1, TCS2 and FR achieve a high TPR  
more quickly without including too many irrelevant variables and thus tend to achieve a small L2 distance.
While the PC-simple algorithm attains a low FPR, its TPR is also low even when the significant level 
for the testing procedure is set to be high. 
For certain set-ups, Lasso or SCAD achieves a high TPR but only at the cost of a high FPR. 

Specifically, for factor models (A)--(C), it can be observed that TCS1, TCS2, FR
(combined with the extended BIC) and SCAD are superior to other methods in terms of achieving small FN, 
especially when $R^2$ is sufficiently high. 
However, the FR and SCAD tend to result in a model with too large an FP in comparison to the TCS algorithm, 
and therefore the L2 distance obtained from TCS2 is often the smallest.
This becomes more obvious as the dimensionality grows and the number of factors increases,
and the ROC curves in Figures \ref{fig:roc:a}--\ref{fig:roc:c} also support this conclusion,
as those from the TCS algorithm attain a higher TPR for a similar level of FPR. 
Note that from our extensive numerical experiments, we observed that 
increasing number of factors led to an increased chance of marginal correlation screening being misleading at the very first iteration in the sense that $\arg\max_j|X_j^T\by|\notin\cS$.
In such set-ups, the adaptive choice of $\cCj$ used by the TCS algorithm turned out to be helpful in
correctly identifying a relevant variable more often than marginal correlation screening.
Between TCS1 and TCS2, while the two perform as well as each other for the two factor models from (A), 
it is TCS2 which outperforms the other for the models with more factors. 
As for the rest of the methods, FS performs as well as FR for lower dimensionality,
and even better in terms of FP, 
but its FN is larger than that of FR as $p$ and the number of factors increase. 
Both PCS algorithm and ISIS return final models which are too small and therefore obtain large FN
and small FP; especially ISIS almost always misses the entire set of variables in $\cS$. 
Lasso is not significantly inferior to, and occasionally better than, TCS1, TCS2 and FR in terms of FN, 
but it tends to select a model with a large FP like SCAD.
While the ROC curves of MC+ and FLASH behave better than that of SCAD for certain set-ups 
(e.g. for two factor models), final selected models for these methods achieve 
larger FN. Finally, in terms of FP, FLASH tends to be better than SCAD, MC+ and Lasso. 

For models (D) and (E), the TCS algorithm and FR outperform the rest when $\varphi=0.5$, 
rapidly identifying all the relevant variables before the FPR reaches $2.5|\cS|/p$ 
(left column of Figure \ref{fig:roc:de}).
However, when correlations among the variables increase with $\varphi=0.95$, 
ROC curves show that TCS1 is the only method that can identify all the relevant variables 
(right column of Figure \ref{fig:roc:de}). 
Other methods, including TCS2 and FR, often neglect to include $X_4$ due to its high correlations with the other variables, $\sqrt{\varphi}$ being almost 0.975.
We note that while the ROC curves indicate that very often all the relevant variables are recovered by TCS1, the models selected by the extended BIC leave out some of them.
Since the final models from TCS1 tend to contain the smallest number of noisy variables, we conclude that 
the extended BIC tends to choose final models which are too small for these particular examples. 
The rest of methods behave similarly as in the case of factor models; 
while Lasso, MC+, SCAD and FLASH achieve relatively small FN, the FP of their final models is too large 
and therefore they end up with a larger L2 distance than that of TCS1. 

For the examples generated from the Leukemia dataset (model (F), Figure \ref{fig:roc:f}), 
the TCS algorithm with either of the rescaling methods always performs the best,
with its ROC curves always dominating those of others.
FR performs the second best and then follows FLASH. The remaining methods
are not able to identify as many relevant variables as the TCS algorithm or FR even for a high FPR.
The results reported in Table \ref{table:sim:f} also support this observation, 
where it is clear that the smallest FP and L2 distance are attained by either TCS1 or TCS2. 
Sometimes FR outperforms the two in terms of FN but TCS1 or TCS2 still achieves a smaller L2 distance,
which implies that TCS algorithm, when combined with the extended BIC, 
can pick up a smaller model that better mimics the true coefficient vector than that yielded by 
FR with the same criterion. 
Interestingly, when it comes to the final model, FLASH achieves similar FN and much smaller FP than FR.

We have observed that the two rescaling methods sometimes select variables in different orders, 
although it does not necessarily imply that the resulting models are different.
Overall, TCS2 performs better than TCS1 except for the examples from (D)--(E). 
In these two models, the variables $X_1, \ldots, X_p$ have a very special correlation structure
in that e.g. $X_4$, a significant variable, can often appear uncorrelated with $\by$ in marginal correlation screening. 
Since TCS1 involves the term $\Vert(\bI_n-\Pi_\cA)X_j\Vert_2^2$ in the denominator of the tilted correlation, as opposed to
the term $\Vert(\bI_n-\Pi_\cA)X_j\Vert_2$ in TCS2, it is better at picking up $X_4$ than TCS2.
In the factor model examples, while the overall correlations among the variables are high,
such ``masking'' does not take place as often among the significant variables. 
Therefore we conclude that unless the correlations are particularly high, 
TCS2 usually performs well.

\section{Boston housing data analysis}
\label{sec:boston}

In this section, we apply the TCS algorithm as well as the methods used in the simulation study in Section \ref{sec:sim} to the Boston housing data, which was previously used to compare the performance of different regression techniques
e.g. in \citet{radchenko2011}.
Originally, the dataset contains $13$ variables which may have influence over the house prices.
As in \citet{radchenko2011}, we include the interaction terms between the variables in the analysis 
such that the data has $p=91$ variables and $n=506$ observations. 
Note that, due to the way the variables are produced, there exist large sample correlations across the columns of the design matrix $\bX$. 
We split the data into three with $n_1=91 (=p)$, $n_2=46$ and $n_3=369$ observations each, and use the first $n_1$ observations as the training data (to compute a solution path for each method), the next $n_2$ observations as the validation data (to choose the solution along the path that minimises the sum of the squared residuals for each method), and the last $n_3$ for computing the test error 
($n_3^{-1}\Vert\by-\hat{\by}\Vert_2^2$).
Random splitting of the data is repeated 20 times and
Table \ref{table:boston} reports the average test error and number of selected variables,
which shows that TCS2 achieves the minimum test error with the fewest variables in the model
(except for the PC-simple algorithm).
TCS1 also performs second best with more variables selected during the validation step. 
FR performs well in terms of both test error and the number of selected variables,
and then follows FLASH. 
We note that the PC-simple algorithm chooses too few variables to describe the data well,
while the non-convex penalty algorithms (MC+, SCAD) tend to include many more variables than the rest. 

\begin{table}
\caption{Boston housing data: test errors and the number of selected variables averaged over 20 test data sets.}
\label{table:boston}
\centering
\begin{tabular}{c|c|c|c|c|c|c|c}
\hline
\hline
 & TCS1 & TCS2 & FR & PC-simple & MC+ & SCAD & FLASH
\\
\hline 
test error & 27.03 & 26.43 & 33.10 & 32.47 & 36.47 & 34.95 & 30.14
\\
number of variables & 19.5 & 13.5 & 16.0 & 2.0 & 83.5 & 36.0 & 26.0
 \\
\hline
\hline
\end{tabular}
\end{table}

\section{Conclusions}
\label{sec:conclusion}

In this paper, we proposed a new way of measuring strength of association between the variables and the response
in a linear model with a possibly large number of covariates, by adaptively taking into account  
correlations among the variables. We conclude by listing the new contributions made in this paper.

\begin{itemize}
\item Although tilting is not the only procedure which measures the association 
between a variable and the response conditional on other variables, 
its selection of the conditioning variables is a step further from 
simply using the current model itself or its sub-models, as is done in existing 
iterative algorithms. The hard-thresholding step in the tilting procedure enables an adaptive choice of 
the conditioning subset $\cCj$ for each variable $X_j$.
Recalling the decomposition of the marginal correlation in (\ref{marginal:corr}),
this adaptive choice can be seen as a vital step in capturing the contribution of each variable to the response.
Also, in the case $\cCj=\emptyset$, tilted correlation is identical to marginal correlation,
which can be viewed as ``adaptivity'' of our procedure.

\item We propose two rescaling factors to obtain the tilted correlation $\cjs$.
Rescaling 1 ($\lamj$) is also adopted by the forward regression and rescaling 2 ($\Lamj$) is also adopted by the PC-simple algorithm,
yet tilting is the only method to meaningfully use both rescaling factors and 
our theoretical results in Section \ref{sec:prop:tilt} are valid for either of the two factors.
It would be of interest to identify a way of combining the two rescaling methods,
which we leave as a topic for future research.

\item The separation of relevant and irrelevant variables,
achieved by tilted correlation (as in our Theorems \ref{thm:one}--\ref{thm:three}), cannot always be achieved by marginal correlation,
and similar results to these theorems have not been reported previously to the best of our knowledge.

\item
The proposed TCS algorithm is designed to fully exploit the theoretical properties of the tilted correlation,
and in particular its asymptotic consistency in separating between the relevant and irrelevant variables.
Although we have not yet been able to demonstrate the model selection consistency of the TCS algorithm, 
numerical experiments confirm its good performance in comparison with other well-performing methods, 
showing that it can achieve high true positive rate without including many irrelevant variables. 
The algorithm is simple, easy to implement and does not require the use of advanced computational
tools.

\end{itemize}

Ending on a slightly more general note, since correlation is arguably the most widely used 
statistical measure of association, we would expect our tilted correlation (which can be viewed 
as an ``adaptive'' extension of standard correlation) to be more widely applicable in various
statistical contexts beyond the simple linear regression model.

\vspace{10pt}

{\bf Acknowledgements}

We would like to thank the Joint Editor, Associate Editor and two Referees for very helpful comments which
led to a substantial improvement of this manuscript.

\vspace{10pt}

\appendix
\section{Proof of Theorem \ref{thm:one}}
\label{appendix:one}

The proof of Theorem \ref{thm:one} is divided into Steps 1--3.
Recalling the decomposition of $X_j^{*T}\by$ in (\ref{decom:tilt:corr}), 
we first control the inner product between $\xjs$ and $\bep$ uniformly over all $j$ in Step 1.
In Steps 2--3, we control the second summand
$I\equiv\sum_{k\in\cS\setminus\cCj, k\ne j}\beta_kX_j^T(\bI_n-\Pij)X_k$ for $j$ falling into two different categories,
and thus derive the result.

\begin{itemize}
\item[Step 1]
For $\bep\sim\cN_n(\bzero, n^{-1}\sig^2\cdot\bI_n)$, we observe that,
with probability converging to 1, $\max_{1\le j \le p}|\langle \bep, Z_j \rangle| \le \sig\sqrt{2\log p/n}$ 
for $Z_1, \ldots, Z_p\in\R^n$ having unit norm as $\Vert Z_j \Vert_2=1$.
From (A2), we have $\sig\sqrt{2\log p/n}\le Cn^{-\gamma}$ for some $C>0$,
and from (A5), $\Vert\xjs\Vert_2>\sqrt{\alpha}>0$.
Therefore by defining $\cE_0=\{\max_j|X_j^{*T}\bep|<C n^{-\gamma}\}$,
it follows that $\p(\cE_0)\to 1$.

\item[Step 2]
In this step, we turn our attention to those $j$ whose corresponding $\cCj$ satisfy 
$\cS\setminus\{j\}\subseteq\cCj$ and thus the corresponding $I=0$ and 
$X_j^{*T}\by=\beta_j(1-\aj)+X_j^{*T}\bep$.
\begin{description}
\item[Rescaling 1.] 
With the rescaling factor $\lamj=(1-\aj)$ which is bounded away from 0 by (A5), 
it can be shown that if such $j$ belongs to $\cS$, its tilted correlation satisfies
$\cjs(\lamj)/\beta_j\to 1$ on $\cE_0$, as $|\beta_j|\gg n^{-\mu}$.
On the other hand, if $j\notin\cS$, we have $\beta_j(1-\aj)=0$ which leads to 
$n^\mu\cdot\cjs(\lamj)\le n^\mu\cdot Cn^{-\gamma}\to 0$ on $\cE_0$.
\item[Rescaling 2.]
Note that $j$ whose $\cCj$ include all the members of $\cS$ cannot be a member of $\cS$ itself,
and in this case, $(\bI_n-\Pij)\by$ is reduced to $(\bI_n-\Pij)\bep$. 
Since (A3) assumes that each $\cCj$ has its cardinality bounded by $Cn^\xi$, 
it can be shown that $\p\left(\max_j\Vert\Pij\bep\Vert_2 \le C'n^{-(\gamma-\xi/2)}\right)\to 1$ 
for some $C'>0$, similarly to Step 1.
Also, Lemma 3 from \citet{fan2008} implies that
$\p\left(\sig^{-2}\cdot\Vert\bep\Vert_2^2 < 1-\omega\right)\to 0$ for any $\omega\in(0, 1)$.
Combining these observations with (A1) and (A4), we derive that
$1-\ajy=\Vert(\bI_n-\Pij)\bep\Vert_2^2/\Vert\by\Vert_2^2\ge Cn^{-\delta}$
with probability tending to 1, and eventually we have $\Lamj\ge C'n^{-\delta/2}$ from (A5).
Therefore, if $\cS\subseteq\cCj$ for some $j\notin\cS$, 
its corresponding tilted correlation satisfies 
$n^\mu\cdot\cjs(\Lamj)\le n^\mu\cdot Cn^{-(\gamma-\delta/2)}\to 0$ on $\cE_0$.

In the case of $\cS\nsubseteq\cCj$, we can derive from (A6) that for such $j$,
$\Vert(\bI_n-\Pij)\by\Vert_2^2/\Vert\by\Vert_2^2=1-\ajy\gg n^{-\kappa}$,
which, combined with (A5), implies that $\Lamj\gg n^{-\kappa/2}$.
Then the following holds for such $j$ on $\cE_0$: 
$n^\mu\cdot|\cjs(\Lamj)|\ge n^\mu\cdot C|\beta_j|\to\infty$ if $j\in\cS$, 
while $n^\mu\cdot\cjs(\Lamj)\le n^\mu\cdot Cn^{-(\gamma-\kappa/2)}\to 0$ if $j\notin\cS$.
\end{description}

\item[Step 3]
We now consider those $j\in\cJ$ for which $\cS\setminus\{j\}\nsubseteq\cCj$ and 
consequently the corresponding term $I \ne 0$ in general.
From (A3) and Condition \ref{cond:one}, we derive that for each $j$, 
there exists some $C>0$ satisfying the following for all $k\in\cS\setminus\cCj, \ k\ne j$,
\begin{eqnarray}
|X_j^T(\bI_n-\Pij)X_k|\le|X_j^TX_k|+|(\Pij X_j)^TX_k|\le Cn^{-\gamma}.
\label{appendix:one:step:three}
\end{eqnarray}
Then from (A1) and (A4), we can bound $I$ as $|I|\le C'n^{-(\gamma-\delta)}$.
Also when $\cS\setminus\{j\}\nsubseteq\cCj$, (A5)--(A6) imply that $\Lamj\gg n^{-\kappa/2}$.
In summary, we can show that the following claims hold on $\cE_0$, similarly as in Step 2:
if $j\notin\cS$, with either of the rescaling factors, 
$n^\mu\cdot\cjs(\lamj) \le n^\mu\cdot Cn^{-(\gamma-\delta-\kappa/2)}\to 0$,
whereas if $j\in\cS$, its coefficient satisfies $|\beta_j|\gg n^{-\mu}$ and
therefore $n^\mu\cdot|\cjs|\ge n^\mu\cdot C|\beta_j|\to\infty$ with $\cjs(\lamj)/\beta_j\to 1$ for $j\in\cS$.
\hfill$\square$
\end{itemize}

\subsection{An example satisfying Condition \ref{cond:one}}
\label{appendix:one:ex}

In this section, we verify the claim made in Section \ref{sec:scen:one},
which states that Condition \ref{cond:one} holds with probability tending to 1
when each column $X_j$ is generated independently as a random vector on an $n$-dimensional unit sphere.
We first introduce a result from modern convex geometry reported in Lecture 2 of \citet{ball1997},
which essentially implies that, as the dimension $n$ grows, 
it is not likely for any two vectors on a $n$-dimensional unit sphere to be within a close distance to each other. 
\begin{lem}
Let $S^{n-1}$ denote the surface of the Euclidean ball $B_2^n=\{\bx\in\R^n: \ \sum_{i=1}^nx_i^2\le 1\}$ 
and $\bu\in\R^n$ be a vector on $S^{n-1}$ such that $\Vert\bu\Vert_2=1$. 
Then the proportion of spherical cone defined as $\{\bv\in S^{n-1}: \ |\bu^T\bv|\ge \omega\}$ for any $\bu$  
is bounded from above by $\exp(-n\omega^2/2)$.
\label{lem:one}
\end{lem}
We first note that any $X_k, \ k\ne j$ can be decomposed as the summation of
its projection onto $X_j$ and the remainder, i.e., $X_k=c_{j, k}X_j+(\bI_n-X_jX_j^T)X_k$. 
Then 
\[
(\Pij X_j)^TX_k=c_{j, k}(\Pij X_j)^TX_j+\left((\bI_n-X_jX_j^T)\Pij X_j\right)^TX_k,
\]
and for $k\in\cS\setminus\cCj, \ k\ne j$, the first summand is bounded from above by $\aj\cdot\pin\le C_1n^{-\gamma}$.
As for the second summand, note that
\[
\Vert(\bI_n-X_jX_j^T)\Pij X_j\Vert_2^2=(\Pij X_j)^T(\bI_n-X_jX_j^T)\Pij X_j=\aj(1-\aj),
\]
and thus $\bw=\left\{\aj(1-\aj)\right\}^{-1/2}\cdot(\bI_n-X_jX_j^T)\Pij X_j$ satisfies $\bw\in S^{n-1}$.
Then the probability of $|\bw^TX_k|>Cn^{-\gamma}$ for any $k\in\cS\setminus\cCj, \ k\ne j$ is 
bounded from above by the proportion of the spherical cone 
$\left\{X_k\in S^{n-1}: \ |\bw^TX_k|>Cn^{-\gamma} \right\}$ in the unit sphere $S^{n-1}$.
Applying Lemma \ref{lem:one}, 
we can show that such proportion is bounded by $\exp\left(-C^2n^{1-2\gamma}/2\right)$ for each $j$ and $k$.
Therefore, we can find some $C>0$ satisfying
\[
\p\left(\max_{j\in\cJ; \ k\in\cS\setminus\cCj, \ k\ne j}|(\Pij X_j)^TX_k| > Cn^{-\gamma}\right)\ge
1-p|\cS|\exp\left(-C'n^{1-2\gamma}/2\right),
\] 
where the right-hand side converges to 1 from assumptions (A1)--(A2).

\section{Proof of Theorem \ref{thm:two}}
\label{appendix:two}

For those $j\in\cK=\cS\cup\left\{\cup_{j\in\cS}\cCj\right\}$, 
Condition \ref{cond:three} implies that $\cC_k\cap\cCj=\emptyset$ if $k\in\cS\setminus\cCj$.
Then from (A3), we have $\Vert\Pij X_k\Vert_2\le Cn^{-(\gamma-\xi/2)}$ and therefore
\begin{eqnarray*}
\left\vert X_j^T(\bI_n-\Pij)X_k \right\vert=\left\vert X_j^TX_k-(\Pij X_j)^T\Pij X_k\right\vert
\le Cn^{-\gamma}+C'n^{-(\gamma-\xi/2)},
\end{eqnarray*}
which leads to 
\begin{eqnarray}
\left\vert\sum_{k\in\cS\setminus\cCj, k\ne j}\beta_kX_j^T(\bI_n-\Pij)X_k\right\vert
=O\left(n^{-(\gamma-\delta-\xi/2)}\right)
\label{cond:two:eq}
\end{eqnarray}
for all $j\in\cK$.
Using Step 1 of Appendix \ref{appendix:one}, we derive that
\[
\cE_{01}=\left\{\max_{j\in\cK}\left\vert
\sum_{k\in\cS\setminus\cCj, k\ne j}\beta_kX_j^T(\bI_n-\Pij)X_k+X_j^T(\bI_n-\Pij)\bep\right\vert
\le Cn^{-(\gamma-\delta-\xi/2)}\right\}
\]
satisfies $\p(\cE_{01})=\p(\cE_0)\to 1$.
Since $\mu+\kappa/2<\gamma-\delta-\xi/2$, we have $n^\mu\cdot\cjs\to 0$ for $j\notin\cS$ on $\cE_{01}$,
whereas $n^\mu\cdot|\cjs|\to\infty$ and $\cjs(\lamj)/\beta_j\to 1$ for those $j\in\cS$. 
Therefore the dominance of tilted correlations for $j\in\cS$ over those for $j\in\cK\setminus\cS$ follows. 
\hfill$\square$

\section{Proof of Theorem \ref{thm:three}}
\label{appendix:three}

Compared to Condition \ref{cond:two}, 
Condition \ref{cond:three} does not require any restriction on $\cCj\cap\cC_k$ when both $X_j$ and $X_k$ are relevant,
although it has an additional assumption (C2).
Since $n^\mu\cdot|\beta_j|(1-\aj)\to\infty$ for $j\in\cS$ from (A4)--(A5),
(C2) implies that for any $j\in\cS$, non-zero coefficients $\beta_k, \  k\in\cS\setminus\cCj$ 
do not cancel out all the summands in the following to 0,
\begin{eqnarray}
X_j^T(\bI_n-\Pij)\bX_\cS\bbeta_\cS=\beta_j(1-\aj)+\sum_{k\in\cS\setminus\cCj, k\ne j}\beta_kX_j^T(\bI_n-\Pij)X_k.
\nonumber 
\end{eqnarray}
If (\ref{cond:two:eq}) in Appendix \ref{appendix:two} holds, (C2) follows and therefore it can be 
seen that Condition \ref{cond:two} is stronger than Condition \ref{cond:three}. 

On the event $\cE_0$ (Step 1 of Appendix \ref{appendix:one}),
$|X_j^T(\bI_n-\Pij)\by| \gg n^{-\mu}$ for $j\in\cS$ under (C2) and therefore
the tilted correlations of relevant variables satisfy $|\cjs| \gg n^{-\mu}$ with either of the rescaling factors.
In contrast, for $j\in\cK\setminus\cS$, we can use the arguments in Appendix \ref{appendix:two} 
to show that $n^\mu\cdot\cjs\to 0$.
\hfill$\square$

\section{Study of the assumptions (A5) and (A6)}
\label{append:one}

In this section, we show that the assumptions (A5) and (A6) are satisfied
under the following condition from \citet{wang2009}.
Let $\lam_*(\bA)$ and $\lam^*(\bA)$ represent the smallest and the largest eigenvalues of an arbitrary positive definite matrix $\bA$, respectively. 
\begin{itemize}
\item Both $\bX$ and $\bep$ follow normal distributions. 
\item There exist two positive constants $0<\tau_*<\tau^*<\infty$ such that
$\tau_*<\lam_*(\bSig) \le \lam^*(\bSig) < \tau^*$, where $\cov(\bx_i)=\bSig$ for $i=1, \ldots, n$. 
\end{itemize}
Then, \citet{wang2009} showed that there exists $\eta\in(0, 1)$ satisfying
\begin{eqnarray}
\tau_* \le \min_{\cD}\lam_*(\bX_{\cD}^T\bX_{\cD}) \le \max_{\cD}\lam^*(\bX_{\cD}^T\bX_{\cD}) \le \tau^*
\label{lem:wang}
\end{eqnarray}
with probability tending to 1, for any $\cD\subset\{1, \ldots, p\}$ with $|\cD| \le n^\eta$.
We use the result from (\ref{lem:wang}) in the following arguments.
\begin{itemize}
\item[(A5)]
Recalling the notations $\tXj=\bX_{\cCj}$ and $\Pij=\tXj(\tXj^T\tXj)^{-1}\tXj^T$, we have
\begin{eqnarray*}
1-X_j^T\Pij X_j = \left\Vert X_j - \tXj(\tXj^T\tXj)^{-1}\tXj^TX_j \right\Vert^2_2.
\end{eqnarray*}
We let $\boldsymbol{\theta}=(\tXj^T\tXj)^{-1}\tXj^TX_j$ and assume that $\xi$ from assumption (A3) satisfies $\xi \le \eta$ such that, by applying (\ref{lem:wang}), we obtain the following;
\begin{eqnarray*}
&&1-X_j^T\Pij X_j = (1, \boldsymbol{\theta}) \left(X_j, \tXj\right)^T\left(X_j, \tXj\right)(1, \boldsymbol{\theta})^T 
\\
&\ge& 
(1, \boldsymbol{\theta}) \lam_*\left((X_j, \tXj)^T(X_j, \tXj)\right)(1, \boldsymbol{\theta})^T
\ge
(1+\Vert\boldsymbol{\theta}\Vert^2_2)\tau_* \ge \tau_*>0.
\end{eqnarray*}

\item[(A6)] 
We note the link between (A6) and the asymptotic identifiability condition 
for high-dimensional problems first introduced in \citet{chen2008}.
The condition can be re-written as
\begin{eqnarray}
\lim_{n\to\infty}\min_{\cD\subset\cJ, |\cD|\le|\cS|, \cD\ne\cS} 
n(\log n)^{-1}\cdot \frac{\Vert(\bI_n-\Pi_\cD)\bX_\cS\bbeta_\cS\Vert_2^2}{\Vert\bX_\cS\bbeta_\cS\Vert_2^2} \to\infty,
\label{asym:iden}
\end{eqnarray}
after taking into account the column-wise normalisation of $\bX$.
Although the rate $n^\kappa$ is less favourable than $n(\log n)^{-1}$,
following exactly the same arguments as in Section 3 of \citet{chen2008},
we are able to show that (A6) is implied by the condition in (\ref{lem:wang}).
That is, letting $\boldsymbol{\theta}=(\tXj^T\tXj)^{-1}\tXj^T\bX_\cS\bbeta_\cS$, we have
\begin{eqnarray}
&& n^\kappa\cdot \frac{\Vert(\bI_n-\Pij)\bX_\cS\bbeta_\cS\Vert_2^2}{\Vert\bX_\cS\bbeta_\cS\Vert_2^2} 
\ge n^\kappa\inf_{j\notin\cS}\frac{\Vert
\bX_{\cS\cap\cCj^c}\bbeta_{\cS\cap\cCj^c}-\tXj\boldsymbol{\theta}\Vert_2^2}{\Vert\bX_\cS\bbeta_\cS\Vert_2^2}
\nonumber \\
&\ge& Cn^{\kappa-2\delta}\inf_{j\notin\cS}\left\{\left(\bbeta^T_{\cS\cap\cCj^c}, -\boldsymbol{\theta}\right)^T
\bX_{\cS\cup\cCj}^T\bX_{\cS\cup\cCj}\left(\bbeta^T_{\cS\cap\cCj^c}, -\boldsymbol{\theta}\right)\right\}
\nonumber \\
&\ge& Cn^{\kappa-2\delta}\lam_*(\cS\cup\cCj)\Vert\bbeta_{\cS\cap\cCj}\Vert_2^2
\label{eq:one}
\end{eqnarray}
for some positive constant $C$, where the second inequality is derived under the assumptions (A1) and (A4).
Then a constraint can be imposed on the relationship between $\kappa$, $\delta$ and $\xi$ such that the right-hand side of the above (\ref{eq:one}) diverges to infinity. 
\end{itemize}

\bibliographystyle{asa}
\bibliography{dbib}

\end{document}